\pdfoutput=1

\documentclass[twocolumn]{aastex701}

\usepackage{amsmath}
\graphicspath{{figs/}{./}}

\newcommand{\IE}{\ensuremath{I_\mathrm{E}}}
\newcommand{\YE}{\ensuremath{Y_\mathrm{E}}}
\newcommand{\JE}{\ensuremath{J_\mathrm{E}}}
\newcommand{\HE}{\ensuremath{H_\mathrm{E}}}
\newcommand{\Pg}{\ensuremath{P_\mathrm{gaia}}}
\newcommand{\Pc}{\ensuremath{P_\mathrm{ctrl}}}
\newcommand{\kms}{km\,s$^{-1}$}
\newcommand{\masyr}{mas\,yr$^{-1}$}

\shorttitle{Brown dwarfs in Blanco 1 with Euclid DR1}
\shortauthors{Mart\'in et al.}

\begin{document}

\title{Euclid DR1 Legacy Science on Ultracool Dwarfs: Brown Dwarfs in the Blanco~1 Open Cluster.}

\author[orcid=0000-0002-1208-4833,gname={Eduardo L.},sname={Mart\'in}]{E.~L.~Mart\'in}
\affiliation{Instituto de Astrof\'isica de Canarias, Calle V\'ia L\'actea s/n, 38204 San Crist\'obal de La Laguna, Tenerife, Spain}
\email[show]{ege@iac.es}
\correspondingauthor{E.~L.~Mart\'in}

\author[sname=Tata]{R.~Tata}
\affiliation{Department of Physics and Astronomy, Ohio University, Athens, OH 45701, USA}
\email{tatar@ohio.edu}

\author[orcid=0000-0001-6023-4974,sname={\v{Z}erjal}]{M.~{\v Z}erjal}
\affiliation{Instituto de Astrof\'isica de Canarias, Calle V\'ia L\'actea s/n, 38204 San Crist\'obal de La Laguna, Tenerife, Spain}
\email{mzerjal@iac.es}

\author[orcid=0000-0003-0316-2956,sname={Olivares Romero}]{J.~Olivares~Romero}
\affiliation{Departamento de Inteligencia Artificial, Universidad Nacional de Educaci\'on a Distancia (UNED), Madrid, Spain}
\email{jolivares@dia.uned.es}

\author[orcid=0009-0001-9574-8585,sname=Sedighi]{N.~Sedighi}
\affiliation{Instituto de Astrof\'isica de Canarias, Calle V\'ia L\'actea s/n, 38204 San Crist\'obal de La Laguna, Tenerife, Spain}
\email{nafise.sedighi@iac.es}

\author[orcid=0000-0002-5378-2240,sname=Tsilia]{S.~Tsilia}
\affiliation{Instituto de Astrof\'isica de Canarias, Calle V\'ia L\'actea s/n, 38204 San Crist\'obal de La Laguna, Tenerife, Spain}
\email{stella@iac.es}

\author[orcid=0000-0002-3612-8968,sname=Lodieu]{N.~Lodieu}
\affiliation{Instituto de Astrof\'isica de Canarias, Calle V\'ia L\'actea s/n, 38204 San Crist\'obal de La Laguna, Tenerife, Spain}
\email{nlodieu@iac.es}

\author[orcid=0000-0003-2150-0787,sname=Vitas]{N.~Vitas}
\affiliation{Instituto de Astrof\'isica de Canarias, Calle V\'ia L\'actea s/n, 38204 San Crist\'obal de La Laguna, Tenerife, Spain}
\email{nikola.vitas@iac.es}

\begin{abstract}
A 35.6\,deg$^2$ field of Euclid Data Release~1 that covers part of the corona and tidal tails of the young (100--130\,Myr) nearby (234\,pc) open cluster Blanco~1  
is presented. We select 22 ultracool dwarfs from  
Euclid photometry and slitless spectra, and measure proper motions over a median baseline of 13\,yr  
by combining Euclid astrometry with VIKING, KiDS and Rubin positions. Ten objects,  are comoving late-M and L dwarf candidates with membership probabilities of at least 0.5 under two field models. Nine of them are typed from their Euclid spectra, and the tenth is a candidate only if its colour-based type is adopted.
Brown dwarf evolutionary models give masses of 0.03--0.06\,$M_\odot$ for the cluster age. The substellar population is about twice that predicted by a lognormal mass function scaled to the cluster's Gaia members.  
The remaining 12 objects are classified as late-M, L and T dwarfs with low probability of membership in the cluster.  
Using the Euclid sample to calibrate a search over 246\,deg$^2$ of VIKING and KiDS photometry and motions, we identify 60 candidates, with about seven expected by chance.
In the core region of Blanco 1 archival Subaru Hyper Suprime-Cam images and motions over 2011--2025 yield 35 candidates of types M7 or later, at least 20 of which are kinematic members at 95\% confidence.  
These results extend the spectroscopic sequence of Blanco~1 into the L-type brown dwarfs and the list of candidate members over the outskirts and tidal tails of the cluster.

\end{abstract}


\section{Introduction}\label{sec:intro}

How many brown dwarfs a star cluster forms and how many it keeps are separate questions. The first concerns the
substellar initial mass function, which nearby young clusters constrain well because their members share an age, a
distance and a composition \citep[e.g.][]{1998Martin,2001Bejar,Moraux2007,2010Bastian,Peña2012}. The second is set by dynamics. Two-body encounters push some stars to the cluster outskirts, and members that escape leave slowly through the Lagrange points, where
the Galactic tide draws them into a leading and a trailing tail \citep{Kupper2008}. As a cluster ages it should
therefore lose a larger share of its brown dwarfs than of its stars, and the lost brown dwarfs should sit in its
outskirts and tails.

Surveys of the substellar population have concentrated on cluster cores, where the surface density of members is
highest and a few deep pointings cover the cluster. The outskirts and tails of nearby clusters have since been mapped
over tens of square degrees with Gaia astrometry \citep{2019Lodieu,Zhang2020,Meingast2021,Pang2021,Kos2024,Risbud2025,Jadhav2025}, but
these maps end at Gaia's faint limit near $G=21$. At a few hundred parsecs that limit falls on late-M stars, leaving the %
evaporated substellar population unmapped. Searching for it means finding objects that are faint and
red, spread over tens of square degrees and outnumbered by field dwarfs and background galaxies, and then showing that
they move with the cluster.

We extend the search for brown dwarfs in the Blanco~1 open cluster. From its discovery \citep{1949Blanco}, it was noted to be a nearby (250\,pc) open cluster well populated with stars with spectral types from B5 to G0 and located at a high Galactic latitude of $-79\arcdeg$.
Lithium
depletion boundary ages of 114--132\,Myr \citep{Cargile2010,Juarez2014} and isochrone ages of 100--105\,Myr %
\citep{CantatGaudin2020,Pang2021} place it at an age where brown dwarfs are still bright enough to detect at this
distance. A deep CFHT survey of 2.3\,deg$^2$ around the core, estimated to contain 30--40 brown dwarf members, %
measured the mass function from 3\,$M_\odot$ down to about 30 Jupiter masses, and the radial distribution of its
candidates already shows mass segregation \citep{Moraux2007}. The core brown dwarfs have since been followed with photometry and astrometry
\citep{Platais2011,Casewell2012}. Gaia revealed two tidal tails, each extending 50--60\,pc from the centre %
\citep{Zhang2020}, later recovered in several independent catalogues \citep{Pang2021,Meingast2021,Risbud2025} and in
stellar rotation periods \citep{Sha2024}. \citet{Vaher2023} traced stars with measured radial velocities back to the
cluster and identified 19 that escaped from it. All of these tail and escape studies stop at the Gaia limit. %

Euclid \citep{Mellier2025} can detect these objects below the Gaia limit with its visible imager (VIS) and Near Infrared Spectrometer and Photometer (NISP). Its VIS imager \citep{Cropper2025} records a single broad
optical band at high spatial resolution, and its NISP instrument \citep{Jahnke2025} records three near-infrared bands and
slitless spectra over the whole field of view. The first Euclid data have already been used to
find ultracool substellar objects: free-floating planetary-mass objects in the nearby star-forming regions in Orion \citep{Martin2025,2026Barrado} and Taurus \citep{2025Bouy} in the Early Release Observations (ERO), and photometric and
spectroscopic samples in the deep fields of the first quick data release \citep{DominguezTagle2025,2026Mohansan,Zerjal2026},
complemented by ground-based spectra \citep{ZhangJY2024} taken in preparation for Euclid. Euclid spectra have also confirmed an L3 dwarf near the planetary-mass limit that lies close to the embedded cluster NGC~2023 but whose proper motion is consistent with the neighbouring cluster NGC~2024 \citep{Tsilia2026}. The cosmological Euclid surveys avoid most nearby young clusters, but Data Release~1 (DR1) includes an area of 36\,deg$^2$ that lies %
 1.0--8.7\arcdeg\ west of the core of the Blanco~1 open cluster.

We use this area to test whether Euclid's deep near-infrared imaging and slitless spectroscopy, combined with ground-based astrometry, can identify substellar members outside the cluster core. We can also address what
such members say about how many brown dwarfs the cluster has kept. Because the field covers none of the core and only part of the region inside the tidal radius (Section~\ref{sec:cluster}), it samples the population in the outskirts.
We also made a spectroscopic selection directly from the NISP spectra, without photometric filtering, using the method of \citet{DominguezTagle2025}.
This selection contains six late-M and L dwarfs and three T dwarfs, including the T0 dwarf R3 (Table~\ref{tab:named}). We selected R1--R7 because their spectral types and magnitudes suggested cluster distances, and T2 and T5 because they are T dwarfs. We test this sample alongside a systematic search of the whole field. In Table~\ref{tab:named} and the figures, R1--R7, T2 and T5 identify the spectroscopic selection, while C1--C9, N1--N3 and B2 identify candidates from the systematic search; B2 is in its bluer boundary slice. Eight of the nine move differently from the cluster, and because an escaped member would
also move differently, Section~\ref{sec:traceback} studies whether any of the eight could have left Blanco~1 at a speed typical of escape.

Sections~\ref{sec:cluster}--\ref{sec:methods} describe the cluster, observations and membership analysis. We assess the candidates and the comoving population in Section~\ref{sec:results}, test an escape origin for the non-comoving objects in Section~\ref{sec:traceback}, and discuss the implications in Section~\ref{sec:discussion}. Section~\ref{sec:prospects} extends the selection to ground-based surveys and deeper Subaru images of the core, followed by the conclusions in Section~\ref{sec:conclusions}.

\section{Blanco~1 and the Euclid field}\label{sec:cluster}

\subsection{Adopted cluster parameters}\label{sec:cluster_params}

We adopt the cluster parameters in Table~\ref{tab:cluster}: the central position, parallax, mean proper motions, tidal radius, Jacobi radius and mass from \citet{HuntReffert2023}, and the radial velocity from \citet{Babusiaux2018}. The Jacobi radius, $r_J=2.60\arcdeg$ (10.7\,pc), is the distance to the Lagrange points for a Jacobi mass of $465\pm46\,M_\odot$. Other published tidal radii lie between 8.5 and 10.2\,pc \citep{Zhang2020,Pang2021,Meingast2021}. %
The proper motions of the \citet{HuntReffert2023} members have an intrinsic dispersion of (0.37, 0.30)\,\masyr, or %
0.38\,\kms\ per component at the cluster distance. Bright members provide an independent estimate of the systemic radial velocity; we fitted their velocities with a broad component for binaries. We measure $5.99\pm0.07$\,\kms\ from the 128 Gaia-ESO members of \citet{Jackson2022}. The 68 members in GALAH DR4 \citep{Buder2025} give $6.06\pm0.15$\,\kms, with a one-dimensional dispersion of 0.4\,\kms. Both estimates fall within the radial-velocity range tested in the traceback (Section~\ref{sec:traceback}). The Gaia DR3 velocities of the members inside $r_t$ give $5.75\pm0.18$\,\kms, 0.03\,\kms\ %
below that range and well within its error (all three errors are bootstrap errors of the fitted mean). We therefore test both the adopted Gaia DR2 mean of 5.78\,\kms, at the low end of these determinations, and 6.23\,\kms\ in the traceback. Because the field lies several degrees from the centre, the motion %
expected for a member differs from the mean cluster motion by up to 0.9\,\masyr\ through projection; at the field %
centroid it is $(\mu_{\alpha*},\mu_\delta)=(19.2,\,1.8)$\,\masyr\ at 236\,pc. All comoving tests below use this %
projected motion at each object's own position.

\begin{deluxetable}{llll}
\tabletypesize{\footnotesize}
\tablecaption{Adopted parameters of Blanco~1.\label{tab:cluster}}
\tablehead{\colhead{Quantity} & \colhead{Value} & \colhead{Unit} & \colhead{Reference}}
\startdata
Centre $\alpha$, $\delta$ (ICRS) & 0.915, $-$30.011 & deg & \citet{HuntReffert2023} \\ %
Galactic $l$, $b$ & 14.76, $-$79.13 & deg & \citet{HuntReffert2023} \\ %
Distance (median, 16--84\%) & 234.4 (234.3--234.5) & pc & \citet{HuntReffert2023} \\ %
Parallax & 4.232 $\pm$ 0.004 & mas & \citet{HuntReffert2023} \\ %
$\mu_{\alpha*}$, $\mu_\delta$ & 18.72, 2.59 & mas\,yr$^{-1}$ & \citet{HuntReffert2023} \\ %
Radial velocity & 5.78 $\pm$ 0.10 & km\,s$^{-1}$ & \citet{Babusiaux2018} \\ %
Half-number radius $r_{50}$ & 0.58 (2.39~pc) & deg & \citet{HuntReffert2023} \\ %
Tidal radius $r_t$ & 1.84 (7.53~pc) & deg & \citet{HuntReffert2023} \\ %
Jacobi radius $r_J$ & 2.60 (10.65~pc) & deg & \citet{HuntReffert2024} \\ %
Jacobi mass $M_J$ & 465 $\pm$ 46 & $M_\odot$ & \citet{HuntReffert2024} \\ %
Tidal tails (leading, trailing) & $\sim$60, $\sim$50 & pc & \citet{Zhang2020} \\ %
Age, lithium depletion boundary & 132 $\pm$ 24 & Myr & \citet{Cargile2010} \\ %
Age, lithium depletion boundary & $126^{+13}_{-14}$; $114^{+9}_{-10}$\tablenotemark{a} & Myr & \citet{Juarez2014} \\ %
Age, isochrone & 105 & Myr & \citet{CantatGaudin2020} \\ %
Age, isochrone & 100 & Myr & \citet{Pang2021} \\ %
Extinction $A_V$ (16--84\%) & 0.057 (0.00--0.11) & mag & \citet{HuntReffert2023} \\ %
\enddata
\tablenotetext{a}{Lithium depletion boundary age before and after a correction for magnetic activity.}
\tablecomments{Radii in parsecs are at 234.4~pc. The traceback of Section~\ref{sec:traceback} uses the centre, distance,
proper motion and radial velocity listed here.}
\end{deluxetable}

We compare the lithium depletion boundary and isochrone ages when choosing the youth correction for Blanco~1. The lithium estimates of \citet{Cargile2010} and \citet{Juarez2014}, including the latter's correction for magnetic activity, are listed in Table~\ref{tab:cluster}. Isochrone fits to the Gaia sequence give 105\,Myr \citep{CantatGaudin2020} and %
100\,Myr \citep{Pang2021}, while the neural-network ages of \citet{HuntReffert2023} give $\log_{10}(t/\mathrm{yr})=8.24$, %
or 174\,Myr. We adopt 100\,Myr for the model youth correction, matching the isochrone ages, and compare it with an empirical correction measured on the AB~Doradus moving group \citep[$149^{+51}_{-19}$\,Myr;][]{Bell2015} in Section~\ref{sec:specphot}. Together the two cases span the published lithium and isochrone ages.

\subsection{The Euclid field}\label{sec:field}

\begin{figure*}
\centering
\includegraphics{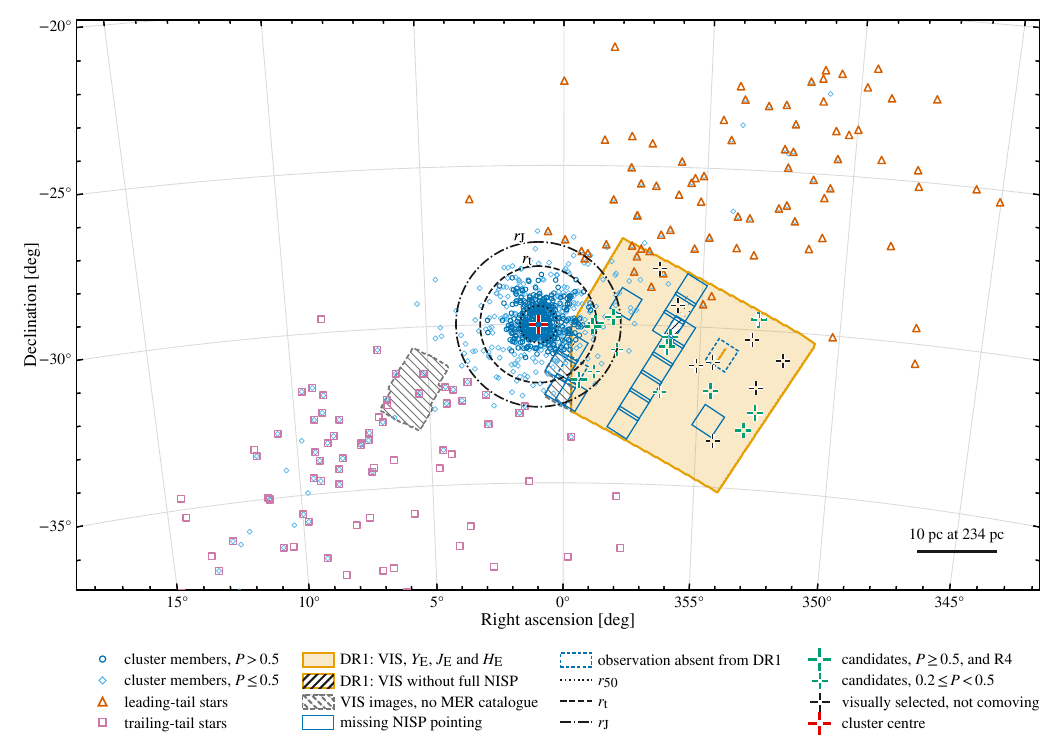}
\caption{Blanco~1, its tidal tails and the Euclid DR1 field in a gnomonic projection. Open circles and small
diamonds are members from \citet{HuntReffert2023} with membership probability above 0.5 and at or below 0.5, respectively. Triangles
and squares are leading- and trailing-tail stars from the four tail catalogues that \citet{Jadhav2025} grade as
reliable \citep{Zhang2020,Meingast2021,Kos2024,Risbud2025}. %
The dotted, dashed and dash-dotted circles mark the half-number radius $r_{50}=0.58\arcdeg$ and tidal radius %
$r_t=1.84\arcdeg$ of \citet{HuntReffert2023} and the Jacobi radius $r_J=2.60\arcdeg$ of \citet{HuntReffert2024}, %
centred on the cluster centre (red crosshair). The DR1 coverage follows the exposure maps of each band, traced on a %
9\arcsec\ grid and simplified for display. The shaded area, where VIS, \YE, \JE\ and \HE\ images all exist and the selection can run, lies 1.04\arcdeg\ to 8.7\arcdeg\ from %
the centre and covers 11\% of the area inside $r_t$. Black hatching marks 0.6\,deg$^2$ with VIS but without all three %
NISP bands, and the gold line is the edge of the VIS coverage in the MER tiles. Grey hatching with a dashed edge
marks sky outside the MER tiles that DR1 covers with VIS images but no MER catalogue, modelled from the pointing
positions: 1.4\,deg$^2$ east of the field and 3.1\,deg$^2$ from 15 VIS-only observations on the trailing tail. %
Blue squares are the nominal fields of the 15 pointings without DR1 images in at least one NISP band, and %
dashed blue squares those of the two observations absent from DR1 in every band. Neighbouring pointings cover most
of the interior fields, with a median NISP depth 0.3\,mag shallower than elsewhere in the field. Crosshairs with a %
central gap mark the 22 named
objects: thick green for R4 and the new candidates with $P\geq0.5$, thin green for $0.2\leq P<0.5$, and black for the
spectroscopically ultracool dwarfs that move unlike the cluster. The bar shows 10\,pc at 234\,pc.} %
\label{fig:skymap}
\end{figure*}

\begin{figure*}
\centering
\includegraphics{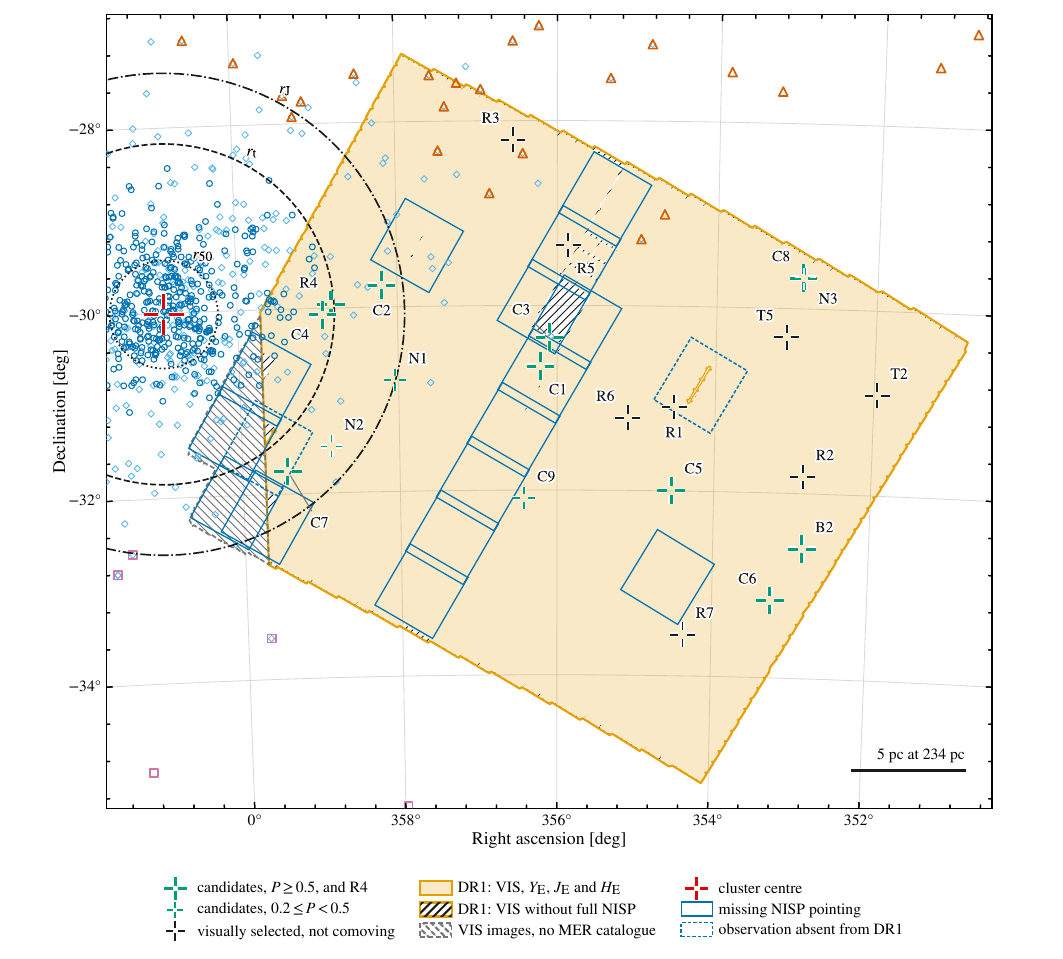}
\caption{The DR1 field and the centre of Blanco~1, with the 22 named objects labelled. Symbols, circles, coverage shading and
crosshair colours as in Figure~\ref{fig:skymap}; the red crosshair marks the cluster centre. The hatched area near
(356.0\arcdeg, $-30.1$\arcdeg) is a hole that no NISP pointing covers; narrow hatched seams and strips lie inside the %
run of missing NISP pointings, along the east edge and along the field boundary. The thin gold outline inside the %
dashed square near (354.2\arcdeg, $-30.9$\arcdeg) encloses a sliver of 0.01\,deg$^2$ that an absent observation %
leaves without VIS. C8 and N3 are close
neighbours on the sky, and their crosshairs overlap at this scale. The bar shows 5\,pc at 234\,pc.}
\label{fig:zoom}
\end{figure*}

The field was observed from December 3 to 10, 2023, during the thermal stabilization periods of the Phase Diversity calibrations of the Euclid telescope optics that preceded the nominal surveys, and %
it is included in the first Data Release (DR1, I. Tereno et al. 2026). It covers 35.6\,deg$^2$ in all four Euclid bands between right ascensions 350.9\arcdeg\ and 359.7\arcdeg\ and %
declinations $-35.1\arcdeg$ and  $-27.3\arcdeg$ (Figures~\ref{fig:skymap} and \ref{fig:zoom}). The cluster centre lies outside it. The field extends from 1.04\arcdeg\ (4.3\,pc) to 8.7\arcdeg\ (35.9\,pc) from the cluster centre. Its centroid is 4.6\arcdeg\ away at position angle 254\arcdeg. %

We restrict the Euclid analysis to the overlap of all four bands, so every area, fraction and count refers to that common coverage. The DR1 exposure maps give 36.2\,deg$^2$ in VIS and 35.7\,deg$^2$ in each NISP band. Two %
of the 135 VIS pointings of the observing sequence are absent from DR1, and 15 of the other 133 lack DR1 images in at %
least one NISP band. The missing images are identified in the DR1 exposure maps within the tiles of the merged (MER) source catalogue. Neighbouring pointings cover most of the fields of the missing NISP
pointings with fewer exposures, so the median NISP depth there is 0.3\,mag shallower than elsewhere. Source %
densities in these fields still match those elsewhere to $\JE=22$, fainter than any late-M or L candidate. Only the %
faint, mostly extragalactic VIS dropouts are about half as numerous there. At $\JE=21$--22.6, where T-dwarf members %
would lie, the ratio is 0.8. The shallower regions occupy 14\% of the area. If their source-density ratio measures T-dwarf completeness, they reduce the expected number across the field by about 3\%; Section~\ref{sec:population} gives the expectation. %

Missing pointings and the slightly wider VIS coverage leave 0.6\,deg$^2$ with VIS images but incomplete NISP coverage (Figure~\ref{fig:zoom}). This includes a 0.24\,deg$^2$ hole near (356.0\arcdeg, $-30.1\arcdeg$) and narrow strips along the field edges. We exclude these gaps from the four-band selection of Section~\ref{sec:selection}. Inside the
convex outline of the catalogue positions, 0.49\,deg$^2$ lacks one or more bands, and the ground-based search of %
Section~\ref{sec:offeuclid_data} covers that sky instead. We also exclude VIS images outside the MER tiles, where DR1 provides no merged catalogue. These cover about 1.4\,deg$^2$ east of the field, as close as 0.92\arcdeg\ to the centre. A further 15 VIS-only observations cover 3.1\,deg$^2$ on the trailing tail, 3.2--5.6\arcdeg\ south-east of the centre. Slitless spectra are available across the four-band area except for an estimated 2\% affected by 18 missing spectroscopic pointings out of 133. %

The Euclid field samples the outskirts and a small part of the tails (Table~\ref{tab:coverage}). It misses the circle of radius $r_{50}=0.58\arcdeg$ that contains half of the \citet{HuntReffert2023} members. In this paper the core is the circle of radius $r_{50}$; the %
brown-dwarf survey of \citet{Moraux2007} covered 2.3\,deg$^2$ around the centre. Of the 841 %
\citet{HuntReffert2023} members, 57 (6.8\%) fall inside it, and 18 of the 422 with membership probability above 0.5. %
Seen from the centre, the field spans position angles 201\arcdeg-- 315\arcdeg. The leading tail leaves the cluster %
towards the northwest, at position angles 305\arcdeg--313\arcdeg, and the trailing tail towards the southeast; the %
field therefore samples the cluster outskirts and the southern flank of the base of the leading tail. The \citet{Jadhav2025} classifications place 7 of 87 leading-tail stars and none of 100 trailing-tail stars in the field. With the assignments of \citet{Risbud2025}, it contains 10 of 68 leading-tail stars and 1 of 82 trailing-tail stars. The two tails extend 50--60\,pc from the %
centre \citep{Zhang2020}, and the field covers 7.8\% of a circle of radius 50\,pc. %

\begin{deluxetable}{lcc}
\tabletypesize{\footnotesize}
\tablecaption{Fraction of circles about the cluster centre covered by the Euclid field.\label{tab:coverage}}
\tablehead{\colhead{Circle} & \colhead{Radius} & \colhead{Covered}}
\startdata
$r_{50}$ \citep{HuntReffert2023} & 0.58\arcdeg & 0.0\% \\ %
$r_t$ \citep{HuntReffert2023} & 1.84\arcdeg &  11.3\% \\ %
$r_t$ \citep{Zhang2020} & 10.0\,pc &  17.8\% \\ %
$r_t$ \citep{Pang2021} & 10.2\,pc &  18.2\% \\ %
$r_J$ \citep{HuntReffert2024} & 2.60\arcdeg &  19.1\% \\ %
5\arcdeg & 5\arcdeg &  23.6\% \\ %
Tail extent, 50\,pc & 12.0\arcdeg &  7.8\% \\ %
Tail extent, 60\,pc & 14.4\arcdeg &  5.5\% \\ %
\enddata
\tablecomments{Coverage is the area  observed in all four Euclid bands inside each circle divided by the area of the circle. Linear radii are
converted to angles at 234.4\,pc.} %
\end{deluxetable}

We use Gaia stars to normalise the expected substellar population and its radial distribution. The published membership lists contain 64 stars in the field, 38 inside $r_J$. Of the listed stars, 58 satisfy a Gaia DR3 comoving test at the cluster distance and have a median distance of 234\,pc. A stricter test selects 53 listed stars and 12 additional stars, giving 65 in all. These two samples set the normalisation in Section~\ref{sec:membership_model}, while the radial prior in Section~\ref{sec:results} follows the 91 listed and additional stars that satisfy either test. Seven Gaia stars have DR1 counterparts, with %
$\IE=19.2$--20.1 and $\IE-\JE=2.2$--2.7: six new stars that pass the Gaia test and one listed member whose %
parallax is too uncertain for that test. They mark the faint end of the Gaia-selected sequence in this field, and they %
serve below as a check on the astrometry (Appendix~\ref{app:astrometry}). The ten ultracool candidates with
membership probabilities above 0.5 (Section~\ref{sec:results}) have $\IE=21.2$--23.1, 1--4\,mag fainter. %

We mask the nearby spiral galaxy NGC~7793 because many of its resolved star-forming regions meet the colour and morphology cuts of Section~\ref{sec:selection}. The DR1 catalogue for this field lacks ground-based optical photometry, so our selection uses Euclid colours alone. Apart from this %
field, the nearest DR1 tiles lie about 22\arcdeg\ from the cluster (Figure~\ref{fig:skymap}). %

\section{Data}\label{sec:data}

\subsection{Euclid DR1 photometry and NISP spectra}\label{sec:euclid_data}

The Euclid photometry comes from the DR1 merged (MER) source catalogue of the field \citep{EuclidDR1}. For each source we use its position, VIS point-spread-function (PSF) flux in the broad \IE\ band, and template-fit fluxes in the NISP \YE, \JE\ and \HE\ bands \citep{Schirmer2022}. We assess morphology with the point-source probability \texttt{POINT\_LIKE\_PROB}, \texttt{KRON\_RADIUS} and \texttt{ELLIPTICITY}. Magnitudes are on the AB system. The MER
positions are tied to Gaia DR3 and provide the most precise epoch of the proper motions (Section~\ref{sec:astrometry}).

NISP also records slitless spectra with its red grism over 1.2--1.9\,\micron\ at a resolving power of about 450. We %
use two extractions: our co-add of the individual dither spectra over all available pointings, which
provides the spectral types (Section~\ref{sec:typing}), and the combined spectrum of the DR1 processing, which serves
as a check where it has unmasked pixels. %

\subsection{Ground-based astrometric epochs}\label{sec:ground_data}

The VISTA Kilo-degree Infrared Galaxy survey \citep[VIKING;][]{Edge2013} imaged the field in $Z$, $Y$, $J$, $H$ and
$K_s$ with the VIRCAM camera on the VISTA telescope. We use the band-merged catalogue of its fifth internal release (ER5).
VIKING provides the earliest epochs and sets most of the 13\,yr median baseline of the proper motions. %

The Kilo-Degree Survey \citep[KiDS;][]{Wright2024} imaged the same area in $u$, $g$, $r$ and $i$ with OmegaCAM on the
VLT Survey Telescope, and its fifth data release includes repeated $i$-band visits. Of the 13,590 sources of the %
photometric selection, 4,405 (32\%) have a KiDS counterpart within 1\arcsec, and 79\% of those with $\JE<20$, so %
KiDS adds epochs mainly for the brighter late-M and early-L dwarfs.

\subsection{Gaia DR3}\label{sec:gaia_data}

Gaia DR3 \citep{Gaia2016,Vallenari2023} defines the reference frame to which the VIKING, KiDS and Euclid positions
are tied. Its faintest stars provide an external check of the proper motions (Appendix~\ref{app:astrometry}), and
its red field stars supply one of the two field models of the membership analysis (Section~\ref{sec:membership_model}).
The cluster members of Section~\ref{sec:field} come from published Gaia catalogues
\citep{CantatGaudin2018,HuntReffert2023,Zhang2020,Pang2021,Meingast2021,Risbud2025,Kos2024}.

\subsection{Rubin Observatory Data Preview 2}\label{sec:rubin_data}

The Vera C. Rubin Observatory \citep{Ivezic2019} observed part of the field during commissioning, and these
observations are included in its Data Preview~2 \citep[DP2;][]{RubinDP2}. We use 26.6 million catalogued detections in %
17 strips across the field. Of the DP2 visits, 30 in $g$ and 39 in $i$ intersect the field; they cover 88\% of it in %
$g$ and 93\% in $i$, with median nominal 5$\sigma$ depths of 25.3 and 24.5\,mag (AB). The visits date from 2025 and %
extend the astrometric baseline by about two years beyond the Euclid epoch. Two candidates, B2 and C6, fall outside the %
DP2 coverage in both bands. %

\subsection{SPHEREx}\label{sec:spherex_data}

SPHEREx, a NASA all-sky survey, observes 0.75--5.0\,\micron\ through linear variable filters at a spectral
resolution of 35--130 \citep{Bock2026}, so each exposure samples one wavelength at a given sky position. The field is %
covered by the Quick Release~2 spectral images \citep{SPHERExQR2}, calibrated by the mission pipeline
\citep{Akeson2026}. For each of the 22 objects in Table~\ref{tab:named}, we jointly fitted the 326--386 exposures containing it. The object and every Gaia, CatWISE and 2MASS source within 130\arcsec\ are modelled as
point sources whose fluxes may be negative, so that faint objects are not biased high; each exposure has its own sky
plane, and each position follows the source's proper motion to the epoch of the exposure. We combine the fitted fluxes into 0.12\,\micron\ bins for the spectra in Figure~\ref{fig:spherex}. For detection, we integrate over the 0.75--2.4 and 3.0--5.0\,\micron\ windows. Eight blank-sky positions per object, fitted in the same exposures, give the noise %
of each window flux, and a detection requires a window flux at least five times that noise.

We estimate the photospheric flux from 669 field M7--L5 dwarfs elsewhere in Euclid DR1 with NISP spectral types and $16.5<\JE<18.0$, using the same extraction. We fit the SPHEREx-to-\HE\ flux ratio as a linear function of spectral type in each window. The scatter is 0.034\,dex over 3--5\,\micron; we normalise to \HE\ because warm dust contributes little at 1.6\,\micron. For most of our sample, which is fainter than $\JE=18$, this relation is an extrapolation in magnitude. For T dwarfs we use the 4.2--5.0\,\micron\ window because their 1--4\,\micron\ spectral shape changes rapidly with type. We scale it to CatWISE $W2$ using 13 field T dwarfs measured in the same way.

\subsection{Other catalogues}\label{sec:other_data}

We used motions from CatWISE2020 \citep{Marocco2021}, based on WISE \citep{WrightEL2010} and NEOWISE \citep{Mainzer2011,Mainzer2014}, to locate the two fast-moving T dwarfs in the KiDS images and to check their measured motions independently. The UltracoolSheet
\citep{Best2025} supplied field ultracool dwarfs with trigonometric parallaxes for the distance calibration
(Section~\ref{sec:specphot}) and the AB~Doradus members for the empirical youth correction. The spectral standards of
the SpeX Prism Library \citep{Burgasser2014}, accessed through SPLAT \citep{Burgasser2017}, were used for spectral typing and gravity indices.
We calibrate the Euclid colour--type relations in Section~\ref{sec:typing} on the 139 ultracool dwarfs with published spectral types in the Quick Release~1 (Q1) deep fields \citep{DominguezTagle2025}. After excluding peculiar types and clipping a few outliers in each colour, this sample defines the photometric type scale.

\section{Methods}\label{sec:methods}

\subsection{Photometric selection}\label{sec:selection}

\begin{figure*}
\centering
\includegraphics{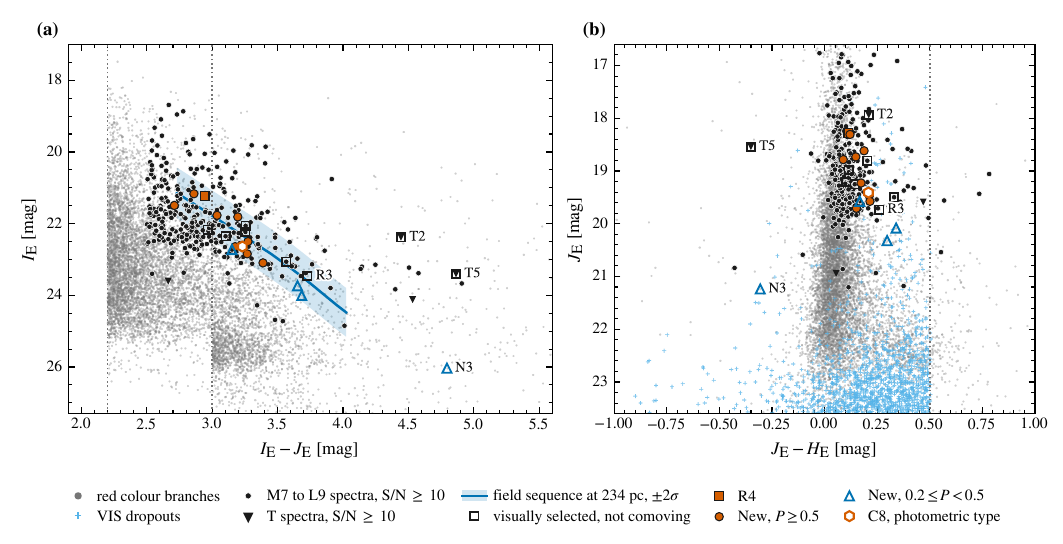}
\caption{Photometric selection in the DR1 field. (a) \IE\ against $\IE-\JE$ for the sources selected by the two
red branches outside the NGC~7793 mask (grey), the 346 NISP spectra of type M7 or later with $\mathrm{S/N}\geq10$ %
(black dots with white edges; triangles for T types), and the named objects. Dotted lines mark the colour limits of the red branches:
$\IE-\JE\geq2.2$ for point-like sources and $\geq3.0$ for sources with a point-like probability of at least 0.1. The %
blue curve and band are the field ultracool sequence placed at 234\,pc and its $\pm2\sigma$ spread, drawn where it is %
defined. (b) \JE\ against $\JE-\HE$, adding the sources selected as VIS dropouts (light-blue plus signs), which requires %
$\JE-\HE<0.5$ (dotted line) and holds the T-dwarf candidates without a VIS detection. Among the named objects, open
squares mark the spectroscopically selected objects that move unlike the cluster and the filled square R4. Filled circles and
open triangles are new candidates with $P\geq0.5$ and $0.2\leq P<0.5$, and the open hexagon is C8, whose type comes
from its colours.}
\label{fig:selection}
\end{figure*}

We select compact sources with colours consistent with late-M, L or T dwarfs, requiring a \JE\ signal-to-noise ratio of at least 10 and one of three criteria:
\begin{enumerate}
\item $\IE-\JE\geq2.2$ and $\texttt{POINT\_LIKE\_PROB}\geq0.5$, for late-M and early-L dwarfs; %
\item a VIS point-spread-function signal-to-noise ratio below 3, below a significant VIS detection, and $\JE-\HE<0.5$, for T dwarfs too faint for VIS; %
\item $\IE-\JE\geq3.0$ and $\texttt{POINT\_LIKE\_PROB}\geq0.1$, for faint red sources that the first criterion %
would reject.
\end{enumerate}
The selection contains 13,590 sources: 10,258 satisfy the colour criteria, 3,369 the VIS-dropout criterion, and 37 both. We mask NGC~7793 with an ellipse of 12\arcmin\ semi-major axis, excluding 2,584 sources. The remaining 11,006 form the sample for the colour--magnitude and kinematic analyses (Figure~\ref{fig:selection}). The mask is applied after %
spectral typing, and the typing statistics below refer to the full selection. Table~\ref{tab:funnel} follows the
sample from this selection to the candidates.

We group sources by Euclid colour into the photometric classes late M, L, L-T (the L/T transition) and T for the membership analysis. It also receives a sequence offset
\begin{equation}
D=\IE-\IE^\mathrm{r}\!\left[(\IE-\JE)_0\right],\label{eq:offset}
\end{equation}
its magnitude difference from a single-star ridge $\IE^\mathrm{r}$ of field dwarfs placed at the cluster distance; %
negative $D$ means brighter. The colour $(\IE-\JE)_0$ is corrected for the extinction of the cluster, the ridge is
calibrated on field dwarfs with parallaxes, and for T dwarfs $D$ is measured in \JE\ against $\JE-\HE$ instead. We brighten the field-dwarf ridge to account for youth and use this youth-corrected ridge as the reference for the cluster sequence. At
$(\IE-\JE)_0\leq2.41$ the shift is 0.56\,mag, the low end of the 0.6--1.2\,mag by which the cluster's own Gaia %
sequence lies above the field ridge there. At colours redder than $(\IE-\JE)_0=3.25$, we use the difference at fixed colour between the 150\,Myr and 1\,Gyr ATMO evolutionary isochrones (Section~\ref{sec:specphot}). The older isochrone extends no bluer than this colour, so we interpolate the shift between the two colour limits. This sequence model uses the 150\,Myr isochrone, while the distance correction of
Section~\ref{sec:specphot} uses 100\,Myr. The model lets the sequence shift by up to $\pm0.85$\,mag from this ridge %
(Section~\ref{sec:membership_model}). A finite sequence offset can be assigned to 4,596 sources, which make up the membership catalogue. Of these, 1,543 are late-M, 2,030 L, 287 L-T and 228 T; the other 508 occupy a slightly bluer boundary slice at $2.6\leq(\IE-\JE)_0<2.7$.

A source %
is compact if $\texttt{POINT\_LIKE\_PROB}>0.5$, $\texttt{ELLIPTICITY}<0.25$ and either $\texttt{KRON\_RADIUS}\leq15$ pixels of 0.1\arcsec %
or $\IE<21$. These cuts keep 78\% of point-like late-M and L dwarfs, 72\% of L-T and 80\% of T dwarfs. %

\begin{deluxetable}{lrc}
\tabletypesize{\footnotesize}
\tablecaption{From the photometric selection to the candidates.\label{tab:funnel}}
\tablehead{\colhead{Step} & \colhead{Sources} & \colhead{Section}}
\startdata
Euclid photometric selection & 13\,590 & \ref{sec:selection} \\ %
\hspace{1em}with a typed NISP spectrum & 3\,002 & \ref{sec:typing} \\ %
\hspace{1em}\hspace{1em}of type M7 or later at S/N\,$\geq$\,10 & 346 & \ref{sec:typing} \\ %
Outside the NGC~7793 mask & 11\,006 & \ref{sec:selection} \\ %
\hspace{1em}with a photometric class and sequence offset$^{a}$ & 4\,596 & \ref{sec:selection} \\ %
\hspace{1em}\hspace{1em}with a spectrum used in the model & 963 & \ref{sec:typing} \\ %
\hspace{1em}\hspace{1em}with a proper motion & 3\,675 & \ref{sec:astrometry} \\ %
\hspace{1em}\hspace{1em}\hspace{1em}late M or L, passing the quality cuts$^{b}$ & 1\,213 & \ref{sec:membership_model} \\ %
\hspace{1em}\hspace{1em}\hspace{1em}\hspace{1em}in the member band, grade A or B$^{c}$ & 193 & \ref{sec:membership_model} \\ %
$P\geq0.5$ under both field models & 10 & \ref{sec:res_new} \\ %
\enddata
\tablecomments{Each indented row is a subset of the row above it at the next lower indentation. Spectral typing precedes
the NGC~7793 mask.}
\tablenotetext{a}{The membership catalogue. The other 6\,410 unmasked sources are 4\,590 stars bluer than the late-M
ridge, mostly mid-M, 1\,622 VIS dropouts redder in $J_\mathrm{E}-H_\mathrm{E}$ than T dwarfs, 119 L, L-T and T sources without a defined offset,
and 79 with colours of no class.}
\tablenotetext{b}{The sample of the likelihood test (Section~\ref{sec:membership_model}).}
\tablenotetext{c}{Compact, with motion errors below 5\,mas\,yr$^{-1}$: the sample of the count test.}
\end{deluxetable}

\subsection{Spectral typing}\label{sec:typing}

Of the selected sources, 3,002 have a co-added NISP spectrum and a typing result. We compare each spectrum with the M5--T9 SpeX standards \citep{Kirkpatrick2010,Burgasser2006} by $\chi^2$ over 1.27--1.83\,\micron, then refine the type using the H$_2$O-H and CH$_4$-H indices of \citet{Burgasser2006}. For M and L dwarfs we adopt the
index-based type of \citet{Allers2013} where it is available and the template type otherwise. The two scales can
differ by several subtypes in faint spectra; for C7 they give M7.0 and L1.1 (Table~\ref{tab:fulla}). Of the 3,002 %
spectra, 981 have a signal-to-noise ratio below 5. Above a signal-to-noise ratio of 10 the sample holds 346 %
ultracool spectra with template types of M7 or later: 268 M, 74 L and 4 T dwarfs. We use a spectrum in the membership model when its signal-to-noise ratio is at least 5 and its type comes from either the indices or a template fit with a fit statistic of at most 2.5 against the best dwarf standard. In all, 963 sources of the membership catalogue %
have such a spectrum.

We compare each adopted spectral type with a photometric type to identify disagreements between the spectrum and colours. It combines three Euclid colours ($\IE-\JE$, $\YE-\JE$ and $\JE-\HE$), calibrated on the Q1 sample %
of Section~\ref{sec:other_data} and counted as about 1.5 independent colours because their residuals correlate, with
the VIKING $J-K_s$ colour, calibrated on 1,694 UltracoolSheet field dwarfs. Across the field, 45 of 690 usable M6--L9 %
spectra (6.5\%) differ from their photometric type by more than two combined standard deviations. The membership model %
uses the spectroscopic types only, because a photometric type would reuse the colour that defines the sequence offset.

The spectral and photometric types agree within 1.3 combined standard deviations for sixteen objects. Among the six disagreements, C8 affects the membership assessment most. Both NISP extractions of its spectrum favour an L7 standard, and the DR1 extraction passes %
the template-fit limit (fit statistic 1.5, against 3.1 for the co-add). Its colours instead give L0, with a 16--84\% %
range of M8.3--L2.1, and favour L0 over L7 by a likelihood ratio of about 30. Its Legacy Surveys and CatWISE colours %
also match an early L dwarf (Section~\ref{sec:res_new}), so we adopt the photometric type, L0, for its distance and %
treat L7 as a disfavoured alternative. Its spectroscopic membership terms are left out of the adopted model, and
Section~\ref{sec:res_new} gives its probability with them. N3 has no usable spectrum, and its colours give T4.9 %
(T3.6--T6.5). R3 keeps its T0 type, which fits its spectrum better than T1, L9 and L3 although its colours resemble %
an L3 dwarf as closely as an early-T one. R1 is red in $\IE-\JE$ (3.22) but blue in the other colours, and subdwarf %
standards fit its spectrum better than dwarf standards; we flag it as possibly metal-poor.

\subsection{Joint astrometry}\label{sec:astrometry}

The proper motions combine VIKING, KiDS and Euclid positions over a median baseline of 13\,yr. The VIKING epochs needed one %
correction before use. In the VIKING ER5 band-merged catalogue, the position $\alpha_{2000}$ of a source is a weighted
mean of its detections in the separate bands, and each band $k$ has an offset $\xi_k$ from the detection in the
first band observed, the master band. That band differs from source to source, so the catalogue position mixes the
epochs of different bands. We recover the master-band position as $p_\mathrm{master}=\alpha_{2000}-\sum_k f_k\,\xi_k$,
with weights $f_k\propto\sigma_{\mathrm{pos},k}^{-2}$, and each band's position as $p_k=p_\mathrm{master}+\xi_k$. Each
of the 403 frames, the images of one band in one VIKING tile observation, is then tied to Gaia DR3 at its own epoch. %
Regressing the tie residuals of each frame on the Gaia proper motions measures its effective epoch.
Using the band positions reduces the fraction of frames with epoch errors above 1\,yr from 13\% to zero, without fitting an epoch. We include the largest residual error, 0.43\,yr, in the uncertainty. %
KiDS gives an independent test: for 216 sources with both a VIKING+Euclid and a KiDS+Euclid motion, the difference %
between the two motions follows the correction applied to the VIKING positions with slope $0.96\pm0.07$ before the %
correction and $-0.04\pm0.04$ after it. %

We calibrate the position errors with 4,016 independent VIKING measurement pairs, normalising each difference by the full error model. With the band positions, faint sources need no inflation of their errors; above a master-band
signal-to-noise ratio of 20 the centroid errors are scaled by 1.28--1.87 (Appendix~\ref{app:astrometry}). %

We fit each object's reference position and proper motion by generalised least squares using the VIKING epoch groups, the KiDS epochs corrected for differential chromatic refraction, and the Euclid position. VIKING detections within 30 days count as one epoch. We fix the parallax at zero for the field hypothesis and at the cluster value for the comoving test. We allow rejection of one epoch with $\chi^2>16$ when at least three other ground-based epochs remain. We multiply the formal errors by the Birge factor $\max(1,\sqrt{\chi^2/\nu})$, scaling them until the reduced $\chi^2$ is one. Fits with all three surveys exist for 891 objects, and with %
VIKING or KiDS the motions extend to 3,651 sources of the membership catalogue. The Rubin epochs below add 24 with only %
Euclid and Rubin positions, baselines of 1.8--2.0\,yr and grades B or C, all of which fail the quality cuts, for %
3,675 in all. For compact sources lacking Gaia astrometry, we obtain median adopted errors of 1.1--1.4\,\masyr\ at $\JE=17$--20.5. At $\JE=20.5$--22, the median error is 2.9\,\masyr. %

Each motion receives a quality grade. Grade A requires a compact source, at least two ground-based epoch groups, a fit
$p$-value above 0.01 and agreement between the motions from VIKING alone and KiDS alone, $\chi^2_{VK}\leq13.8$ (the %
99.9\% point for two degrees of freedom). Grade B
is compact but fails one of these conditions, and grade C is not compact. Tested on 145 faint Gaia DR3 stars, the %
joint motions show no zero-point offset larger than about 0.3\,\masyr, and their formal errors run 7--15\% low %
(Appendix~\ref{app:astrometry}).

We include Rubin DP2 $i$-band positions as an additional epoch group for 958 candidates. Relative to the VIKING, KiDS and Euclid frame, 361 field objects outside the candidate sample give a mean offset of $(-12.6,\,+5.3)\pm(1.7,\,1.5)$\,mas in $(\alpha*,\delta)$. We subtract this offset, which agrees within the errors across all populated $i-\JE$ colour bins. A common floor of 20\,mas per coordinate, correlated across an object's DP2 %
visits, absorbs residual visit-to-visit structure. The DP2 $g$-band positions scatter more than their stated centroid %
errors and are not used. Because of the generous floor, extra DP2 visits would lower a fit's Birge factor mainly by
adding degrees of freedom. We therefore use the larger Birge factor of the fits with and without DP2; the fit without DP2 determines it for 281 of the 958 sources. We assign quality grades from the fits without DP2. %

\subsection{Spectrophotometric distances and gravity indices}\label{sec:specphot}

We calibrated absolute magnitudes in \JE\ and \HE\ against NISP spectral type on DR1 spectra of ultracool dwarfs with
trigonometric parallaxes. We express spectral type numerically as $N_\mathrm{sp}$, with 10 at M0, 20 at L0 and 30 at T0. For the index types, 25 M6--L1 calibrators give $M_{\JE}=12.78+0.462\,(N_\mathrm{sp}-20)$, with an intrinsic scatter of 0.37\,mag. The template types give 0.52\,mag and the \HE\ relations 0.34--0.47\,mag, and the field relations give %
distances to about 20--25\%.

For types from L1.5 to L9.5 we use the relation of 964 UltracoolSheet field dwarfs with parallaxes better than 10\%, %
transformed to \JE\ with synthetic colours of the SpeX standards. Over M6--L1 it agrees with the NISP relations to
within 0.03\,mag in the median. The four NISP T-dwarf calibrators have an unexplained offset of 0.5--1.3\,mag below the UltracoolSheet relation. Their fit gives $M_{\JE}=17.23+0.362\,(N_\mathrm{sp}-35)$, with 0.28\,mag scatter and a 0.3\,mag systematic. For the four T dwarfs we therefore quote distances on the UltracoolSheet relation and give the NISP T
relation as the alternative, which places T2, T5 and N3 closer and R3 slightly farther. We obtain distances for 1,748 of the 3,002 typed objects. %

The adopted youth correction comes from the ATMO 2020 evolutionary models \citep{Phillips2020,Chabrier2023} at
100\,Myr. We place a young object at the field effective temperature of its spectral type \citep{Filippazzo2015} and take the correction to be the difference between the 100\,Myr and field-age magnitudes at that temperature.
It brightens a member by 0.14--0.18\,mag at M7--M8 and by 0.55\,mag at L0--L1, with an uncertainty of half the %
difference between this placement and one at the cooler effective temperature of young objects of the same type, %
plus 0.15\,mag in quadrature. Because the field reference of the models changes with type, the correction steps across %
M9--L2. The ATMO atmospheres are valid only below about 2000\,K, because they omit some opacity sources that %
form at higher temperatures \citep{Phillips2020}, so for the warmer late-M and early-L types the correction
extrapolates the models; the empirical alternative below avoids them.

\begin{figure*}
\centering
\includegraphics{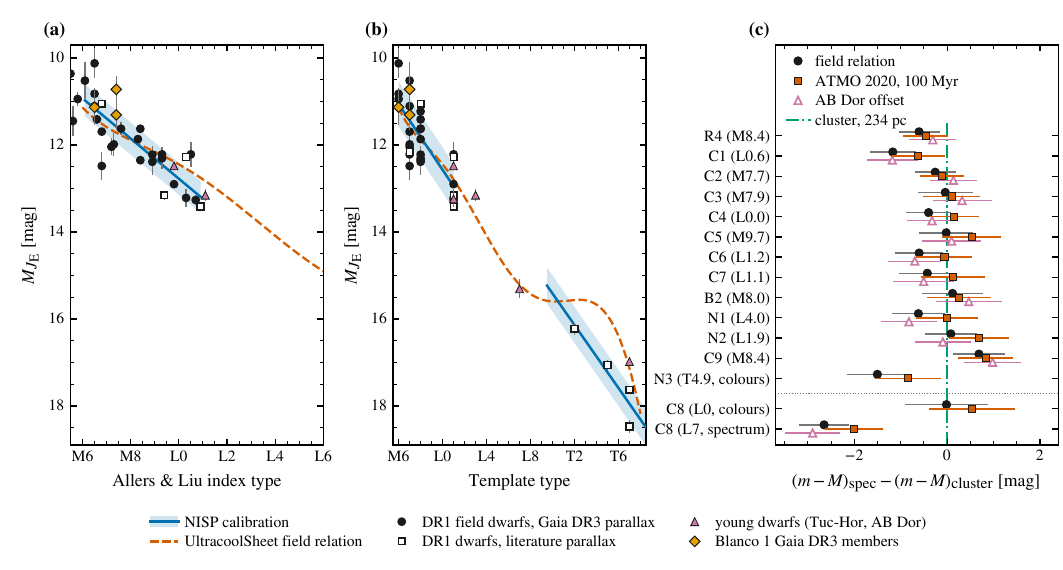}
\caption{Spectrophotometric distances. (a, b) Absolute \JE\ magnitude against the index type of
\citet{Allers2013} and against the template type. Blue lines and bands are the NISP calibration relations and their
intrinsic scatter (for T types the band adds 0.3\,mag in quadrature); the dashed curve is the UltracoolSheet field %
relation converted to \JE. Symbols: DR1 field dwarfs with Gaia DR3 parallaxes (filled circles), DR1 dwarfs with
literature parallaxes (open squares), young dwarfs in Tucana-Horologium and AB~Doradus (triangles) and Blanco~1 Gaia
DR3 members (diamonds). (c) Spectrophotometric minus cluster distance modulus for R4 and the new candidates under the
field relation (circles), with the adopted ATMO~2020 youth offset at 100\,Myr (squares), and with the empirical %
AB~Doradus offset as a sensitivity case (open triangles). N3 uses its photometric type, T4.9, on the UltracoolSheet T %
relation. C8 appears at its adopted photometric type, L0 (233\,pc on the field relation), and at the disfavoured L7 %
type of its spectra (69\,pc); N3 and C8 at L0 lack an AB~Doradus point. %
}
\label{fig:specphot}
\end{figure*}

As the alternative we use an empirical offset measured on members of the AB~Doradus moving group in the
UltracoolSheet, whose age lies at the old end of the published range for Blanco~1. The 20 young late-M dwarfs are $0.40\pm0.20$\,mag brighter in $J$ than the field relation. The 16 early-L dwarfs are instead $0.17\pm0.16$\,mag fainter, following the pattern described by \citet{Liu2016}. The two corrections agree at late M and differ in sign at %
L0--L1. We compare both corrections with the three Gaia-selected Blanco~1 members having NISP spectra, of types M6.5--M7.4 (Figure~\ref{fig:specphot}). Their offsets from the field relation are $-0.85$, $-0.27$ and $-0.03$\,mag. Parallax errors of 0.3--0.4\,mag allow either correction at these types.

Low surface gravity is the other spectroscopic sign of youth. NISP covers the KI$_J$ and $H$-cont indices of
\citet{Allers2013} but not FeH$_J$, whose blue continuum lies below the NISP cutoff. With two indices only, the %
published classes flag 22.5\% of 1,908 NISP field spectra (signal-to-noise ratio of at least 20, M6--L5) as intermediate %
(INT-G) or very low (VL-G) gravity. We therefore express each index as a z-score against the NISP field locus,
positive for lower gravity, and combine the mean z-score with the published class. We identify low gravity when the mean z-score exceeds 2 and the published class is INT-G or VL-G. With the same class but a mean above 1.5, we assign a weak low-gravity hint. It is inconclusive if the class condition holds alone or the mean exceeds 1 without it. On nine young %
ultracool dwarfs with independent evidence of youth, most of them $\sim$45\,Myr old, this rule classes six as low %
gravity or a weak hint while flagging 4.2\% of the field locus. The three Blanco~1 Gaia members with usable spectra are all classed as field-like, so at %
the age of the cluster a field-like class counts little against membership (Appendix~\ref{app:gravity}).

\subsection{Membership model}\label{sec:membership_model}

The membership probability of each source combines its proper motion with its position in the colour--magnitude
diagram and, where a usable spectrum exists, its spectral type, distance and gravity. At a given shift $s$ of the
cluster sequence from the youth-corrected ridge,
\begin{align}
P&=\frac{\pi\,\mathrm{LR}}{\pi\,\mathrm{LR}+1-\pi},\label{eq:posterior}\\
\mathrm{LR}&=\mathrm{LR}_\mu\,\mathrm{LR}_\mathrm{type}\,\mathrm{LR}_d\,\mathrm{LR}_\mathrm{grav}\,\mathrm{LR}_\varpi,
\end{align}
where the five factors are the likelihood ratios, member over field. They are those of the proper motion, the
spectral type, the spectrophotometric distance, the gravity indices and a Gaia parallax. We average $P$ over the posterior of $s$, using a flat prior over $\pm0.85$\,mag. This allows the data to determine the member-sequence location. The prior $\pi$ is the expected fraction of members among the sources at the same %
sequence offset. It is built from the expected number of members of the photometric class, the fraction of members
that pass the quality cuts, and the member and total densities at offset $D$. The expected numbers come from the
\citet{Chabrier2003} system initial mass function scaled to the 58--65 Gaia stars of Section~\ref{sec:field}, with %
class boundaries in mass from the 150\,Myr ATMO models. They sum to 10.8 members before the quality cuts and 8.3 %
after them, of which 5.2 are late-M and L dwarfs. %

Each likelihood ratio compares the member hypothesis with a field hypothesis. For the proper motion, a member moves
with the projected cluster motion at 234.4\,pc at the source's position, with a 1\,\kms\ allowance per component and a %
5\% spread in distance, convolved with an error model calibrated on the Gaia validation stars. The cluster space %
velocity is computed from the mean motion and parallax of \citet{HuntReffert2023} and the radial velocity of %
\citet{Babusiaux2018}, and the motion it implies is evaluated at the median distance of 234.4\,pc, 0.8\% faster than %
the catalogue mean motion. We evaluate membership with two estimates of the field-star motion distribution. The Gaia model uses Gaia DR3 field stars with %
$G_\mathrm{BP}-G_\mathrm{RP}>2.5$ and $G=18$--20.9, each weighted by how well its parallax matches the distance the
source would have as a field dwarf; it gives \Pg. The velocity-shifted control model, giving \Pc, estimates the density of field sources at the same sequence offset by displacing the source's motion by 10 and 14\,\kms\ in eight directions. Both include a 5\% component of distant or nearby %
field stars and a term for chance associations in VIKING. Because both models use the same functional form and most of the same inputs, this comparison tests the choice of field-motion distribution; we discuss the shared systematic uncertainties in Section~\ref{sec:disc_limits}.

The spectroscopic terms follow the same logic. $\mathrm{LR}_\mathrm{type}$ compares the likelihood of the NISP type for a member with that for a field source of the same colour. The member distribution allows 1.5 subtypes about the field type--colour relation. $\mathrm{LR}_d$ compares the spectrophotometric distance modulus with the
cluster's after the youth correction. It is conditioned on the sequence offset, because both use the same \JE\
photometry, and it includes the brightening of an unresolved binary that the offset implies. $\mathrm{LR}_\mathrm{grav}$
compares the gravity z-scores with the field distribution, allowing members to be shifted by 0 to 1.2 locus widths, %
and $\mathrm{LR}_\varpi$ uses a Gaia parallax where one exists. Sources without a usable spectrum, among them N3 and
the spectra of R1, T2 and T5 that fail the template-fit limit or prefer subdwarf standards, have all spectroscopic
terms set to 1, as does C8, whose NISP type disagrees with its colours. %

We test for a population excess by comparing the number of comoving late-M and L dwarfs with the field-model predictions.
For the first test, which is independent of the prior, we count compact late-M and L dwarfs in the colour--magnitude member band with grade A or B motions, errors below 5\,\masyr, and comoving velocities.
The band extends to
twice the combined intrinsic, distance, age and photometric scatter on either side of the field ridge at the cluster %
distance, and 0.75\,mag further on the bright side for unresolved binaries. The test requires the %
$\chi^2$ against the cluster motion, minimised over 200--280\,pc with a 1\,\kms\ allowance, to be below 5.99, the 95\% %
point for two degrees of freedom. The count is compared with the predictions of the weighted Gaia field model and of a
fixed-window Gaia model. For the second test, we fit the amplitude of the prior and the sequence location to all late-M and L sources with motions that satisfy the quality cuts, keeping the shape of the prior fixed. The fitted number of members is the %
amplitude times the expected number at the fitted location; its interval, from the profile likelihood of the
amplitude, is approximate, because the expected number changes with the location. The likelihood-ratio statistic is %
compared with those of 64 control fits in which the cluster motion is displaced by 8--22\,\kms\ in eight directions %
and both parameters are refitted.

\subsection{Proper-motion-only classification}\label{sec:pmonly}

An independent classification by one of us (J.O.) applies Miec, the Bayesian hierarchical model of
\citet{Olivares2021}, to the proper motions alone. Miec treats a catalogue as a mixture of cluster and field
populations, models the cluster proper motions as a Gaussian mixture, and gives each source a membership probability
that accounts for its measurement uncertainties. Here the cluster is described by the Gaia DR3 motions of the 494 %
members of \citet{Meingast2021} and the field by the motions of the DR1 sources. A source passes the 95\% (or 99.7\%)
recovery threshold if its probability exceeds the value that 95\% (or 99.7\%) of the \citet{Meingast2021} members
exceed, so the 99.7\% threshold is the looser of the two. These are low probabilities (R4 passes both with 0.12), %
because the cluster is a small fraction of the field sample. We estimate contamination over the whole field with a Gaussian mixture deconvolved for each source's errors. Its classifications agree with Miec for all 22 objects in Table~\ref{tab:named}. We apply this classifier to the 3,651 sources with motions from VIKING, KiDS and Euclid. Displacing the cluster motion by 8--14\,\kms\ in eight directions then gives the number of field sources selected by chance. %

\section{Results}\label{sec:results}

\begin{figure*}
\centering
\includegraphics{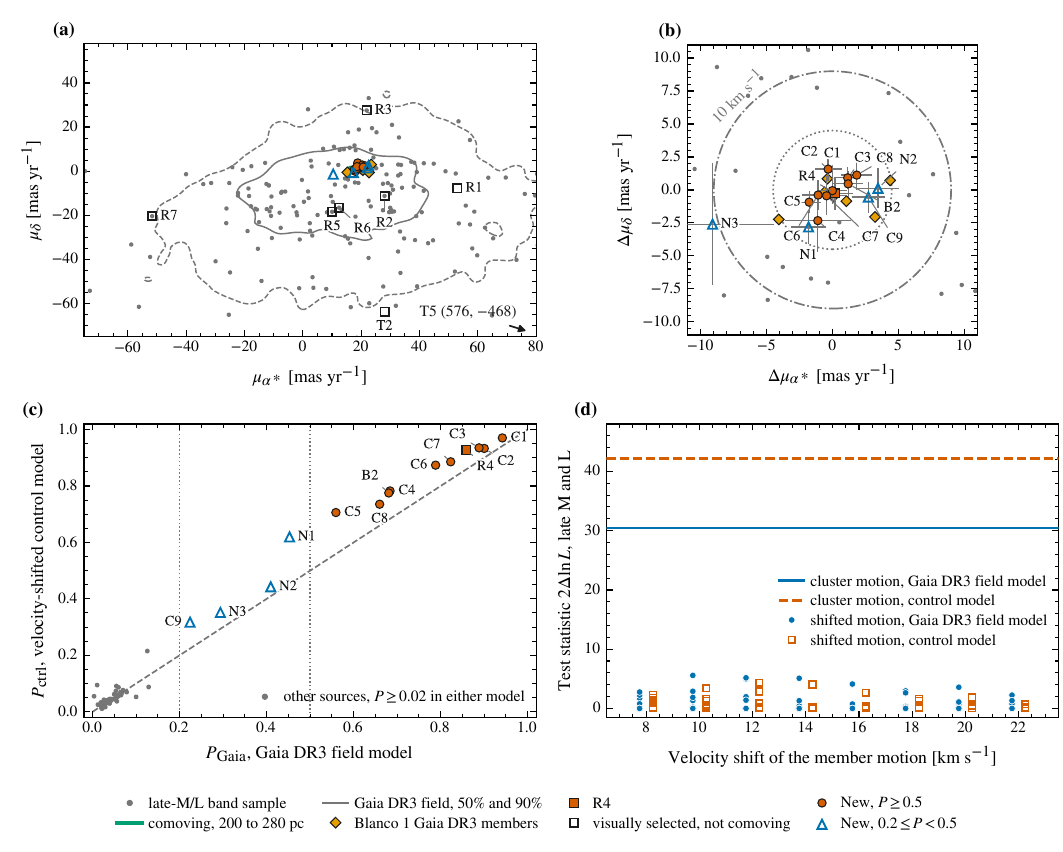}
\caption{Proper motions and the membership test. (a) Vector-point diagram of the 193 compact late-M and L band sources with grade A or B motions (grey dots). Solid and dashed contours enclose 50\% and 90\% of the Gaia DR3 field model. Green indicates the motion of a comoving member at 200--280\,pc; diamonds are Gaia-selected members and candidates with DR1 counterparts. The labelled objects are those of Table~\ref{tab:named}. T5 lies outside the panel. (b) Motion relative to the comoving prediction at
234\,pc for the candidates, with $1\sigma$ errors; the circles mark relative tangential velocities of 5 and %
10\,\kms. (c) Membership probability under the Gaia DR3 field model against that under the velocity-shifted control
model; dotted lines mark $P=0.2$ and 0.5. (d) Likelihood test statistic for late-M and L members at the cluster
motion (horizontal lines) and at 64 velocity-shifted control motions (symbols); no control reaches the value at the %
cluster motion in either field model. Motions and probabilities include the Rubin DP2 $i$-band astrometry.}
\label{fig:vpd}
\end{figure*}

\begin{figure*}
\centering
\includegraphics{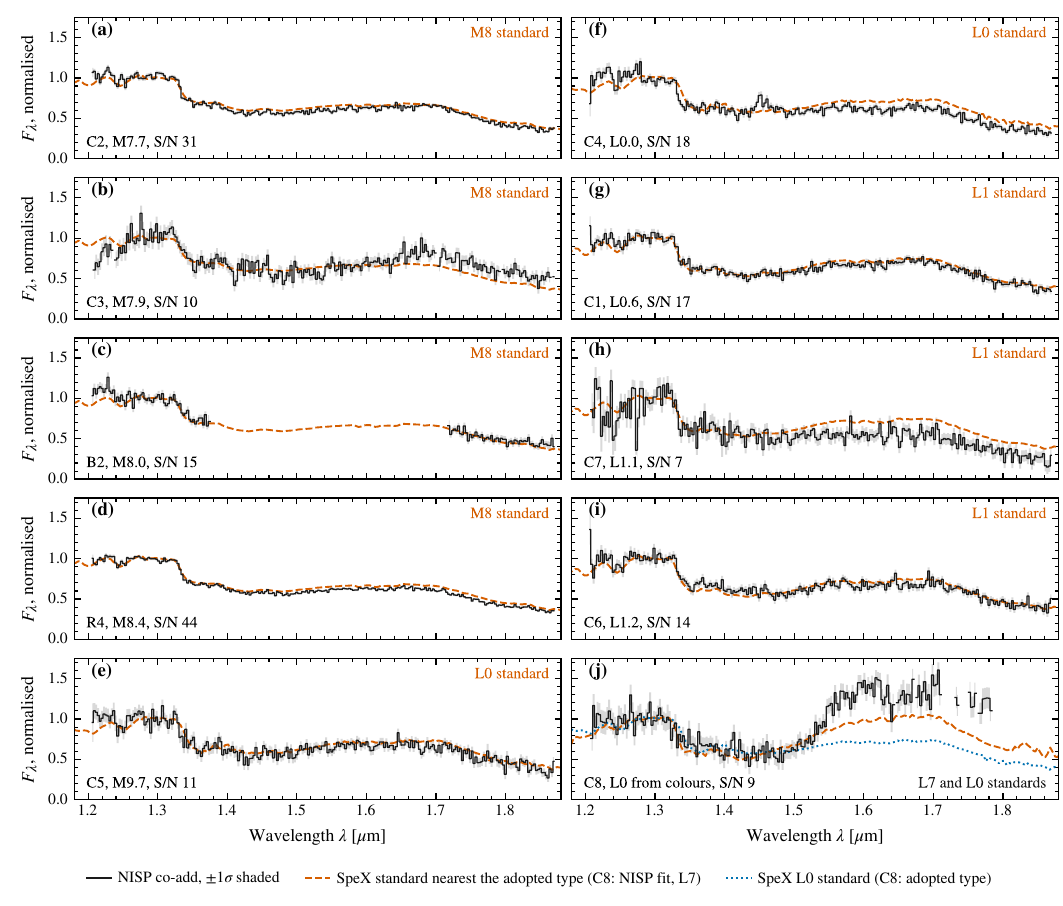}
\caption{Euclid NISP co-added spectra (black, $\pm1\sigma$ shaded) of R4 and the nine new candidates with
$P\geq0.5$, ordered by adopted spectral type and normalised between 1.25 and 1.32\,\micron. Dashed lines are the %
SpeX spectral standards nearest the adopted types, and each panel gives the adopted type and the S/N of the co-add.
The gap in the B2 spectrum marks pixels without valid data in its co-add. (j) C8 has the photometric type L0$\pm$1.9
(dotted line, L0 standard). Both NISP extractions match the L7 standard best (dashed), although even that standard %
fits the co-add poorly (reduced $\chi^2$ of 14). The spectrum departs from both standards in the $H$ band.} %
\label{fig:spectra}
\end{figure*}

We assess the 22 objects in Table~\ref{tab:named} using both field models. A candidate has $P\geq0.5$ under both models; a possible member has $\Pg\geq0.2$ but falls short of that criterion. The object labels were introduced in Section~\ref{sec:intro}, and B2 occupies the boundary slice at $2.6\leq(\IE-\JE)_0<2.7$. Appendix~\ref{app:table} gives the photometry, positions and likelihood terms, and Appendix~\ref{app:finders} the finder charts. The motions are compared with the field distribution in Figure~\ref{fig:vpd}, while Figure~\ref{fig:spectra} shows the spectra of the ten objects with $\Pg\geq0.5$, including C8 conditional on its photometric type.

\begin{deluxetable*}{lllcrrcrccccl}
\tabletypesize{\scriptsize}
\tablecaption{The nine spectroscopically selected ultracool dwarfs (top) and the thirteen new candidates (bottom).\label{tab:named}}
\tablehead{\colhead{Label} & \colhead{Selection} & \colhead{Adopted} & \colhead{$J_\mathrm{E}$} & \colhead{$\mu_{\alpha*}$} &
\colhead{$\mu_\delta$} & \colhead{$Q$} & \colhead{$\chi^2_\mathrm{c}$} & \colhead{$d_\mathrm{f}/d_\mathrm{y}$} &
\colhead{$P_\mathrm{gaia}$} & \colhead{$P_\mathrm{ctrl}$} & \colhead{PM-only} & \colhead{Status}\\
\colhead{} & \colhead{type} & \colhead{type} & \colhead{(AB)} & \colhead{(mas\,yr$^{-1}$)} & \colhead{(mas\,yr$^{-1}$)} &
\colhead{} & \colhead{} & \colhead{(pc)} & \colhead{} & \colhead{} & \colhead{95/99.7} & \colhead{}}
\startdata
R4 & M7 & M8.4$^{i}$ & 18.28 & 19.3 $\pm$ 1.4 & 2.0 $\pm$ 1.3 & B$^{r}$ & 0.04 & 177/189 & 0.86 & 0.93 & Y/Y & candidate \\ %
R1 & L3 & L1$^{t,m}$ & 18.98 & 52.9 $\pm$ 1.0 & $-$7.8 $\pm$ 0.9 & A$^{r}$ & 675 & 161/207 & $<10^{-4}$ & $<10^{-4}$ & N/N & rejected (motion) \\ %
R2 & L5 & L2.4$^{i}$ & 19.50 & 28.0 $\pm$ 0.9 & $-$11.2 $\pm$ 0.9 & A$^{r}$ & 142 & 168/224$^{c}$ & $<10^{-4}$ & $<10^{-4}$ & N/N & rejected (motion) \\ %
R3 & T0 & T0$^{t}$ & 19.73 & 21.9 $\pm$ 1.8 & 27.6 $\pm$ 1.8 & A & 171 & 67/92$^{a}$ & $<10^{-4}$ & $<10^{-4}$ & N/N & rejected (motion) \\ %
R5 & L1 & M9.1$^{i}$ & 19.21 & 10.0 $\pm$ 1.7 & $-$18.4 $\pm$ 1.6 & A$^{r}$ & 141 & 234/261 & $<10^{-4}$ & $<10^{-4}$ & N/N & rejected (motion) \\ %
R6 & L0 & M8.4$^{i}$ & 19.25 & 12.5 $\pm$ 1.0 & $-$16.6 $\pm$ 0.9 & A$^{r}$ & 238 & 277/296 & $<10^{-4}$ & $<10^{-4}$ & N/N & rejected (motion) \\ %
R7 & L0 & L0.8$^{i}$ & 18.81 & $-$51.7 $\pm$ 1.1 & $-$20.4 $\pm$ 1.0 & A$^{r}$ & 2\,836 & 135/174 & $<10^{-4}$ & $<10^{-4}$ & N/N & rejected (motion) \\ %
T2 & T2 & T2$^{t}$ & 17.94 & 28.0 $\pm$ 4.8 & $-$63.6 $\pm$ 4.8 & B & 179 & 30/42$^{a}$ & $<10^{-4}$ & $<10^{-4}$ & N/N & rejected (motion) \\ %
T5 & T5 & T5$^{t}$ & 18.54 & 575.9 $\pm$ 2.5 & $-$468.3 $\pm$ 2.1 & C & 53\,756 & 33/45$^{a}$ & $<10^{-4}$ & $<10^{-4}$ & N/N & rejected (motion) \\ %
\hline
C1 & \nodata & L0.6$^{i}$ & 18.73 & 19.3 $\pm$ 1.1 & 1.8 $\pm$ 1.1 & A$^{r}$ & 0.00 & 137/176 & 0.94 & 0.97 & Y/Y & candidate \\ %
C2 & \nodata & M7.7$^{i}$ & 18.31 & 18.8 $\pm$ 1.0 & 3.7 $\pm$ 0.9 & A$^{r}$ & 1.6 & 208/223 & 0.90 & 0.93 & Y/Y & candidate \\ %
C3 & \nodata & M7.9$^{i}$ & 18.62 & 20.5 $\pm$ 0.9 & 2.7 $\pm$ 0.9 & A$^{r}$ & 1.4 & 230/246 & 0.89 & 0.94 & Y/Y & candidate \\ %
C7 & \nodata & L1.1$^{i}$ & 19.71 & 18.6 $\pm$ 1.5 & 2.1 $\pm$ 1.5 & A & 0.13 & 193/248 & 0.82 & 0.89 & Y/Y & candidate \\ %
C6 & \nodata & L1.2$^{i}$ & 19.58 & 18.4 $\pm$ 1.2 & 1.1 $\pm$ 1.2 & A & 0.61 & 177/228 & 0.79 & 0.87 & Y/Y & candidate \\ %
B2 & \nodata & M8$^{t}$ & 18.78 & 20.7 $\pm$ 1.2 & 1.8 $\pm$ 1.0 & A & 0.79 & 247/264 & 0.68 & 0.77 & Y/Y & candidate \\ %
C4 & \nodata & L0$^{i}$ & 19.23 & 18.0 $\pm$ 2.5 & 0.0 $\pm$ 2.4 & B$^{r}$ & 1.0 & 195/252 & 0.68 & 0.78 & N/Y & candidate \\ %
C8 & \nodata & L0$^{p}$ & 19.41 & 21.3 $\pm$ 1.2 & 2.4 $\pm$ 1.1 & A$^{r}$ & 2.2 & 233/300$^{b}$ & 0.66 & 0.73 & N/Y & candidate$^{e}$ \\ %
C5 & \nodata & M9.7$^{i}$ & 19.47 & 17.6 $\pm$ 0.9 & 0.7 $\pm$ 0.9 & A$^{r}$ & 2.5 & 232/300 & 0.56 & 0.70 & Y$^{d}$/Y & candidate \\ %
N1 & \nodata & L4$^{t}$ & 20.32 & 17.3 $\pm$ 1.3 & $-$0.6 $\pm$ 1.3 & A & 4.5 & 176/235$^{c}$ & 0.45 & 0.62 & N/Y & possible \\ %
N2 & \nodata & L1.9$^{i}$ & 20.08 & 22.5 $\pm$ 1.6 & 2.5 $\pm$ 1.5 & A & 3.6 & 243/321$^{c}$ & 0.41 & 0.44 & N/Y & possible \\ %
N3 & \nodata & T4.9$^{p}$ & 21.24 & 10.4 $\pm$ 4.6 & $-$1.3 $\pm$ 4.6 & C & 4.0 & 117/159$^{a}$ & 0.29 & 0.35 & N/N & possible \\ %
C9 & \nodata & M8.4$^{i}$ & 19.57 & 22.0 $\pm$ 1.2 & 1.5 $\pm$ 1.2 & A$^{r}$ & 3.3 & 321/343 & 0.22 & 0.32 & N/Y & possible \\ %
\enddata
\tablecomments{Selection: spectral type assigned at the spectroscopic selection (Section~\ref{sec:intro}). Adopted: NISP type from the
\citet{Allers2013} indices ($^i$) or template fit ($^t$); $^p$ photometric type;
$^m$ low-metallicity flag (Section~\ref{sec:typing}). Proper motions are the joint fits (cluster parallax fixed), including the Rubin epochs where marked, with
adopted errors. $Q$: astrometric quality grade of the fit without the Rubin epochs (Section~\ref{sec:astrometry}); $^r$ Rubin DP2
$i$-band positions included in the motion. With the Rubin epochs only the grade of R4 would change, from B to A, because its fit
$p$-value rises from 0.0005 to 0.025, above the limit of 0.01.
$\chi^2_\mathrm{c}$: $\chi^2$ (2 degrees of freedom) against the comoving motion at 234.4~pc.
$d_\mathrm{f}/d_\mathrm{y}$: spectrophotometric distance on the field relation ($d_\mathrm{f}$) and with the ATMO youth offset at 100~Myr ($d_\mathrm{y}$).
$P_\mathrm{gaia}$, $P_\mathrm{ctrl}$: membership probability under the Gaia and control field models.
PM-only: passes the proper-motion-only classification at 95\% and 99.7\% (Section~\ref{sec:pmonly}).
Status: candidate, $P\geq0.5$ under both field models; possible, $P_\mathrm{gaia}\geq0.2$ otherwise; rejected (motion), the spectroscopically
selected objects whose motions exclude membership.}
\tablenotetext{a}{UltracoolSheet relation for T dwarfs; the four-calibrator NISP T relation gives 18, 23 and 73~pc for T5, T2 and R3 (Section~\ref{sec:specphot}).}
\tablenotetext{b}{For the adopted photometric L0 type. Both NISP extractions give L7, which the Legacy Surveys and CatWISE colours disfavour (Section~\ref{sec:res_new}); as an L7 it would lie at 69/93~pc.}
\tablenotetext{c}{UltracoolSheet relation (types later than L1.5).}
\tablenotetext{d}{Falls below the 95\% threshold when the classifier is evaluated with the proper motion listed here.}
\tablenotetext{e}{A candidate only if its colours give its type: with the L7 spectroscopic terms switched on, $P_\mathrm{gaia}=0.42$ and $P_\mathrm{ctrl}=0.51$ (Section~\ref{sec:res_new}).}
\end{deluxetable*}

\subsection{The spectroscopically selected ultracool dwarfs}\label{sec:res_report}

Only R4 of the spectroscopic selection moves with the cluster. Its motion gives $\chi^2=0.04$ against the projected cluster motion, and membership probabilities of $\Pg=0.86$ and $\Pc=0.93$ (Table~\ref{tab:named}). Its M8.4 type places it at 189\,pc with the ATMO youth correction. The distance modulus differs from the cluster's by $-0.46\pm0.48$\,mag with that correction and $-0.31\pm0.51$\,mag with the AB~Doradus correction, so both estimates agree with membership. Its field-like gravity indices are also expected for a late-M member at this age (Section~\ref{sec:specphot}). We assign grade B to its astrometry because the joint fit has $p=0.0005$, and include the Birge factor of 1.70 in its errors. Taken together, the motion and distance make R4 a good candidate member.

The motions of the other eight exclude present-day membership: their $\chi^2$ values range from 141 for R5 to 53,756 for T5, for two degrees of freedom, and both field models give probabilities below $10^{-4}$. The proper-motion-only classification %
rejects the same eight (Section~\ref{sec:pmonly_res}). Three of them are foreground T dwarfs. For T5, we measure a total motion of 744\,\masyr\ at an assumed parallax of 31.25\,mas. This agrees with $736\pm23$\,\masyr\ from \citet{Karpov2025}; our fit with a free parallax gives $21.8\pm8.7$\,mas. The rounded motion components are $(576,\,-468)$\,\masyr. Both distance calibrations place these T dwarfs in the foreground (Table~\ref{tab:named} and its note a). %
The remaining five (R1, R2, R5, R6 and R7) are late-M and L dwarfs at 135--277\,pc on the field relation, and for R5 %
and R6 the distance matches the cluster's. The motions therefore separate these field dwarfs from members even where their spectra and distances agree with the cluster. %

\subsection{New candidates}\label{sec:res_new}

We identify nine new candidates with $P\geq0.5$ under both field models: C1--C8 and B2, with C8 included on its photometric type. With R4, ten of the
4,596 sources have $\Pg\geq0.5$, and the model expects 2.2 of them to be field dwarfs. The control model places %
eleven sources above 0.5, adding N1, with 1.9 expected field dwarfs. Apart from C8, their NISP types run from M7.7 to %
L1.2, and their \JE\ magnitudes from 18.3 to 19.7. With the ATMO youth correction, all nine with spectroscopic types %
lie within 1.1 standard deviations of the cluster distance modulus. The probabilities come mainly from the
proper motion ($\mathrm{LR}_\mu=31$--134), with the spectroscopic terms between 0.69 and 2.5 %
(Table~\ref{tab:fullb}). R4, C2, C4 and C7 are inside $r_J$, at 1.7--2.4\arcdeg\ from the centre. The other six candidates are 4.1--7.4\arcdeg\ away, or 17--30\,pc in projection. %

C1 provides the strongest individual evidence for membership: it has the highest probability, $\Pg=0.94$, and a weak low-gravity hint. Its KI$_J$ and $H$-cont z-scores are 2.0 and 1.0, with a mean of 1.54 and a VL-G class. Omitting the model's binary component leaves its membership probability unchanged (Table~\ref{tab:fullb}). The distance depends on the youth correction: the field relation places it $2.4\sigma$ in front of the cluster, the ATMO correction reduces the difference to $1.1\sigma$, and the AB~Doradus correction gives 136\,pc, or $2.2\sigma$ in front. An unresolved binary of nearly equal mass would reconcile the field-relation distance with the cluster distance.

Blending is the main concern for C4. Both VIKING and KiDS classify it as extended (Section~\ref{sec:offeuclid_purity}), and its grade-B astrometry has a Birge factor of 2.55 and errors of 2.5\,\masyr. In the Legacy Surveys \citep{Dey2019}, a source 3.4\,mag fainter in $z$ lies 1.8\arcsec\ away, close enough to blend in the ground-based images and plausibly account for the large Birge factor. This neighbour also contaminates the CatWISE photometry, making the WISE colours unusable. We regard C4's weak low-gravity hint as an artefact because it comes from $H$-cont alone, with a z-score of 2.9 caused by a spike in the continuum window. C4 consequently has the weakest astrometry of the five objects selected under every model variant below.

For C6, the distance favours membership while the gravity indices weakly disfavour it (Table~\ref{tab:fullb}). C3 is unusually red in the mid-infrared: $W1-W2=0.50\pm0.06$, redder than both field and young M8 dwarfs. The corresponding M8 colours are $0.23\pm0.04$ for field dwarfs and $0.26\pm0.04$ for young dwarfs, differences of 4 and 3 standard deviations. Neither Legacy Surveys nor CatWISE has a neighbour within 6\arcsec, and the Euclid sources there are at least 4\,mag fainter in \JE. An unresolved cooler companion would produce this red colour.

The co-added spectra of C3, C7 and C8 have signal-to-noise ratios of 7.1--10, so their types are less secure than Table~\ref{tab:named} suggests. C3 lies at the edge of the NISP coverage gap (Section~\ref{sec:field}), where its photometry comes from one exposure with errors of 0.007--0.010\,mag. Its spectrum likewise comes from one dither; every pixel of the DR1 combined spectrum is masked, leaving only our extraction for typing. The motion and photometry alone give C3 $\Pg=0.84$ (Table~\ref{tab:sens}). %

C8 qualifies as a candidate only under its colour-based L0 classification. Both NISP extractions favour an L7 standard, but the Euclid colours favour L0 (Section~\ref{sec:typing}). The Legacy Surveys and CatWISE colours support the earlier type, matching M9--L0 within 1.8 standard deviations. We measure $z_\mathrm{PS1}-J=2.64\pm0.04$, where PS1 denotes the Pan-STARRS1 photometric system, and $J-W2=1.26\pm0.20$. They differ from L7 by 4.1 and 3.5 standard deviations, respectively, with no neighbour within 6\arcsec\ to blend them \citep{Dey2019,Marocco2021}. We therefore set the spectroscopic terms to 1 in the adopted model, giving $\Pg=0.66$ from the motion and prior. The L0 photometric distance of 233\,pc uses the same colour information as the sequence offset. Adopting L7 and its 69\,pc distance instead gives $\Pg=0.42$ and $\Pc=0.51$: the type term disfavors membership ($\mathrm{LR}_\mathrm{type}=0.13$), while the distance term conditioned on the offset favours it ($\mathrm{LR}_d=2.9$). Thus the disagreement in type changes C8's membership classification.

We classify N1, N2, N3 and C9 as possible members because $0.2\leq\Pg<0.5$ (Table~\ref{tab:named}). N1 is the only one that meets the candidate threshold under the control model. The spectra of N1 and N2 have signal-to-noise ratios %
of 6.8 and 8.9, and the Legacy Surveys $z-\JE$ and $z-\YE$ colours of N2 are 0.3 and 0.25\,mag redder than %
those of the other named dwarfs at its type, closer to L3--L4. N3 is the faintest object ($\IE=26.0$, %
$\JE=21.2$). It is an extended source with a grade-C motion ($\pm4.6$\,\masyr), and its motion adds little %
information ($\mathrm{LR}_\mu=3.6$). Its photometric distance places it in front of the cluster, at 117\,pc on the %
UltracoolSheet relation ($2.3\sigma$) or 64\,pc on the NISP T relation. Its absence from the Legacy Surveys
($z>22.7$) and CatWISE ($W2>17.3$) is expected at its $J$ for either a T dwarf or a galaxy. Every T-class source %
with a reliable motion has $P<0.3$, and the fitted member amplitude of the L-T and T classes is zero. %

\begin{deluxetable}{llcccccccc}
\tabletypesize{\scriptsize}
\tablecaption{Membership probabilities of the comoving candidates under variants of the model.\label{tab:sens}}
\tablehead{\colhead{Label} & \colhead{Type} & \multicolumn{2}{c}{Adopted} & \multicolumn{2}{c}{No spectra} &
\multicolumn{2}{c}{Radial prior} & \multicolumn{2}{c}{Fixed centre}\\
\colhead{} & \colhead{} & \colhead{$P_\mathrm{gaia}$} & \colhead{$P_\mathrm{ctrl}$} & \colhead{$P_\mathrm{gaia}$} &
\colhead{$P_\mathrm{ctrl}$} & \colhead{$P_\mathrm{gaia}$} & \colhead{$P_\mathrm{ctrl}$} & \colhead{$-0.56$} & \colhead{$-1.17$}}
\startdata
C1 & L0.6$^{i}$ & 0.94 & 0.97 & 0.82 & 0.90 & 0.94 & 0.97 & 0.92 & 0.87 \\ %
C2 & M7.7$^{i}$ & 0.90 & 0.93 & 0.75 & 0.82 & 0.94 & 0.96 & 0.92 & 0.85 \\ %
C3 & M7.9$^{i}$ & 0.89 & 0.94 & 0.84 & 0.90 & 0.87 & 0.93 & 0.96 & 0.96 \\ %
R4 & M8.4$^{i}$ & 0.86 & 0.93 & 0.76 & 0.87 & 0.98 & 0.99 & 0.91 & 0.88 \\ %
C7 & L1.1$^{i}$ & 0.82 & 0.89 & 0.71 & 0.80 & 0.90 & 0.93 & 0.46 & 0.46 \\ %
C6 & L1.2$^{i}$ & 0.79 & 0.87 & 0.70 & 0.81 & 0.64 & 0.76 & 0.37 & 0.36 \\ %
C4 & L0$^{i}$ & 0.68 & 0.78 & 0.59 & 0.70 & 0.96 & 0.97 & 0.54 & 0.53 \\ %
B2 & M8$^{t}$ & 0.68 & 0.77 & 0.56 & 0.67 & 0.52 & 0.62 & 0.21 & 0.01 \\ %
C8 & L0$^{p}$ & 0.66 & 0.73 & 0.67 & 0.74 & 0.48 & 0.56 & 0.30 & 0.27 \\ %
C5 & M9.7$^{i}$ & 0.56 & 0.70 & 0.60 & 0.74 & 0.37 & 0.51 & 0.15 & 0.10 \\ %
\hline
N1 & L4$^{t}$ & 0.45 & 0.62 & 0.24 & 0.38 & 0.60 & 0.74 & 0.11 & 0.11 \\ %
N2 & L1.9$^{i}$ & 0.41 & 0.44 & 0.47 & 0.51 & 0.56 & 0.59 & 0.16 & 0.16 \\ %
N3 & T4.9$^{p}$ & 0.29 & 0.35 & 0.29 & 0.35 & 0.15 & 0.19 & 0.29 & 0.29 \\ %
C9 & M8.4$^{i}$ & 0.22 & 0.32 & 0.36 & 0.48 & 0.22 & 0.30 & 0.02 & 0.01 \\ %
\enddata
\tablecomments{Named objects (Table~\ref{tab:named}) with a probability of at least 0.3 in any variant, above the line those
with $P_\mathrm{gaia}\geq0.5$. Each variant is applied on its own. No spectra: spectral-type, distance and gravity terms
removed; the Gaia parallax term is kept. Radial prior: prior multiplied by the radial density of the 91 Gaia DR3 stars that pass either version of the comoving test of Section~\ref{sec:field}. Fixed
centre: the member sequence held at the youth-corrected ridge, 0.56 or 1.17\,mag brighter than the field ridge at the blue end, instead of the
fitted location (Gaia field model; the control model gives higher values). Types as in Table~\ref{tab:named}.}
\end{deluxetable}

We assess how the prior, youth correction and sequence location affect individual membership probabilities (Table~\ref{tab:sens}). All ten candidates remain above $\Pg=0.5$ when the spectroscopic terms are omitted, showing that their selection rests on the motions and photometry. Halving the expected member count lowers B2, C8 and C5 below this threshold, while a prior following the Gaia radial density lowers C8 and C5 and raises N1 and N2 above it. R4, C1--C4, C6 and C7 meet the threshold under both field models in all three tests. B2 also meets it with the spectra omitted and with the radial prior. %

The sequence location matters most for the fainter L candidates. We fit a location $0.53\pm0.16$\,mag fainter than the youth-corrected ridge, within about 0.2\,mag of the field-dwarf ridge at the cluster distance (Section~\ref{sec:disc_field}). Holding the sequence on the youth-corrected ridge lowers C7, C6, B2, C8 and C5 below $P=0.5$, while R4 and C1--C4 exceed it at every tested location. Five of the ten candidate classifications therefore depend on fitting the sequence location.

Changing the distance correction has a smaller effect on the classifications. Replacing ATMO with the AB~Doradus correction lowers C1 and C6 by 0.12, to 0.82 and 0.67. N2 rises from 0.41 to 0.57 and is the only object to cross $\Pg=0.5$. This comparison uses motions without Rubin, whose inclusion changes every probability by 0.05 or less (Section~\ref{sec:rubin_res}). Removing the binary component changes the ten candidates' probabilities by at most 0.04; the larger change for N3 is listed in Table~\ref{tab:fullb}. %

\subsection{The proper-motion-only classification}\label{sec:pmonly_res}

The independent classification from proper motions alone rejects the same eight spectroscopically selected field dwarfs. Within the colour--magnitude band, its 95\% recovery threshold selects 8 sources against 1.1--1.5 expected from the shifted-velocity controls, giving about 85\% purity. This threshold recovers a true member with grade-A errors in only 50--80\% of simulated trials, so we also keep candidates selected at the looser 99.7\% threshold and flag both results in Table~\ref{tab:named}. The additional group contains 7--8 sources against 3.1--3.5 expected by chance.

Outside the objects in Table~\ref{tab:named}, the classifier selects 3 sources at the 95\% threshold against $6.2\pm3.6$ expected by chance. At the 99.7\% threshold, it selects 119 against $104\pm38$. These counts span all colours and magnitudes and exclude the candidates responsible for the excess within the member band. We find the excess only among the objects already identified in that band. %

\subsection{Population test}\label{sec:population}

We find an excess of comoving late-M and L dwarfs in the colour--magnitude band without using individual membership probabilities. Of 193 compact late-M and L dwarfs in the band with grade A or B motions, 14 meet the comoving criterion. The weighted Gaia field model predicts 7.1 (Poisson $p=0.015$), and the fixed-window model predicts 6.3 ($p=0.006$). We assess the field prediction behind the band, 1.0 to 2.5\,mag fainter than the ridge, where few members are expected. Of 598 sources there, 65 meet the motion criterion, close to the weighted model's prediction of 68; the fixed-window model predicts 40.
Inside $r_J$, 6 of 34 band objects pass against 1.1 predicted ($p=0.001$). On a finer grid of 441 velocity shifts at %
2\,\kms\ spacing, the count is 14 at the cluster motion and at most 17 anywhere. All 14 other velocity shifts selecting at least 14 sources are within 8.2\,\kms\ of the cluster motion. These nearby shifts still select real members, so the counts peak around the cluster velocity.

Fitting the population uses all 1,213 late-M and L sources with motions that satisfy the quality cuts. Its statistic is 30.4 under the %
Gaia field model and 42.2 under the control model. The largest values among the 64 velocity-shifted controls are 5.6 %
and 4.3. The observed statistic is more than five times the largest Gaia-model control value. In both models, the fraction of trials at least as extreme, including the cluster itself, is $1/65=0.015$, the smallest value available with 64 controls; Section~\ref{sec:disc_limits} discusses its calibration. We fit 11 late-M and L members under the Gaia model, with an approximate interval of 8--15 (Section~\ref{sec:membership_model}). The control model gives 12 (9--17), while the summed probabilities of the same sources are 8.7. Scaled to the 58--65 %
Gaia stars of the field, the \citet{Chabrier2003} system mass function predicts 5.2 late-M and L members after the %
quality cuts, so the fitted number is 2.2 (1.5--2.9) times this prediction under the Gaia model. %

\begin{figure}
\centering
\includegraphics{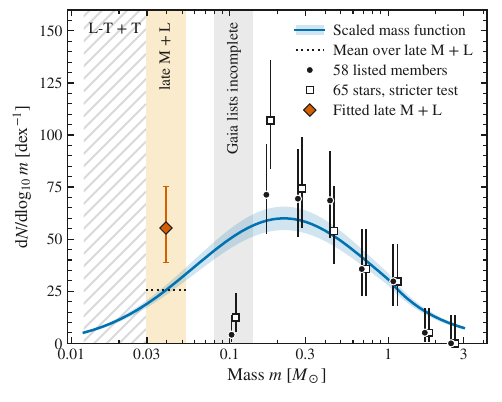}
\caption{Mass function of Blanco~1 in the Euclid field. The curve is the \citet{Chabrier2003} system mass function, %
lognormal below 1\,$M_\odot$. Its integral over 0.08--3\,$M_\odot$ is set to the 58 listed members that pass the Gaia %
test, or its integral over 0.10--3\,$M_\odot$ to the 65 stars that pass its stricter version; the band covers the two %
scalings and the solid line is their mean. Circles and squares show the two overlapping samples, with masses from
their Gaia $G$ magnitudes on the 150\,Myr PARSEC isochrone \citep{Bressan2012,Chen2014} at 234\,pc; error bars are %
68\% Poisson intervals. The listed members end at $G=19.8$, or 0.14\,$M_\odot$, so the bin below that mass is %
incomplete; one star of 3.3\,$M_\odot$, in both samples, is not shown. The gold and hatched intervals denote the late-M and L classes and the L-T and T classes, respectively. The 150\,Myr ATMO mass boundaries are 0.0294--0.0532\,$M_\odot$ for late-M and L dwarfs. The L-T and T classes span 0.0120--0.0294\,$M_\odot$, divided at 0.0205\,$M_\odot$; the lower limit of 0.012\,$M_\odot$ corresponds to the selection limit. Over the L-T and T classes the curve integrates to 4.2 members, %
or 3.2 after the quality cuts, which keep 72\% of L-T and 80\% of T dwarfs; the fitted number is zero, against 1.9 predicted among the quality-passing L-T and T %
sources (The difference is explained in Section~\ref{sec:population}). The diamond, on the scale of the curve before the quality cuts, is the fitted number of late-M and L %
members divided by the fraction of members that pass the quality cuts, 0.78, and by the 0.26\,dex width of the %
class, with its approximate interval (Section~\ref{sec:membership_model}). The dotted line is the mean of the curve %
over the class, where the curve integrates to 6.6 members, or 5.2 after the quality cuts; the ratio of the diamond %
to the dotted line is 2.2 (1.5--2.9).} %
\label{fig:imf}
\end{figure}

We assess the normalisation by assigning masses to the field's Gaia stars from their $G$ magnitudes (Figure~\ref{fig:imf}). Scaling to the 58 listed members satisfying the Gaia test predicts 47 stars in the mass range from 0.14 to 3\,$M_\odot$. The lower boundary is approximately the mass of the faintest listed member. We count 56 stars in this range (Poisson $p=0.20$). %
The number of these stars and their mass distribution (Kolmogorov--Smirnov $p=0.27$) are consistent with the scaled %
mass function. The adopted scalings integrate the mass function from 0.08 and 0.10\,$M_\odot$, below the faintest listed member, which lowers the prediction if the lists are incomplete there. Normalising within the observed stellar mass range gives 5.0--6.1 late-M and L members after the quality cuts, depending on the sample, stellar models and age. The fitted population is then 1.8--2.2 times the expectation. The prediction at brown-dwarf masses is the extension of the %
lognormal shape below the stellar mass range.

We fit all photometric classes separately on either side of $r_J$. Inside, we obtain 5.4 members with an interval of 3.4--7.6. Outside, the fit gives 4.9 (2.6--7.9). The fraction inside $r_J$ is therefore $53\pm17$\%. The prior is the same at every position, and its %
sums in the two regions, 1.4 and 5.3, only follow the distribution of all selected sources. For comparison, 45--59\% of the field's Gaia members lie inside $r_J$: 45\% of the 91 stars satisfying either Gaia comoving test and 59\% of the 64 listed members. The two fractions therefore agree within the errors.

For the L-T and T classes the fitted number is zero, against 1.9 predicted among the quality-passing L-T and T sources, with an upper end %
of its interval at 0.9, whereas the mass function predicts 3.2 after the quality cuts (Figure~\ref{fig:imf}). The expectation of 1.9 comes from summing the prior over the 38 selected L-T and T sources with motions (Section~\ref{sec:membership_model}), whereas the mass-function integral predicts members independently of their detection. The 27 L-T sources account for 1.6 expected members, close to the integral prediction of 1.7 after the quality cuts. For the 11 T sources, the corresponding expectations are 0.34 and 1.5. With no members fitted the sequence shift is unconstrained, and over the allowed shifts the sum is 1.5--1.9. %

\subsection{SPHEREx spectra}\label{sec:spherex_res}

\begin{figure*}
\centering
\includegraphics{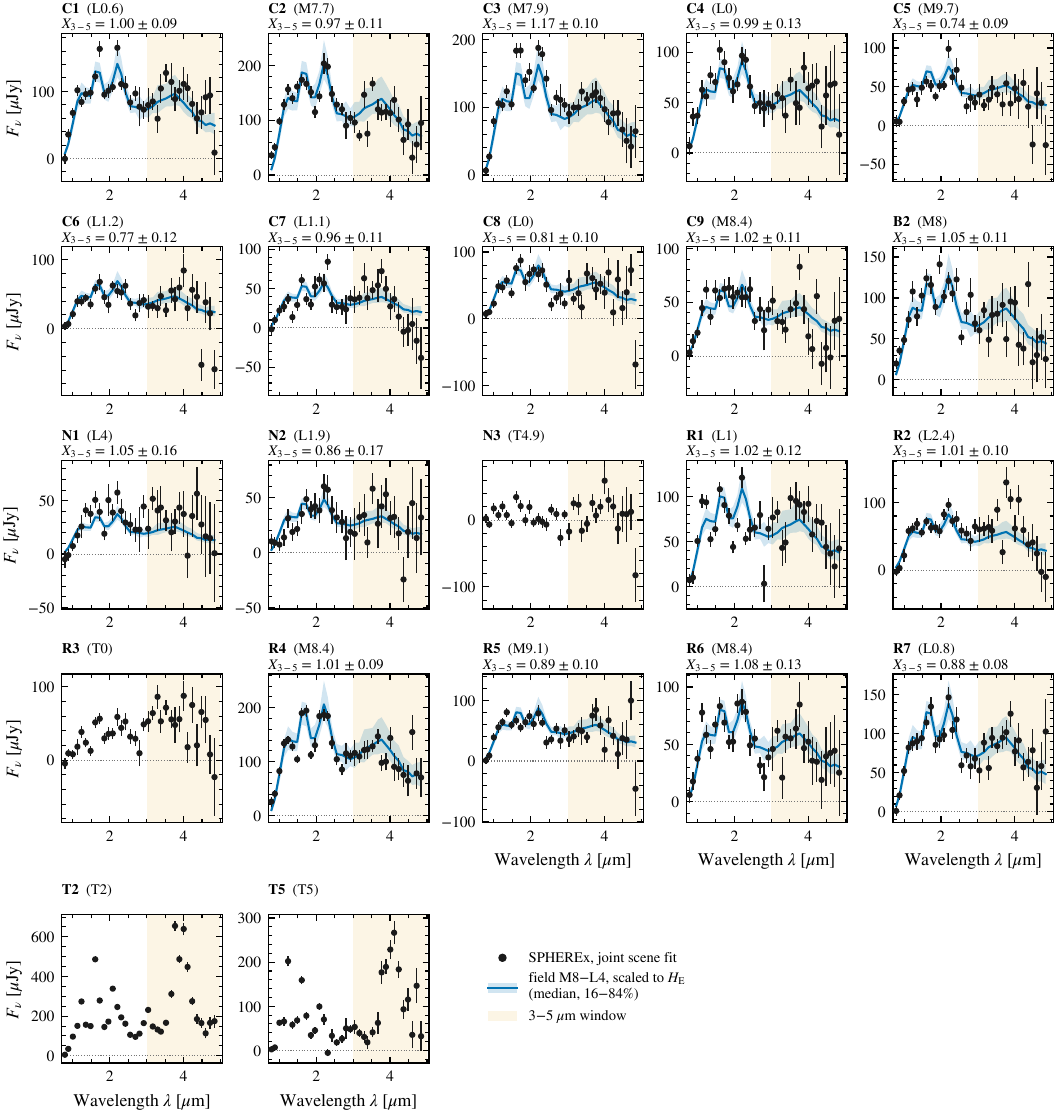}
\caption{SPHEREx spectra of the 22 named objects of Table~\ref{tab:named} (points, $\pm1\sigma$), from a joint fit %
of each object and its catalogued neighbours in all exposures that cover it (Section~\ref{sec:spherex_data}). For late-M and L objects, the blue line and band give the median and 16--84th percentile range of the M8--L4 reference dwarfs, each scaled to the object's \HE\ flux. The flux ratio $X_{3-5}$ is computed from the type-dependent photospheric relation. The shaded region marks the
3--5\,\micron\ window. Each panel gives the object, its adopted spectral type and, for the late-M and L objects,
$X_{3-5}$ with its error. The T-class objects R3, N3, T2 and T5 have no photospheric curve.}
\label{fig:spherex}
\end{figure*}

We use SPHEREx independently of the membership model to test for 3--5\,\micron\ emission above the photosphere, as a warm disk or cool companion would produce, and to compare the 1--5\,\micron\ spectra with field dwarfs of the same type (Figure~\ref{fig:spherex}). We detect all 22 objects except N3 in both wavelength windows. The signal-to-noise ratios are 11--70 in the 0.75--2.4\,\micron\ window. Over 3--5\,\micron, they are 5.4--47. N3, the faintest, reaches 2.9 and 3.2. We define $X_{3-5}$ as the observed-to-photospheric flux ratio over 3--5\,\micron. For the 18 late-M and L objects, including candidates and rejected field dwarfs, it spans 0.74--1.17, with a median of 1.00 and a typical error of 0.11. No object lies three %
standard deviations above the photosphere. Taking the larger of $X_{3-5}$ and 1 and adding three standard deviations,
the data exclude an excess above 23--53\% of the photosphere, 35\% for a typical object. The absence of a disk at this level agrees with the expectation for Blanco~1's age. The lowest ratios, 0.74--0.81 for C5, C6 and C8, belong to objects with
$\JE=19.4$--19.6, 1.4--1.6\,mag fainter than any reference dwarf, so they more likely reflect the extrapolated
photospheric relation than a flux deficit.

C3 has the largest ratio, $1.17\pm0.10$. Over 4.2--5.0\,\micron, we measure $1.29\pm0.25$ times the photospheric flux for C3. In the nearby CatWISE $W2$ band the ratio is $1.44\pm0.17$. The two agree, but the SPHEREx value is
also consistent with no excess, so it does not test the cooler companion suggested by the $W1-W2$ colour of C3
(Section~\ref{sec:res_new}). The two brightest T dwarfs, T2 and T5, show the expected 3.3\,\micron\ methane trough and the peak near 4\,\micron. Their 4.2--5.0\,\micron\ fluxes are $0.82\pm0.06$ and $1.06\pm0.14$ times the $W2$-scaled T-dwarf reference, providing a check on the extraction at the red end. Over 1--5\,\micron, the late-M and L spectra follow those of field dwarfs at their adopted types.

\subsection{Rubin's contribution}\label{sec:rubin_res}

\begin{figure*}
\centering
\includegraphics{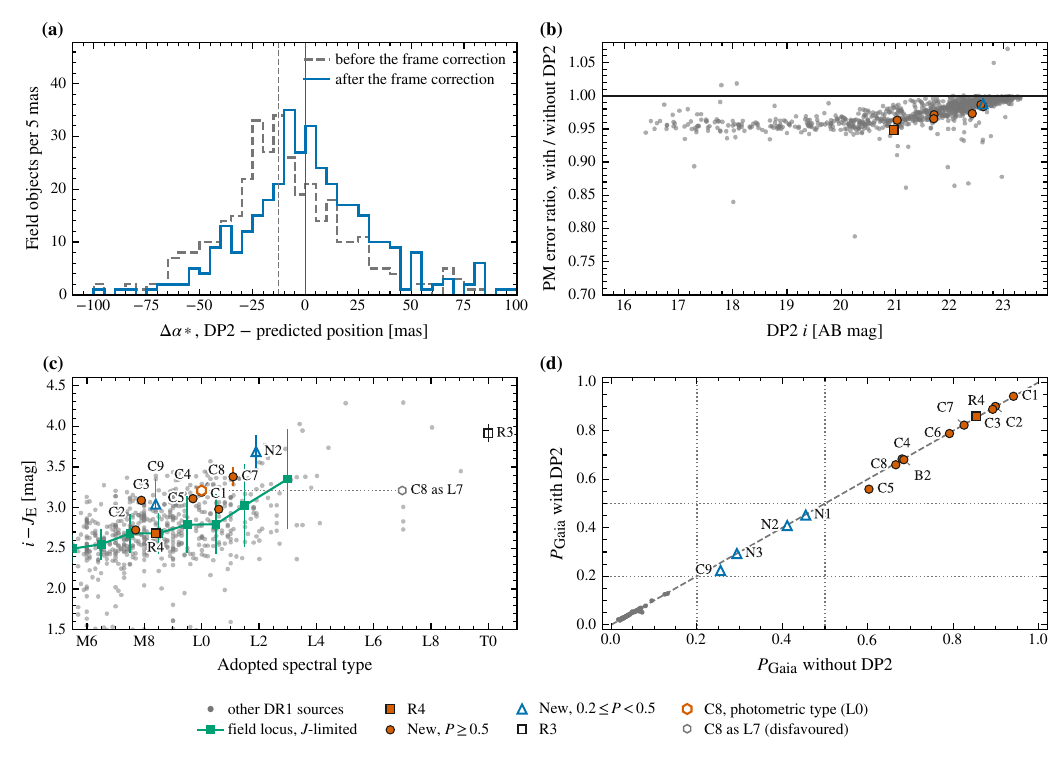}
\caption{Contribution of the Rubin DP2 $i$-band data. (a) DP2 position minus the position predicted by the joint
motion without DP2, in right ascension, for the 361 field objects that define the frame correction, before (dashed) and %
after (solid) removing the common offset of $-12.6$\,mas; the declination offset is $+5.3$\,mas. (b) Ratio of the %
formal proper-motion error with and without the DP2 epochs, against DP2 $i$ magnitude; the median ratio is 0.978. %
(c) $i-\JE$ against adopted spectral type, with the locus of a $J$-limited field sample and its scatter (green). Grey
dots in (b)--(d) are other DR1 sources. C8
is plotted at its adopted L0 type and at the disfavoured L7 type of its spectra; its colour agrees with the locus near
L0. The locus ends before the T types and before R3 (T0). (d) Membership probability under the Gaia
DR3 field model with the DP2 astrometry against the probability without it; no object crosses $P=0.2$ or 0.5. %
}
\label{fig:rubin}
\end{figure*}

The Rubin DP2 positions extend the astrometric baseline by a median of two years for 958 candidates, including 13 %
of the 22 named objects. They shrink the formal proper-motion errors by a median of only 2\%, because the DP2 epochs %
are recent and few compared with the 13\,yr VIKING, KiDS and Euclid baseline. Every source keeps its probability %
category at 0.5 and 0.2 in both field models, and its comoving classification. Among the objects, C5 and C9 %
lose 0.03--0.04 in \Pg\ (C5 from 0.60 to 0.56, Figure~\ref{fig:rubin}). Archival DECam images give the objects %
an independent test. Adding 92 DECam epochs from 2015--2026 for 21 objects changes the motions by at most 0.4 times their adopted errors and the membership probabilities by at most 0.03. The late-M and L likelihood statistic increases from 30.4 to 30.8. %

DP2 also adds an optical colour that does not depend on spectral type. For 443 compact field sources, the DP2 %
$i-\IE$ colour follows a smooth relation with $\IE-\JE$ with a robust scatter of 0.069\,mag. All 16 objects %
with DP2 $i$-band means lie within 1.6 standard deviations of it, which supports the identification of their optical %
counterparts. The $i-\JE$ colours of C8 and R3 resemble those of L0--L2 and about L3 dwarfs, respectively (Table~\ref{tab:fullb}), adding an independent colour comparison to the type disagreements in Section~\ref{sec:typing}.

\section{Tracing the motions back in time}\label{sec:traceback}

Eight of the nine spectroscopically selected ultracool dwarfs move differently from Blanco~1 today. A member that escaped would
also move differently, so a mismatched motion alone does not rule out an origin in the cluster. If any of the eight
left Blanco~1 during its lifetime, its track traced backwards should pass close to the cluster centre, and the
passage should be slow, at a relative speed comparable to those at which members escape. We traced each of the eight back to test for such a passage, and applied the same calculation to R4 and the comoving candidates.

\subsection{Encounter geometry and reference speeds}\label{sec:tb_geometry}

The radial velocities of the ultracool dwarfs are unknown, and their distances are uncertain by 0.34--0.89\,mag in %
distance modulus. The encounter geometry lets us bound the relative speed despite these uncertainties. Let $\hat{\boldsymbol r}$ be the
direction to the object and $\boldsymbol\Pi=\mathsf{I}-\hat{\boldsymbol r}\hat{\boldsymbol r}^{\mathsf T}$, with $\mathsf{I}$ the
identity matrix, the projection across the line of sight. The velocity of the object relative to the cluster has a part along $\hat{\boldsymbol r}$,
set by the unknown radial velocity, and a part across it,
\begin{equation}
\boldsymbol v_\perp=\boldsymbol v_\mathrm{tan}-\boldsymbol\Pi\,\boldsymbol v_\mathrm{cl},
\label{eq:vperp}
\end{equation}
where $\boldsymbol v_\mathrm{tan}$ follows from the proper motion and the distance and $\boldsymbol v_\mathrm{cl}$ is the
cluster velocity. On a straight-line track the relative speed at any past encounter is at least
$v_\perp=|\boldsymbol v_\perp|$, whatever the radial velocity. Minimising $v_\perp$ over distances from 5 to %
2000\,pc gives a floor that depends on neither the radial velocity nor the distance. Because Blanco~1 lies at %
$b=-79\arcdeg$, the line of sight is within 11\arcdeg\ of the vertical, and the free radial velocity mostly sets the %
vertical relative motion.

Slow evaporation and few-body ejection set different reference speeds. A \citet{Plummer1911} sphere of $465\,M_\odot$ with scale radius %
2.39\,pc has a central escape speed of 1.3\,\kms. In the same model a star needs 0.2\,\kms\ from $r_t$, and %
1.1\,\kms\ from the centre, to reach the distance of the Lagrange points, through which evaporating stars leave
\citep{Kupper2008}. These speeds are comparable to the one-dimensional velocity dispersion of 0.4\,\kms\ measured from %
the radial velocities of bright members (Section~\ref{sec:cluster_params}). Few-body interactions typically eject
members at 3--4\,\kms\ \citep{Sterzik1998,Reipurth2001}. In $N$-body models of young regions of 500--700\,$M_\odot$ %
with stars above 0.1\,$M_\odot$ and no primordial binaries, at most 3 stars in 20,000 exceed 30\,\kms\ in projection %
\citep{Schoettler2019}. Field ultracool dwarfs, for comparison, have radial-velocity dispersions of 13--20\,\kms\ %
\citep{Hsu2021,Hsu2024}.

\subsection{Method}\label{sec:tb_method}

We traced the cluster and each object back 150\,Myr, spanning the lithium depletion boundary and isochrone ages of Section~\ref{sec:cluster_params}. For each object, 2000 to 5000 Monte Carlo draws sample the proper-motion errors and adopted distance (Table~\ref{tab:traceback}). That distance comes from the field relation, and for R3, T2, T5 and
N3 from the UltracoolSheet T relation of Section~\ref{sec:res_report}. The four-calibrator NISP T relation places T2 %
and T5 closer and R3 farther; with it, none of these objects has a slow passage either, and the distance-free floor
below does not depend on the choice. C8 is traced at the distance of its adopted photometric L0 type, with the %
disfavoured L7 type of its spectra as a variant. In each draw the radial velocity was scanned from $-150$ to %
$+150$\,\kms. We use the cluster parameters of Table~\ref{tab:cluster} and the proper-motion fits including Rubin (Section~\ref{sec:astrometry}). %

We compare two independent dynamical calculations. The first integrates orbits with galpy \citep{Bovy2015} in the MWPotential2014
model, with the potential of \citet{McMillan2017} as a sensitivity case. The second propagates straight lines and the
linearised tidal (Hill) equations about the cluster, with the Oort constants of \citet{Bovy2017}. A draw passes the %
cluster if it comes within $r_J=10.65$\,pc of the centre, and the passage is slow if the relative speed there is %
below 5\,\kms. For each object we record the fraction of draws with any passage for some radial velocity, the %
fraction with a slow passage, the median over draws of the slowest passage inside $r_J$, and the floor of $v_\perp$.

We calibrate the traceback statistics with synthetic escapees. They were launched from the cluster centre at
0.5--10\,\kms\ in random directions and at random times 0.2--150\,Myr ago, and propagated both in the Hill model and %
in a separate nonlinear Galactic potential. We assigned them the proper-motion errors of the real candidates and distance-modulus errors of 0.55\,mag, omitted their radial velocities, and traced them back with the same code. Random field stars, and field stars selected to pass a comoving proper-motion cut, provide the controls.

\subsection{The eight rejected ultracool dwarfs}\label{sec:tb_rejected}

\begin{figure*}
\centering
\includegraphics{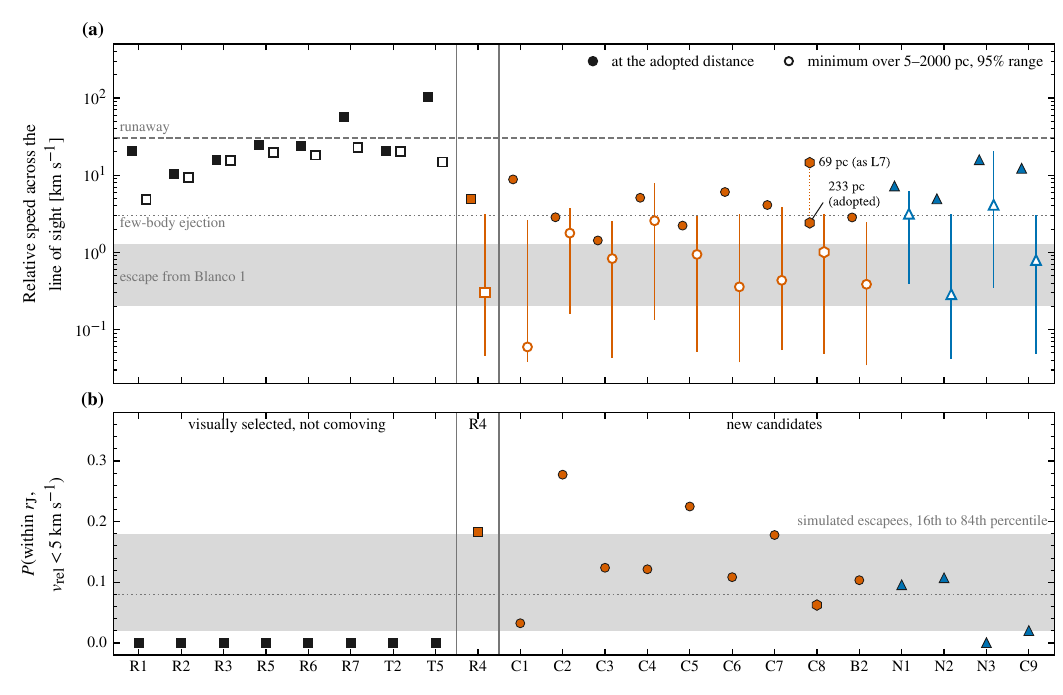}
\caption{Kinematic traceback. (a) Relative speed across the line of sight between each object and the cluster,
at the adopted distance used in the traceback (filled symbols) and minimised over all distances from 5 to
2000\,pc (open symbols, with the 95\% range from the proper-motion errors). This component does not depend on the %
radial velocity, so on a straight-line track it bounds the speed of any past encounter from below. Grey band:
0.2 to 1.3\,\kms, from the speed needed to reach the Lagrange points from $r_t$ to the central escape speed of a %
Plummer model of the cluster. Dotted line: 3\,\kms, the low end of typical few-body ejection speeds. Dashed line,
labelled runaway: 30\,\kms, which at most 3 in 20,000 stars exceed in $N$-body models of young regions \citep{Schoettler2019}. For T2, %
T5 and R3 the filled symbols use the distances of the UltracoolSheet T relation (30, 33 and 67\,pc). C8 is shown at %
the 233\,pc distance of its adopted photometric L0 type and at the 69\,pc distance of the disfavoured L7 type of its %
spectra. The 95\% ranges of the minimum come from the closed-form minimum over 20\,000 proper-motion draws. (b) %
Fraction of Monte Carlo draws, with the radial velocity free between $-150$ and $+150$\,\kms, that pass within $r_J$ %
of the cluster centre in the last 150\,Myr at a relative speed below 5\,\kms\ (galpy, MWPotential2014). The grey band
and dotted line give the 16th to 84th percentiles and the median of the same statistic for simulated escapees in the
field, none of which exceeds 0.35. %
}
\label{fig:traceback}
\end{figure*}

The tracebacks exclude slow evaporation for all eight objects within the tested assumptions: in either code, every passage within $r_J$ is faster than 10\,\kms\ (Table~\ref{tab:traceback}, Figure~\ref{fig:traceback}). Across the eight objects and both codes, the median speed of the slowest encounter is 52--149\,\kms, or 40--115 times the escape speed. These encounters require present-day radial velocities of about 45--130\,\kms. %
The distance-free floor gives the same answer without the orbit integration. At any distance between 5 and 2000\,pc,
seven of the eight move across the line of sight at 9--23\,\kms\ relative to the cluster, with 95\% lower limits of %
at least 8\,\kms. The eighth, R1, reaches 4.9\,\kms\ only at 82\,pc, half its spectrophotometric distance and %
$2.2\sigma$ from it, and at that distance its closest approach misses the cluster by 22\,pc. T5 reaches its floor of %
15\,\kms\ only at the 5\,pc lower end of the distance range.

We tested the sensitivity to distance, cluster velocity and Galactic dynamics. Multiplying the distances by $\sqrt2$ for equal-mass binaries, applying the youth correction, or adopting a cluster radial velocity of 6.23\,\kms\ \citep{Tarricq2021} still excludes slow evaporation. The same conclusion follows with either Galactic potential, the solar motion of \citet{Schonrich2010}, and the bias of at most 0.7\,\kms\ from fitting proper motions at the fixed cluster parallax.

\begin{figure*}
\centering
\includegraphics{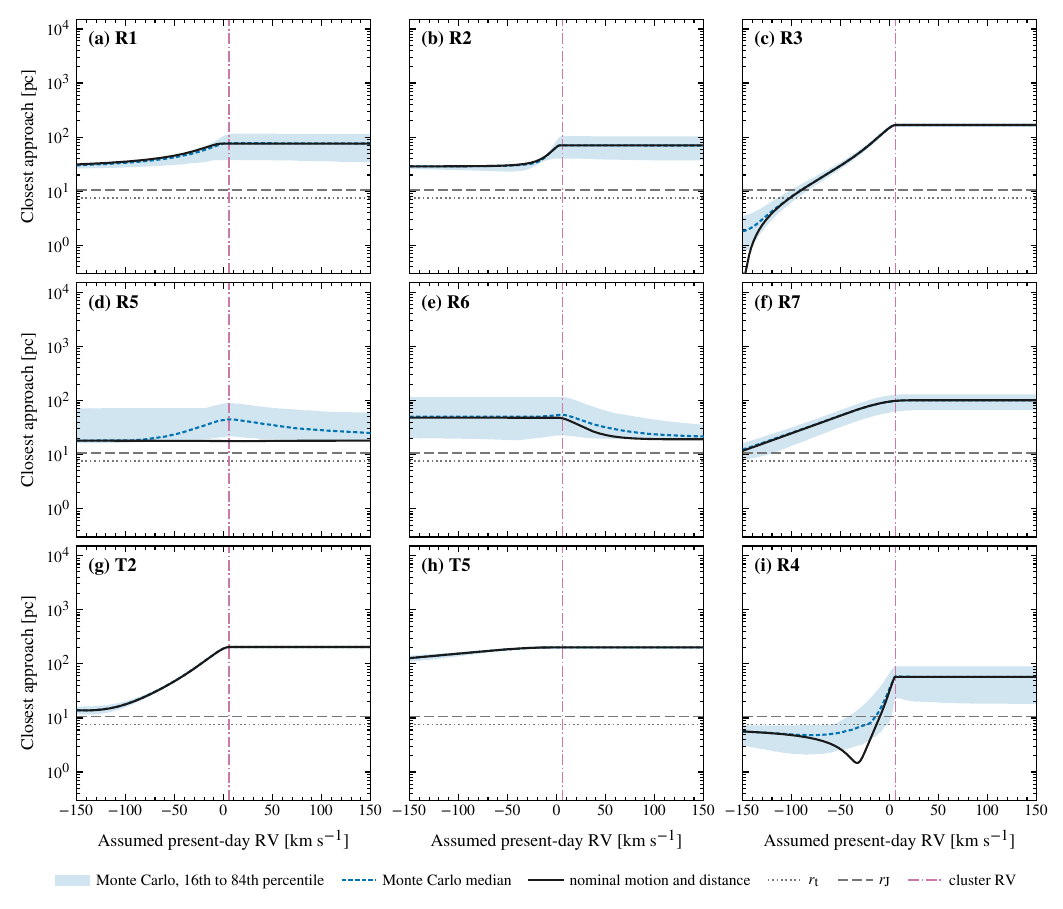}
\caption{Closest approach to the centre of Blanco~1 within the last 150\,Myr against the assumed present-day %
radial velocity, for the eight spectroscopically selected objects that move unlike the cluster (a to h) and for R4 (i). Black:
nominal proper motion and distance; blue dashed line and band: median and 16th to 84th percentiles of the Monte Carlo
draws over proper motion and distance (galpy, MWPotential2014). Distances of T2, T5 and R3 are those of the
UltracoolSheet T relation. Horizontal dotted and dashed lines mark $r_t=7.5$\,pc and $r_J=10.7$\,pc, and the vertical dash-dotted line %
marks the cluster radial velocity of 5.78\,\kms. R3 and R7 enter $r_J$ only for radial velocities of about $-90$ %
to $-150$\,\kms, as fast crossings along the line of sight: R3 in the median curve and R7 only at the 16th percentile. %
T2 comes no closer than 14\,pc in the median curve. %
}
\label{fig:minsep}
\end{figure*}

When any radial velocity is allowed, R3 and R7 pass within $r_J$ in 100\% and 43\% of the draws, and T2 in 9\%. %
These are fast crossings along the line of sight about 1\,Myr ago at about 150\,\kms, with radial velocities %
near the $-150$\,\kms\ bound of the scan. Field stars placed at the same positions show such crossings in 7--45\% of %
draws, and one object above 0.5 among eight, against 0.9--1.3 expected by chance, is what the geometry alone %
produces. Figure~\ref{fig:minsep} shows how the closest approach of each object depends on the assumed radial
velocity. Under the tested potentials, distances and radial-velocity range, the eight objects are therefore
inconsistent with slow evaporation from Blanco~1, and their fast crossings are those expected of field ultracool dwarfs.

\begin{deluxetable*}{lcccccccc}
\tabletypesize{\footnotesize}
\tablecaption{Traceback summary. The eight spectroscopically selected ultracool dwarfs that move unlike the cluster are listed first.\label{tab:traceback}}
\tablehead{\colhead{Label} & \colhead{$d$} & \colhead{$v_\perp(d)$} & \colhead{$v_{\perp,\min}$ (95\%)} & \colhead{$d$ at $v_{\perp,\min}$} &
\colhead{$P_{<r_J}$} & \colhead{$P_{<r_J,\,<5}$} & \colhead{$P_{<r_J,\,<5}$} & \colhead{$v_{\min}(<r_J)$}\\
\colhead{} & \colhead{(pc)} & \colhead{(km\,s$^{-1}$)} & \colhead{(km\,s$^{-1}$)} & \colhead{(pc)} & \colhead{orbit} &
\colhead{orbit} & \colhead{analytic} & \colhead{(km\,s$^{-1}$)}}
\startdata
R1 & 161 & 20.4 & 4.9 (4.1--5.6) & 82 & 0.01 & 0 & 0 & 94 \\ %
R2 & 168 & 10.5 & 9.4 (8.2--10.5) & 136 & 0.01 & 0 & 0 & 59 \\ %
R3 & 67 & 15.8 & 15.4 (13.9--16.8) & 88 & 1.00 & 0 & 0 & 105 \\ %
R5 & 234 & 24.5 & 19.6 (18.0--20.7) & 86 & 0.01 & 0 & 0 & 125 \\ %
R6 & 277 & 24.2 & 18.2 (17.2--19.1) & 115 & 0.04 & 0 & 0 & 102 \\ %
R7 & 135 & 56.6 & 22.8 (22.7--22.8) & 5 & 0.43 & 0 & 0 & 149 \\ %
T2 & 30 & 20.4 & 20.2 (19.0--21.1) & 23 & 0.09 & 0 & 0 & 136 \\ %
T5 & 33 & 102.6 & 14.8 (14.7--14.9) & 5 & $<$0.01 & 0 & 0 & \nodata$^{c}$ \\ %
\hline
R4 & 177 & 4.9 & 0.30 (0.05--3.2) & 231 & 0.99 & 0.18 & 0.18 & 19 \\ %
C1 & 137 & 8.8 & 0.06 (0.04--2.6) & 233 & 0.97 & 0.03 & 0.03 & 34 \\ %
C2 & 208 & 2.8 & 1.8 (0.2--3.8) & 233 & 0.91 & 0.28 & 0.27 & 9 \\ %
C3 & 230 & 1.4 & 0.83 (0.04--2.5) & 218 & 0.43 & 0.12 & 0.10 & 9 \\ %
C7 & 193 & 4.1 & 0.43 (0.05--3.9) & 239 & 0.95 & 0.18 & 0.17 & 22 \\ %
C6 & 177 & 6.1 & 0.36 (0.04--3.1) & 247 & 0.65 & 0.11 & 0.095 & 14 \\ %
B2 & 247 & 2.8 & 0.39 (0.03--2.5) & 219 & 0.34 & 0.10 & 0.09 & 8 \\ %
C4 & 195 & 5.1 & 2.6 (0.1--7.9) & 247 & 0.98 & 0.12 & 0.13 & 21 \\ %
C8 & 233$^{b}$ & 2.4 & 1.0 (0.05--3.2) & 211 & 0.36 & 0.06 & 0.05 & 13 \\ %
C5 & 232 & 2.2 & 0.94 (0.05--2.9) & 256 & 0.61 & 0.23 & 0.20 & 8 \\ %
N1 & 176 & 7.2 & 3.2 (0.4--6.2) & 256 & 0.94 & 0.096 & 0.09 & 24 \\ %
N2 & 243 & 4.9 & 0.29 (0.04--3.2) & 197 & 0.68 & 0.11 & 0.11 & 13 \\ %
C9 & 321 & 12.3 & 0.80 (0.05--3.0) & 204 & 0.15 & 0.02 & 0.02 & 23 \\ %
N3 & 117$^{a}$ & 15.9 & 4.1 (0.3--20.8) & 425 & 0.85 & $<$0.01 & $<$0.01 & 62 \\ %
\enddata
\tablecomments{$d$: adopted distance (field relation; for R3, T2, T5 and N3 the literature T relation). $v_\perp(d)$: speed relative to the cluster across
the line of sight at that distance; $v_{\perp,\min}$: its minimum over all distances from 5 to 2000~pc, with the 95\% range
from the proper-motion errors (closed form over 20\,000 draws). $P_{<r_J}$: fraction of draws passing within the Jacobi radius for some radial velocity in
$\pm150$~km\,s$^{-1}$; $P_{<r_J,\,<5}$: the same with a relative speed below 5~km\,s$^{-1}$, from the orbit integration
(MWPotential2014) and the analytic tidal model; 0 means that no draw qualifies, and three decimals mark values just below 0.1. $v_{\min}(<r_J)$: median over draws of the
slowest passage inside $r_J$.}
\tablenotetext{a}{Photometric type T4.9 and a grade-C motion; the four-calibrator NISP T relation gives 64~pc.}
\tablenotetext{b}{Distance of its photometric L0 type. At the 69~pc distance of the L7 type of its spectra, $v_\perp(d)$ is 14.5~km\,s$^{-1}$, $P_{<r_J}$ is 1.00, no draw passes within $r_J$ slower than 5~km\,s$^{-1}$, and $v_{\min}(<r_J)$ is 65~km\,s$^{-1}$.}
\tablenotetext{c}{One of the 2000 draws passes within $r_J$, at any speed; no median is defined.}
\end{deluxetable*}

\subsection{R4 and the comoving candidates}\label{sec:tb_comovers}

For the comoving objects, we ask whether the motions allow a slow passage and whether its time can be measured. R4 allows a slow passage within $r_J$ in 18\% of draws in both codes; Table~\ref{tab:traceback} gives its encounter speeds and transverse-velocity floor. Thirteen objects have comoving motions and spectroscopic or %
photometric distances: R4, C1--C8, B2, N1, N2 and C9. Of these, 9 have slow-passage fractions of 0.1 or more in the %
orbit integration and 7 in the analytic model, against about 1.2 expected for field stars that pass a comoving %
proper-motion cut. The candidates were also selected for distances near the cluster's, which this statistic rewards,
so the contrast restates their membership evidence.

The simulated escapees and comoving field stars overlap in the traceback statistics. Slow-passage fractions reach at most 0.35 for injected escapees in the orbit integration and 0.37 in the analytic model, below 0.5, while comoving field stars reach 0.31. The fractions for R4 and the candidates reach 0.28, within the range of both current members and recent escapees. Their velocity floors of 0.06--3.2\,\kms\ are also compatible with both populations (Table~\ref{tab:traceback}). Thus, with 0.5\,mag distance errors and unknown radial velocities, we measure compatibility with slow escape but cannot distinguish it from present membership or date an individual escape.

The type disagreement for C8 also changes its traceback. At the distance implied by its L0 colours, it allows slow passages like the other comoving candidates (Table~\ref{tab:traceback}). At the 69\,pc distance implied by its L7 spectra, its transverse relative speed is 14.5\,\kms\ and it has no slow passage. For N3, the photometric T4.9 type gives a distance of 117\,pc and a transverse relative speed of 15.9\,\kms, based on its uncertain grade-C motion. Only 1 of its 2000 draws allows a slow passage, and the median of its slowest passages is 62\,\kms. Taken with its distance modulus 1.5\,mag in front of the cluster, this weak kinematic evidence counts against a Blanco~1 origin.

\subsection{What would make the traceback decisive}\label{sec:tb_future}

In the injection tests, the distance errors limit the method most, followed by the missing radial velocities.
Distance-modulus errors of 0.15\,mag, as trigonometric parallaxes would give, let 18\% of the injected escapees reach a %
slow-passage fraction of 0.5, against 1.5\% of the comoving field stars. With radial velocities good to $\pm2$\,\kms\ %
as well, a slow-passage fraction of 0.1 or more keeps 31\% of the escapees and 1\% of the comoving field stars, a %
factor of about 30. For the eight rejected objects the present data suffice: their relative speeds are already far
above evaporation speeds.

The approach follows earlier work on brighter stars. \citet{Vaher2023} traced Gaia stars with measured radial velocities
back to Blanco~1 and other young clusters, and \citet{Heyl2022} used the same free-radial-velocity geometry for stars
escaping from the Pleiades. We apply this geometry below the Gaia limit, where the transverse motion bounds the encounter speed even without radial velocities.

\section{Discussion}\label{sec:discussion}

\subsection{What the field shows}\label{sec:disc_field}

We find a population of late-M and L dwarfs moving with Blanco~1 in the Euclid field (Section~\ref{sec:population}). Under the adopted model, which includes C8 on its photometric type, about eight of the ten candidates should be members and about two field dwarfs. On the ATMO models %
the ten have masses of 0.030--0.061\,$M_\odot$ at 100--150\,Myr if they are single, and less if they are binaries: %
the members among them are brown dwarfs. R4 and C1--C4 stay above $\Pg=0.5$ in every variant of Table~\ref{tab:sens}, %
and among them C4, blended and with the poorest astrometry, is the weakest case. C1 is the strongest, with the highest
probability and the only low-gravity hint among the candidates that is not an artefact. The other five depend on the
fitted location of the member sequence, and C8 also on its photometric type: with the L7 type of its spectra it would %
fall to $\Pg=0.42$.

The fitted sequence location has two possible interpretations, with different implications for the faint candidates. The sequence lies $0.53\pm0.16$\,mag fainter than the %
youth-corrected ridge and within about 0.2\,mag of the field ridge at the cluster distance, whereas the cluster's Gaia %
stars at the blue end of the same diagram lie 0.6--1.2\,mag above that ridge. If most of the candidates are members, %
as an expected contamination of about two implies, the late-M and early-L dwarfs of Blanco~1 are as faint at fixed
colour as field dwarfs at its distance. Alternatively, contamination would pull the fitted sequence toward the field-dwarf ridge and raise the probabilities of the five candidates that depend on the sequence location. Independent membership measurements are needed to distinguish these interpretations. The first interpretation resembles the empirical AB~Doradus offsets of Section~\ref{sec:specphot}, in which early-L members are 0.17\,mag fainter in $J$ than field dwarfs of the same type, and the near-infrared photometry of young L dwarfs \citep{Liu2016}. That comparison is at fixed spectral type, whereas our fitted offset is at fixed colour. The interpretation nevertheless conflicts with the adopted ATMO distance correction, which brightens an L0--L1 member by 0.55\,mag using atmospheres outside their valid temperature range. As discussed in Section~\ref{sec:disc_limits}, the inferred distances therefore depend on how youth changes the luminosity.

The excess relative to the mass-function prediction raises the question of whether brown dwarfs have moved outward from the core. The fitted number of late-M and L members is 2.2 (1.5--2.9) times the %
number that the system mass function predicts when it is scaled to the field's Gaia stars. That ratio is relative to a
model normalisation (Section~\ref{sec:disc_limits}), and only the same measurement in the core would
separate an outward redistribution of brown dwarfs from a normalisation error. Taken at face value it would be the
one surprising result, because an outer field should hold more brown dwarfs per star than the mass function predicts
only if they have moved outward or the cluster formed more of them.

In the core region, the deep search of
Section~\ref{sec:deep_core} finds fewer late-M candidates than the same mass function predicts there, 27 against 49, %
while its L count agrees (8 against 11). The core result allows either a real late-M deficit, as mass segregation or evaporation would produce, or an overestimate of photometric completeness. Distinguishing them requires the common core and outskirts measurement above, with the completeness established for the late-M population.

Within the Euclid field, $53\pm17$\% of the fitted members of all classes lie inside the projected %
Jacobi radius, against 45--59\% of the field's Gaia members, so within the errors the brown dwarfs lie inside $r_J$ in the same %
proportion as the stars. A cluster that had lost its brown dwarfs preferentially through the Lagrange points
would hold more of them than of its stars beyond $r_J$, near the base of the leading tail. With 4.9 members %
fitted there, this field could detect only a large difference of that kind. For L-T and T dwarfs, the fitted member count is zero. A Poisson count gives this result 15\% of the time for the expectation of 1.9 among selected sources with reliable motions, or 4\% of the time for the mass-function prediction of 3.2 after the quality cuts. %
The census
of the tails that would show such a loss lies beyond this field (Section~\ref{sec:prospects}).

Six of the ten candidates lie 17--30\,pc from the centre in projection, beyond the Jacobi radius even before the %
unknown line-of-sight offset is added, in the region of the stellar corona of \citet{Meingast2021} and near the base
of the leading tail. If they are members, the tidal model behind $r_J$ \citep{HuntReffert2024} makes them unbound,
although their unknown line-of-sight positions and velocities leave that unestablished. At evaporation speeds of
order 1\,\kms, or 1\,pc\,Myr$^{-1}$, they would have crossed $r_J$ at least 6--19\,Myr ago, a time that the traceback %
leaves undetermined (Section~\ref{sec:tb_comovers}).

The spectroscopic selection identifies ultracool dwarfs near the cluster distance, while the motions determine which are comoving. Of its nine
objects, R4 is a probable member. The other eight are field ultracool dwarfs, three of them foreground T dwarfs at 30--67\,pc, and %
their tracebacks are inconsistent with slow evaporation from Blanco~1. Most late-M and L dwarfs at the cluster distance are unrelated field objects, so a spectrophotometric distance accurate to 20--25\% needs a proper motion to establish membership.

\subsection{The potential of Euclid for nearby young open clusters and associations}\label{sec:disc_euclid}

Euclid provides the photometry, spectra and precise recent positions needed to search for brown dwarfs beyond the core of Blanco~1. Euclid photometry selects late-M and L dwarfs at 234\,pc with $\IE=21.2$--23.1 and reaches T dwarfs: %
N3, at $\IE=26.0$, is typed from its Euclid colours alone. The VIS point-source probability removes most of the %
background galaxies that are hard to separate from ultracool dwarfs in ground-based images of this depth. NISP
slitless spectra typed 3,002 of the 13,590 selected sources and gave types for nine of %
the ten candidates, good to 20--25\% in distance, enough to reject foreground T dwarfs at a glance. The gravity %
indices add little at this age: among the candidates only C1 gives an informative result. Combining Euclid positions on the Gaia frame with VIKING and KiDS gives us a 13\,yr baseline. The proper-motion errors for $\JE\leq20.5$ are 1.1--1.4\,\masyr, corresponding to 1.2--1.6\,\kms\ at the cluster distance. That is well below the 10--103\,\kms\ relative speeds of %
the eight rejected objects, and the motions separate members from the field, though not yet escapees from members. Rubin adds
epochs and an optical colour, and its later releases will lengthen the baseline. %

The published reference footprint of the Euclid Wide Survey includes Blanco~1 and the nominal directions of its tidal
tails (\citealt{Scaramella2022}, their figure~46), so later Euclid data could extend this test through the core and to %
fainter objects, subject to the delivered footprint and release schedule. A circle of 50\,pc radius around Blanco~1, which contains both tails to that distance, has about 13 times the area of this field. %
Many nearby young clusters and associations lie at Galactic latitudes that the Euclid
survey avoids. This work shows that complete substellar censuses of cluster and association tails will need targeted observations with Euclid and will benefit from the combination of
existing near-infrared surveys plus Rubin (Section~\ref{sec:prospects}).

\subsection{Limitations}\label{sec:disc_limits}

The field covers none of the cluster core,  11.3\% of the circle of radius $r_t$ and  7.8\% of a circle of radius %
50\,pc. It samples the outskirts and one flank of the leading tail and supports no census of the cluster. The inferred contamination and member counts depend on the models. Because the field models and count test share most inputs, their agreement does not constrain common systematic errors. Likewise, 0.015 is the smallest empirical tail fraction available with 64 overlapping velocity shifts; its calibration is limited to these controls. The
mass-function ratio depends on a system mass function normalised to the field's Gaia stars, on quality cuts
calibrated on Gaia M dwarfs and on class boundaries from the 150\,Myr ATMO models. At 100\,Myr the boundaries fall by 12--17\%. Holding the fitted number at its 150\,Myr value, without refitting, then gives a ratio of about 2.3 (1.6--3.1). The DR1 footprint masks, unlike the exposure maps, also blank the pixels around bright stars. Eight %
listed members of $G=6.8$--12.8 fall inside the masks around their own images, so the Gaia counts use the exposure %
maps. Counted on the footprint masks, the field holds 50--57 Gaia stars and the ratio becomes 2.5 (1.7--3.4). The ten %
candidates keep $P\geq0.5$ under both field models, B2 drops to $\Pg=0.49$ with the radial prior, and C9 to %
$\Pg=0.195$, just below the limit of 0.2 for a possible member. %

The youth correction dominates the distance systematic for L dwarfs, on top of the 20--25\% precision of the field relations. At types later than L1, the UltracoolSheet relation has not been tested against NISP calibrators, and the four T-dwarf calibrators disagree with it by 0.5--1.3\,mag. The ATMO correction and the empirical AB~Doradus one differ in sign
at L0--L1: C1, C6 and N1 lie 1.2--2.2$\sigma$ in front of the cluster with the empirical correction and within %
$1.1\sigma$ with the model one. A model age of 150\,Myr in place of 100\,Myr moves every distance by at most 6.3\%. %
Spectral youth diagnostics add less information at this age because the comparison objects are mostly younger than Blanco~1 and the cluster's own late-M members have field-like indices. For the faintest spectra, template and index classifications differ by up to four subtypes; C8 therefore remains conditional on its colour-based type.

Distance uncertainty also limits the traceback: with errors of 0.5\,mag and no radial velocities, it cannot confirm or date escape for individual candidates. The astrometric systematics are smaller but affect membership probabilities. Joint proper-motion errors run 7--15\% low on faint Gaia stars and were not rescaled for the candidates, while the catalogue-derived VIKING position weights leave epoch errors of up to 0.43\,yr per frame. Rubin colours and positions are catalogue means, and we allow a 20\,mas floor in the astrometry. %

For the deep core search (Section~\ref{sec:deep_core}), photometric completeness has the largest bearing on the late-M deficit. We simulated recovery in $J$, $K_s$ and $z$ with Gaussian noise at the local depth; losses from blending and false associations, and tests using negative images or variability, are absent from that estimate, making it an upper limit. In addition, the Legacy Surveys DR10 $z$ photometry uses model fluxes for extended sources, without forced fits to the native DECam exposures. Five alternative assumptions lower the M7--L6 bound. The bound falls to 18 if only the 32 candidates whose motions stay comoving when each %
epoch group is left out are counted. It is also 18 when each candidate is weighted by its probability of a type of M7 %
or later, which counts losses from the subset but no gains, or with the three new candidates removed. Using a nearest-rank percentile gives 19; calibrating the stellar-profile interval only on injected sources gives 15 (selection 2 in Appendix~\ref{app:deep_versions}). A boundary at M6.75 or M7.5 would %
change the subset from 35 to 43 or 33 objects. Typing with VIKING $Z$ in place of the Legacy Surveys $z$, whose $z-Y$ %
is 0.1--0.2\,mag redder than that of the spectral standards, moves two candidates below M7 and gives 33 objects and %
a bound of 19. Weighting all 74 candidates and the control selections by the same type probability, which counts %
gains as well as losses, raises it to 27. Because the Poisson limit already includes fluctuations in chance selections, subtracting a control-count percentile counts that fluctuation twice and gives coverage above 95\%. The spatial comparison has a different limitation: its fields cannot reproduce the epoch set or the per-exposure half-stack systematics, and two have seeing of 0.88 and 0.93\arcsec\ compared with 0.81\arcsec\ in Blanco~1. These differences limit how closely the comparison fields can represent the photometric selection in the core. %

\section{Ultracool dwarf candidates beyond the Euclid field}\label{sec:prospects}

Most of Blanco~1 lies outside the Euclid field, including the core, almost all of the trailing tail and all but the base of the %
leading tail (Table~\ref{tab:coverage}). Ground-based near-infrared and optical surveys cover the rest of the cluster, at lower depth and
without spectra. We transfer the Euclid selection of Section~\ref{sec:methods} to these surveys to identify spectroscopic targets elsewhere in the cluster, using archival images back to 1999 to improve their motions. We assign follow-up priorities according to the strength of the kinematic evidence, as defined in Section~\ref{sec:offeuclid_pm}. In the core region, deeper archival Subaru images extend the search to fainter candidates and bound
the number of kinematic members among them (Section~\ref{sec:deep_core}).

\subsection{Surveys and search area}\label{sec:offeuclid_data}

The search covers 246\,deg$^2$ within 16\arcdeg\ of the cluster centre, outside the Euclid field, where VIKING and %
KiDS overlap (Figure~\ref{fig:offeuclid}a). The Euclid field excluded here is the sky observed in all four Euclid %
bands (Section~\ref{sec:field}), so the search includes the 0.49\,deg$^2$ that lacks one or more bands inside the %
convex outline of the Euclid catalogue positions. Nine sources there pass the selection of %
Section~\ref{sec:offeuclid_selection}, none of them inside $r_t$, and none reaches priority~1 or~2 %
(Section~\ref{sec:offeuclid_pm}). VIKING provides $ZYJHK_s$ photometry, morphology and the earliest astrometric epochs for 15.0 million sources, while KiDS DR5 adds $ugri$ photometry and $r$-band epochs. VIKING magnitudes in this wide-area search are on the Vega system. Both surveys %
end near declinations of $-36\arcdeg$ and $-26.7\arcdeg$, so the leading tail, which runs north, is covered along its axis only to %
about 7\arcdeg\ from the centre and the trailing tail to 8--10\arcdeg. Rubin DP2 $i$-band visits from September to %
November 2025 add a late epoch: two or more visits cover 119\,deg$^2$, including nearly all of the area inside $r_J$. Gaia %
DR3 provides the reference frame, parallaxes and motions of the brightest candidates and one of the two field models,
and CatWISE2020 \citep{Marocco2021} measures $W1$ and $W2$ over the whole area. %

Six zones divide the area (Table~\ref{tab:offeuclid_zones}). Two are the regions inside $r_t$ and between $r_t$ and
$r_J$. We define the tail zones from $r_J$ to 15\arcdeg, along position angles of 306\arcdeg\ and 133\arcdeg. Their 30\arcdeg\ half-width contains 90\% and 86\% of the leading- and trailing-tail stars of \citet{Jadhav2025} over this range. The corona is the rest of the annulus out to 10\arcdeg, and the outer field extends from 10\arcdeg\ to
16\arcdeg. The cluster parameters are those of Table~\ref{tab:cluster}.

\begingroup\setlength{\tabcolsep}{2.6pt}
\begin{deluxetable*}{lrrrrrrrrrrrr}
\tabletypesize{\scriptsize}
\tablecaption{Zones of the search beyond the Euclid field.\label{tab:offeuclid_zones}}
\tablehead{\colhead{Zone} & \colhead{Area} & \colhead{DP2 area} & \colhead{Gaia} & \colhead{Prior} &
\colhead{Candidates} & \colhead{Density} & \colhead{Excess} & \colhead{Measured} & \colhead{Comoving} &
\colhead{Comoving,} & \colhead{Priority 1} & \colhead{Priority 1,}\\
\colhead{} & \colhead{(deg$^2$)} & \colhead{(deg$^2$)} & \colhead{members} & \colhead{$\pi$} & \colhead{} &
\colhead{(deg$^{-2}$)} & \colhead{} & \colhead{motions} & \colhead{} & \colhead{chance} & \colhead{} &
\colhead{chance}}
\startdata
Inside $r_t$ & 9.4 & 9.4 & 632 & 0.280 & 213 & 22.7 & 74 $\pm$ 15 & 137 & 46 & 6.7 & 45 & 3.8 \\ %
$r_t$ to $r_J$ & 7.8 & 7.8 & 42 & 0.032 & 125 & 16.1 & 10 $\pm$ 11 & 74 & 4 & 3.6 & 4 & 0.9 \\ %
Leading tail & 18.0 & 12.5 & 17 & $6.6\times10^{-3}$ & 242 & 13.5 & $-$25 $\pm$ 17 & 85 & 5 & 4.4 & 1 & 0.4 \\ %
Trailing tail & 56.9 & 46.1 & 60 & $7.0\times10^{-3}$ & 814 & 14.3 & $-$32 $\pm$ 34 & 366 & 18 & 17.9 & 7 & 1.9 \\ %
Corona & 69.3 & 30.0 & 27 & $2.5\times10^{-3}$ & 1005 & 14.5 & \nodata & 436 & 24 & 21.3 & 0 & 0.1 \\ %
Outer field & 85.2 & 13.6 & 3 & $2.2\times10^{-4}$ & 1292 & 15.2 & \nodata & 136 & 4 & 6.7 & 0 & 0.0 \\ %
\hline
All & 246.4 & 119.3 & \nodata & \nodata & 3691 & 15.0 & \nodata & 1234 & 101 & 60.7 & 57 & 7.2 \\ %
\enddata
\tablecomments{Areas are true sky areas. DP2 area is the part with two or more DP2 $i$-band visits. Gaia members are
the members of \citet{HuntReffert2023} and \citet{Jadhav2025} in the zone, which scale the prior $\pi$. Density and
excess count the photometric candidates; the excess is the zone count minus the density of the corona and outer
field, 14.9\,deg$^{-2}$, times the zone area, with Poisson errors on both. Measured motions have
both components better than 5\,\masyr; comoving means $\chi^2<6.18$ against the cluster motion. The chance numbers
are the larger of the spatial control (corona and outer-field candidates evaluated at the cluster motion with the
zone's prior) and the mean of 32 controls with the cluster motion displaced by 14--20\,\kms. The corona and the
outer field form the spatial control sample, so their comoving chance numbers are the pooled comoving rate of that
sample applied to each zone. All counts use the search motions; with the archival epochs of
Section~\ref{sec:offeuclid_archival} the priority-1 counts of the four inner zones become 47, 4, 2 and 7 (60 in all), and the chance numbers remain those of the search motions.}
\end{deluxetable*}
\endgroup

\subsection{Transfer of the Euclid selection}\label{sec:offeuclid_selection}

The membership analysis places each source on the cluster sequence through its Euclid colour and its offset $D$ from
the field ridge (Section~\ref{sec:selection}). To apply the same sequence without Euclid photometry, we predict both
quantities from VIKING and KiDS photometry, calibrated on Euclid-field sources that have VIKING counterparts. A %
quadratic regression on $Z-J$, $Y-J$, $J-H$, $J-K_s$ and, where available, KiDS $i-J$ gives $c_\mathrm{pred}$, the
predicted $(\IE-\JE)_0$, with a cross-validated robust scatter of 0.062\,mag with $i-J$ and 0.088\,mag without it. A %
cubic ridge in VIKING $J$ against $c_\mathrm{pred}$ then gives $D_V$, the VIKING counterpart of $D$. For compact Euclid sources, we recover $D$ from $D_V$ with a robust scatter of 0.24\,mag at $(\IE-\JE)_0=2.7$--3.05. The scatter increases to 0.46\,mag at colours of 3.05--3.5. The normalised scatters, 0.94 and 1.13, agree with the propagated errors. At colours redder than 3.9, the calibration sample is sparse and the scatter reaches 1.1\,mag; the transfer is consequently most useful for the bluer candidates. %

We require a VIKING $J$ signal-to-noise ratio of at least 10 and colours satisfying $c_\mathrm{pred}+\sigma_c\geq2.6$ and $c_\mathrm{pred}\leq4.3$, where $\sigma_c$ is the colour error. The offset $D_V$ must be within twice its error of the sequence band, which extends from 1.5\,mag above to 0.8\,mag below the field-dwarf ridge. Colour floors at the 1st percentile of the typed Euclid-field ultracool
dwarfs remove bluer stars, and the CatWISE colours must satisfy $W1-W2\leq0.73$ and $J-W2\geq1.06$ where they are %
measured. A source is rejected as a galaxy when both VIKING and KiDS find it extended, or VIKING alone where KiDS has no
counterpart, a rule that loses 4\% of %
the typed ultracool dwarfs and removes half of the galaxies that Euclid resolves. These criteria select 3691 of the 8.9 million sources with VIKING $J$ measurements in the area. %

The cuts on $c_\mathrm{pred}$ and $D_V$ include the colour errors through one random draw, and a different draw %
changes the list at its margins. Across 20 draws, 448 of the 3691 candidates are omitted at least once, including two of priority~2 (targets 68 and 71) and none of priority~1. The draws add 537 sources, of which 536 lack a measured motion and one has a non-comoving motion. Target numbers here and below refer to the rank column of Table~\ref{tab:offeuclid}. We group the candidates by $c_\mathrm{pred}$, using boundaries at 2.5, 2.7, 3.05, 3.9 and 4.3\,mag. In the first three bins, the Euclid-field dwarfs have median NISP types of M7.6, M8.4 and M9.9. %

\subsection{Completeness and purity}\label{sec:offeuclid_purity}

\begin{figure*}
\centering
\includegraphics{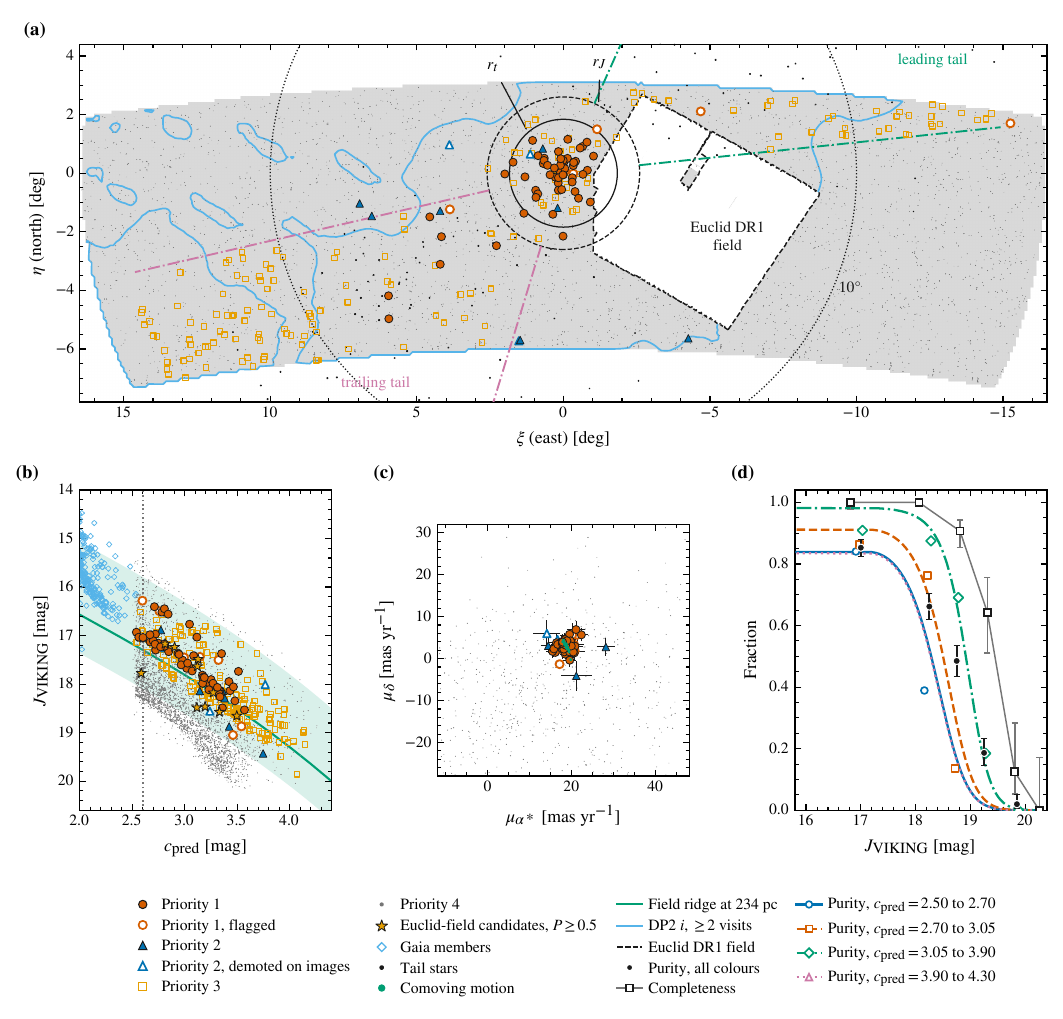}
\caption{The photometric search beyond the Euclid field. (a) Candidates in a gnomonic projection about the cluster
centre, east to the left. Grey shading is the area searched with VIKING and KiDS, the dashed outline %
the Euclid DR1 sky observed in all four bands,  %
and the blue contour the area with two or more DP2 $i$-band visits. The
solid, dashed and dotted circles mark $r_t$, $r_J$ and 10\arcdeg, the dash-dotted lines the two tail zones, and black
dots the tail stars of \citet{Jadhav2025}. Filled circles are priority-1 targets and open circles the seven with
notes b--e in Table~\ref{tab:offeuclid}. Filled and open triangles are priority-2 targets kept and set aside after image inspection, squares priority 3 and grey dots priority 4. For the 73 inspected targets, priorities and the motions in panel (c)
include the archival epochs (Section~\ref{sec:offeuclid_archival}). (b) VIKING $J$ against $c_\mathrm{pred}$, the $(\IE-\JE)_0$ colour
predicted from VIKING and KiDS photometry, with the field ridge at 234\,pc (green line), the cluster band (shading), %
the colour limit (dotted line), Gaia members of Blanco~1 (diamonds) and the ten Euclid-field candidates with
$P\geq0.5$ (stars). (c) Proper motions measured to better than 5\,\masyr, with errors for priorities 1 and 2 and the
comoving motion expected at each of those targets (green dots). (d) Purity and completeness against $J$, both measured %
in the Euclid field. Filled circles give the purity of all colour bins with 68\% intervals, open markers the purity of
single colour bins with at least eight selections, and lines the purity model of each bin. Open squares give the
completeness for compact Euclid sources inside the cluster band, with Jeffreys 68\% intervals.}
\label{fig:offeuclid}
\end{figure*}

We measure completeness and photometric selection purity by comparing the VIKING and KiDS selection with Euclid in the overlap area. Applied to Euclid-field sources, it recovers 97\% of the 204 NISP-typed ultracool dwarfs inside the Euclid %
cluster band and 90\% of the 208 compact Euclid sources in the band (Figure~\ref{fig:offeuclid}d). For compact sources on the Euclid sequence, completeness is 100\% to $J=18.5$ and 91\% at 18.5--19.0 (Figure~\ref{fig:offeuclid}d). It falls to 64\% at 19.0--19.5 and 12\% at 19.5--20.0 because the VIKING $Z$ and $Y$ measurements become too noisy for the colour. Nine of the ten Euclid-field candidates with $P\geq0.5$ are recovered; C4 %
is lost because VIKING and KiDS both class it as extended. On the ridge brightened for youth that
Section~\ref{sec:deep_core} adopts,  L3 members have $J\approx19.7$, where the completeness is 12\%. The search therefore misses %
most mid- and late-L members, and T dwarfs at the cluster %
distance ($J\approx21.3$) lie below the VIKING limit. %

Here, photometric selection purity means the fraction of VIKING-selected sources that Euclid identifies as compact ultracool dwarfs on the cluster sequence. Applied to the  1.33 million VIKING sources with $J$ inside the Euclid field (within the outline of %
Section~\ref{sec:offeuclid_data}), the
selection returns 525 sources outside the NGC~7793 region and the gaps in the Euclid coverage (averaged over ten %
random draws of the colour errors). Euclid places 56\% of them (68\% interval 54--58\%) on the cluster sequence as compact ultracool %
dwarfs. Of the rest, 26\% are compact ultracool dwarfs off the sequence, field dwarfs at other distances, 14\% have no %
Euclid counterpart, and 4\% are resolved or bluer sources. Photometric selection purity decreases from 85\% at $J<18$ to 2\% at $J\geq19.5$. It also decreases toward redder colours, from 69\% in the bluest bin to 33\% in the reddest. We model this fraction as a smooth function of $J$ with an offset for each colour bin and assign the corresponding value to each target (Table~\ref{tab:offeuclid}). For the 51 priority-1 targets inside $r_J$ (Section~\ref{sec:offeuclid_targets}), the model mean is 0.90, compared with 0.85 from the binned measurements. We adopt these photometric selection rates across the search area because it lies at the same high Galactic latitude as the Euclid field. The proper-motion analysis then estimates membership within this photometric sample.

\subsection{Proper motions and membership screening}\label{sec:offeuclid_pm}

Kinematic screening needs motions precise enough to separate the cluster, which moves at about 19\,\masyr, from field
dwarfs at other distances. The positions are tied to Gaia DR3 as in Section~\ref{sec:astrometry}: VIKING band
positions frame by frame, and KiDS $r$-band positions tile by tile with a correction for differential chromatic
refraction. DP2 $i$-band positions are tied visit by visit and detector by detector, with the same frame offset and %
20\,mas floor. Gaia DR3 motions are combined where they exist. On 8364 Gaia DR3 stars at $G=17$--20.7 held out of %
every tie, the normalised residuals have robust widths of 1.02 and 0.95. With a DP2 epoch the median error is %
1.1\,\masyr\ over a 12.3\,yr baseline; without one it is 7.9\,\masyr\ over 3.2\,yr, too large to separate members %
from the field. Of the 3691 candidates, 1234 have both components measured to better than 5\,\masyr. We recover comoving motions for 531 of the 575 Gaia members measured here (92\%). Using VIKING, KiDS and DP2 alone also gives comoving motions for all six measured Euclid-field candidates: R4, C1, C2, C3, C5 and C8.

The motion expected for a member at each position is computed as in Section~\ref{sec:membership_model}; at the %
centre it is $(18.88,\,2.61)$\,\masyr. Each candidate then receives a comoving $\chi^2$ and a membership probability built as in
Section~\ref{sec:membership_model}, from a proper-motion term with the Gaia and control field models, a Gaia parallax
term where one exists, and a photometric term in $D_V$. In the photometric term, members follow the Euclid-field
sequence 0.21\,mag above the ridge, rising towards the 0.58\,mag offset of the Gaia members at the blue end. A %
fifth of them are unresolved binaries 0.75\,mag brighter, and the field follows the $D_V$ distribution of all %
candidates in the same colour bin. The prior of each zone is its expected number of selected late-M and L members %
divided by its number of candidates. The expectation scales the mass-function estimate for the Euclid field, taken
before the Euclid quality cuts, by the zone's share of Gaia members and by the 90\% completeness of the VIKING
selection. The resulting priors run from 0.28 inside $r_t$ to $2\times10^{-4}$ in the outer field %
(Table~\ref{tab:offeuclid_zones}). These probabilities give the order of targets for follow-up, with member counts dependent on the zone priors.

Targets fall into four priorities. Priority~1 requires both motion components measured to better than 5\,\masyr\ and
$P\geq0.5$ under both field models. Priority~2 marks partial kinematic support, either $P\geq0.1$ under one field
model with a comoving $\chi^2<25$ or a comoving motion with $P\geq0.02$. Priority~3 is a candidate on the sequence
inside $r_J$ or in a tail zone whose motion is not yet measured. Priority~4 holds the rest: candidates whose motions %
do not support membership, and those in the corona and outer field without a measured motion. The search gives 57,
19, 162 and 3453 candidates in priorities 1 to 4, before the archival epochs of Section~\ref{sec:offeuclid_archival}. %

\subsection{An excess inside the tidal radius}\label{sec:offeuclid_core}

We find a photometric excess inside $r_t$, where the cluster contribution should be strongest. The candidate
density there is 22.7\,deg$^{-2}$, against 14.9\,deg$^{-2}$ in the corona and outer field, where few members %
are expected. The excess, $74\pm15$ sources, is comparable to the 54--65 late-M and L members that the mass-function %
estimate predicts there after selection. Between $r_t$ and $r_J$ the excess is $10\pm11$, and both tail zones lie at %
or below the field density (Table~\ref{tab:offeuclid_zones}).

The proper motions identify a corresponding excess of comoving sources. Of the 137 candidates inside $r_t$ with measured motions, 46 move with the cluster %
($\chi^2<6.18$ for two degrees of freedom). To estimate how many field dwarfs pass by chance, we evaluate the 572 %
corona and outer-field candidates with measured motions at the cluster motion, using each zone's prior. We multiply the fraction selected in this spatial comparison sample by the number of candidates with measured motions in each zone. The result assumes that all are field objects, so it bounds the chance count from above where members are present. This spatial control predicts 6.7 comoving candidates inside $r_t$. Controls %
with the cluster motion displaced by 14--20\,\kms\ in eight directions predict 3.7, and each zone adopts the larger %
of the two. Smaller displacements re-select real members, 24.2 of the priority-1 targets inside $r_t$ passing at %
8\,\kms\ against 2.2 at 14\,\kms, and are not used. The region inside $r_t$ therefore holds about 39 more comoving candidates than chance predicts. Scaled to the %
137 of its 213 candidates with measured motions and to the 92\% recovery of known members, the photometric excess predicts %
about 44 comoving members, so the two excesses agree.

\subsection{Images of the targets}\label{sec:offeuclid_vet}

We inspected nine images of each of the 76 priority-1 and priority-2 targets to identify artefacts, blends and stars with the wrong colours before spectroscopy (Figure~\ref{fig:offeuclid_finders} and its figure set). They are
VIKING $J$ at two epochs and $K_s$, KiDS $r$ and both $i$ passes, and Legacy Surveys DR10 $g$, $r$ and $z$ %
\citep{Dey2019}, resampled onto one 20\arcsec\ grid and spanning 2009.9--2021.7. All 57 priority-1 targets are point sources at their %
catalogue positions, the Legacy Surveys model all 76 targets as point sources, and none has a neighbour within %
2\arcsec. Bright-star ghosts cover the KiDS images of rank 42, which is a point source in VIKING and in Legacy %
Surveys $z$. A member moves by about 0.1\arcsec\ over the typical five-year span of the images, and they test only %
fast or erroneous catalogue motions. Four targets (ranks 3, 7, 22 and 69) move as their catalogue motions predict, %
and for 67 of the others the image motions agree with the catalogue motions without excluding zero. The images of %
ranks 63 and 68 contradict catalogue motions of 199 and $-69$\,\masyr\ in declination and are consistent with both the %
cluster motion and zero. So is the image motion of rank 65, whose catalogue motion is uncertain by more than %
5\,\masyr, and rank 71 lacks a usable VIKING position. %

We set aside three priority-2 targets after inspecting their images. Ranks 59 and 71 have the Legacy Surveys colours of mid-M stars, and rank 59 %
also has a Gaia DR3 colour $G_\mathrm{BP}-G_\mathrm{RP}=2.10$; the VIKING and KiDS images of rank 71 lie in light %
scattered by a bright star. Rank 64 moves at $(69,\,-27)\pm10$\,\masyr\ over 5\,yr, 5.8 standard errors from the %
cluster motion, a result that rests mainly on the Legacy Surveys epoch. That leaves 73 targets: all 57 of priority~1 %
and 16 of priority~2.

\subsection{Archival epochs}\label{sec:offeuclid_archival}

The search motions span 12\,yr, and for five priority-1 targets their early end is a single VIKING epoch group. %
To lengthen the baselines and test those motions, we measured the 73 targets kept after the inspection on archival
images that predate or fill the VIKING to DP2 interval. DECam images from 2012--2026 \citep{Flaugher2015} cover all 73 %
targets, and Subaru Hyper Suprime-Cam (HSC) $Y$-band images from 2016 \citep{Miyazaki2018,Komiyama2018,Kawanomoto2018}, obtained from %
SMOKA \citep{Baba2002}, cover 31. CFHT CFH12K images from 1999--2000 \citep{Cuillandre2000} provide earlier epochs for 18 targets, and CTIO Mosaic-II images from 2006 \citep{Muller1998} for eight. Target 45 also has a VISTA Hemisphere Survey (VHS) epoch from 2016 \citep{McMahon2013}. We measure each position against Gaia DR3 stars on the same detector, propagated to the image epoch, using a polynomial plate model of up to third order and a correction for differential chromatic refraction. HSC and VHS provide the most precise individual epochs, with median errors of 4 and 8\,mas. The corresponding errors are 13\,mas for Mosaic-II, 20\,mas for DECam and 30\,mas for CFH12K. %

The archival positions share the frame of the search. Relative to the survey-only fits, we measure a mean offset of $(-0.8\pm0.9,\,-0.5\pm0.9)$\,mas for the 637 DECam positions. For the 31 HSC positions it is $(0.3\pm2.1,\,-2.9\pm2.1)$\,mas. Fitted from the archival epochs alone, the motions of the three targets with Gaia
DR3 five-parameter solutions, ranks 1, 20 and 45, agree with Gaia DR3 to within 1.6 standard errors in each %
component. We exclude the Spitzer IRAC mosaics from 2007 because their reference stars scatter by 79--92\,mas about the plate model, giving a weaker constraint than the existing motions.

\begin{figure*}
\centering
\includegraphics{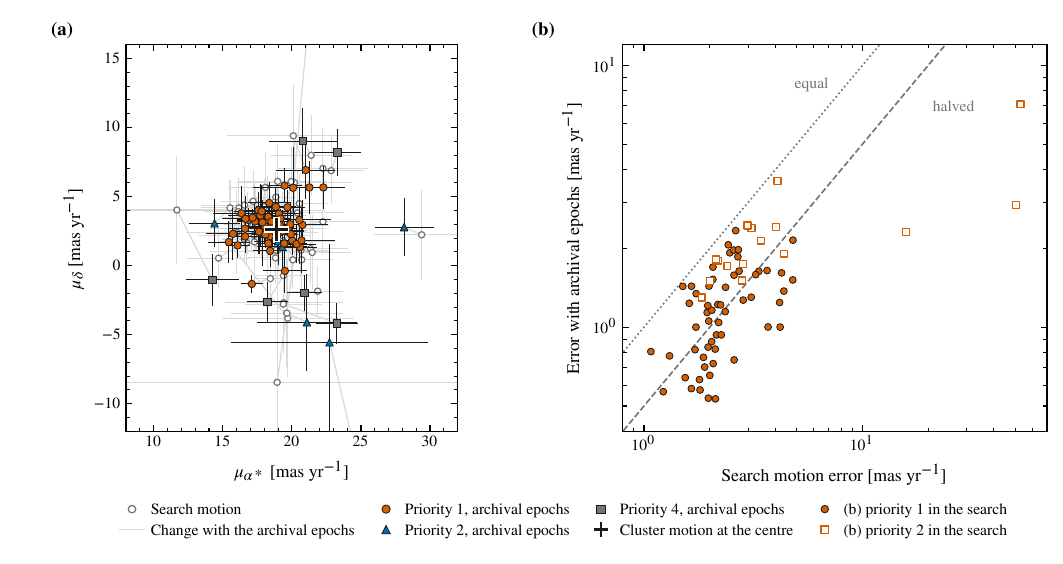}
\caption{Proper motions of the 73 inspected targets before and after the archival epochs. (a) Search motions (open grey
circles with grey errors) and motions with the archival epochs (filled symbols with black errors), joined by thin
lines. The symbol of the filled points gives the priority with the archival epochs, and the cross marks the cluster
motion at the centre, $(18.88,\,2.61)$\,\masyr. The search motions of ranks 63 and 68, which exceed 50\,\masyr\ in %
declination, lie outside the panel; their archival motions are plotted. (b) Error of the motion (the larger of the two components) with the archival epochs
against the error of the search. Filled circles were priority~1 in the search and open squares priority~2; the
dotted and dashed lines mark equal and halved errors.}
\label{fig:offeuclid_archival}
\end{figure*}

With the archival positions, the median baseline is 14\,yr and the earliest epoch is 1999.7. The median proper-motion error of priority-1 targets falls from 2.1 to 1.2\,\masyr\ (Figure~\ref{fig:offeuclid_archival}). All 57 retain priority~1, including the five whose survey motions rested on one VIKING epoch group: omitting VIKING gives $P_\mathrm{min}=0.75$--0.99, where $P_\mathrm{min}$ is the smaller probability under the two field models. Target 30 has the most precise motion and is one of 18 targets with a 26\,yr baseline (Table~\ref{tab:offeuclid}).

The longer baselines also resolve uncertain motions among the priority-2 targets. Targets 58 and 65 have cluster-like motions with errors of 2.4 and 2.3\,\masyr, while DECam and HSC determine the motion of target 63, whose survey estimate was uncertain by 51\,\masyr. Its probability increases from 0.28 to 0.91, and all three qualify for priority~1. Six targets (66, 70 and 72--75) have motions inconsistent with the cluster and are assigned priority~4; seven retain priority~2. The three set aside after image inspection, 59, 64 and 71, retain their formal priority~2 classification. The full list contains 60, 10, 162 and 3459 objects in priorities 1 to 4; the chance estimates in Table~\ref{tab:offeuclid_zones} use the survey motions.

\subsection{Priority-1 targets}\label{sec:offeuclid_targets}

Sixty candidates reach priority~1 (Table~\ref{tab:offeuclid}): the 57 of the search and ranks 58, 63 and 65. The 51 %
inside $r_J$, 47 of them inside $r_t$, are the strongest targets. The search controls predict 3.8 chance passes inside %
$r_t$ and 0.9 between $r_t$ and $r_J$, so about 46 of the 51 are not expected by chance, an estimate made with the %
search motions; spectra must show how many of them are members. Their $J=16.4$--19.0 magnitudes and $c_\mathrm{pred}=2.5$--3.6 colours make them accessible late-M and early-L spectroscopic targets, with a mean photometric selection purity of 0.90. Rank 20 has an %
astrometric fit with $\chi^2=92$ for 36 degrees of freedom, plausibly caused by a saturated star 11\arcsec\ away %
whose halo shifts the centroids at some epochs, and rank 42 has an image defect found in the inspection of %
Section~\ref{sec:offeuclid_vet}. For target 50 between $r_t$ and $r_J$, the zone prior determines whether it qualifies for priority~1 (note e of Table~\ref{tab:offeuclid}). %

The seven targets in the trailing-tail zone (51--57), at 3.4--7.7\arcdeg\ from the centre, provide weaker membership evidence than those inside $r_J$. Chance predicts %
1.9 of them, but the looser comoving count in that zone, 18, matches its spatial-control expectation. With the archival %
motions six of them stay above $P=0.5$ when the tail prior is halved, and rank 57, which also lies near the edge of the %
zone, falls below it. These targets test whether the trailing tail holds substellar members. The two leading-tail
targets compare with 0.4 expected by chance in the search. Target 45 is 14.99\arcdeg\ from the centre, at the corner of its zone, and lacks a DP2 epoch. Its VHS and DECam measurements span 8\,yr; combining the fit with Gaia DR3 gives the motion in Table~\ref{tab:offeuclid}. Rank 58, 5.1\arcdeg\ from the centre, reaches priority~1 only with %
the archival epochs. Their classification depends on the leading-tail prior: halving it lowers both probabilities below $P=0.5$. Priority~1 %
is reached only inside $r_J$ and in the tail zones. Over the whole area the search controls predict 7.2 chance passes %
among the priority-1 targets.

Twenty-one of the 76 priority-1 and priority-2 targets of the search appear in earlier lists of the cluster, and all %
21 are among the 60 priority-1 targets (Table~\ref{tab:offeuclid}). We propagate each target from its VIKING epoch to the epoch of each earlier list and match within 2\arcsec, except for the object discussed below. Nineteen priority-1 targets, among them rank 65, are candidates of \citet{Moraux2007}. %
Once the median offset of these 19 matches, 0.59\arcsec\ in right ascension, is removed, 18 of them lie %
0.11--0.89\arcsec\ from their candidates. %
The exception, rank 30, lies 2.2\arcsec\ from CFHT-BL-52, or 2.6\arcsec\ after %
that correction. In the CFH12K images of \citet{Moraux2007} it is the only source within 12\arcsec, at the position %
its motion predicts, and its magnitudes match those of CFHT-BL-52: the offset lies in the catalogued position. %
\citet{Moraux2007} list CFHT-BL-52 as a suspected binary, and rank 30 lies 1.1\,mag above the sequence. Rank 1 is a %
Gaia member of \citet{HuntReffert2023}. The 44 members listed by \citet{Casewell2012} in their Table 3 include 14 of the %
targets, 13 of them also \citet{Moraux2007} candidates; the fourteenth, rank 63, appears only there. All 21 lie %
inside $r_t$. Of the 26 targets inside the outline of the complete \citet{Moraux2007} candidate lists, 20 are known: %
the 19 \citet{Moraux2007} candidates and rank 63. The proper-motion catalogues of \citet{Platais2011}, which reach
$V=22$, contain none of the 76. We identify 55 targets absent from these earlier lists, including 39 of priority~1.

Four priority-1 targets already have optical spectra. Targets 2, 10, 20 and 23 correspond to CFHT-BL-49, -34, -41 and -51. \citet{Moraux2007} classified them as M7--M8, with the low surface gravity expected at the cluster age. Target 2 also shows lithium \citep{Cargile2010}, with an equivalent width of 1.93\,\AA\ \citep{Juarez2014}. %

\subsection{A deeper search in the core region}\label{sec:deep_core}

The VIKING-based selection loses completeness fainter than $J\approx19.5$ (Vega), about L3 at the distance of the cluster %
(Section~\ref{sec:offeuclid_purity}), so in the core it recovers late-M and early-L members but few later ones. HSC,
the wide-field imager of the 8.2-m Subaru Telescope \citep{Miyazaki2018,Komiyama2018}, observed the %
core in its $Y$ band \citep{Kawanomoto2018} in 2016, and these exposures supplied the 2016 epochs of %
Section~\ref{sec:offeuclid_archival}. Their coadd reaches 1.7\,mag deeper in $Y$ than VIKING. We use it to measure the kinematic population of M7 and later objects in the core region and to assess whether T members are detectable.

We reduced the six 150-s exposures from the Subaru archive SMOKA \citep{Baba2002}, corrected for fringing, and coadded them. The coadd covers 1.8\,deg$^2$ within 0.75\arcdeg\ of a field centre offset by 0.13\arcdeg\ from the cluster centre. %
This area contains the circle of radius $r_{50}$. The median %
5$\sigma$ point-source depth of the coadd is $Y_\mathrm{HSC}=23.5$ (AB), and 23.6 where all six exposures overlap. %
We detected sources at 4$\sigma$ in a PSF matched-filter significance map and measured them in the full coadd and in two independent half-stacks of three exposures each. Colours combine HSC $Y$ with $z$ from the DESI Legacy Imaging Surveys
DR10 \citep{Dey2019} and with $J$ and $K_s$ from VIKING \citep{Edge2013}. Each source receives probabilities of lying
in late-M, L, early-T and mid-T colour boxes, which are defined in the HSC system from synthetic colours of SpeX
spectral standards \citep{Burgasser2014}. We estimate a photometric type and its error from the same standards, treating non-detections as limits.

The main list requires a $Y_\mathrm{HSC}$ %
signal-to-noise ratio (S/N) of at least 8, $J$ S/N\,$\geq5$, detection in both half-stacks, a stellar profile and a box %
probability of at least 0.8. The stellar-profile interval combines the scatter of injected point sources with a term %
for the mismatch between the local and the modelled PSF, measured on Gaia DR3 stars \citep{Gaia2016,Vallenari2023}.
The profile criterion keeps 97--99\% of Gaia stars from S/N 60 to above 3000. For stars excluded from the PSF models, recovery is 96--98\% from S/N\,100. Below S/N\,13, galaxies with effective radii under 0.15\arcsec\ can satisfy the profile criterion; their colours and lack of motion provide the additional discrimination. The main %
list holds 235 sources: 202 in the late-M box, 30 in the L box and 3 in the early-T box. A low-S/N list applies %
the same criteria at a $Y_\mathrm{HSC}$ S/N of 5--8, and a T list takes sources with an early- or mid-T box %
probability of at least 0.5 and $J$ S/N\,$\geq3$. %

\begin{figure*}
\centering
\includegraphics{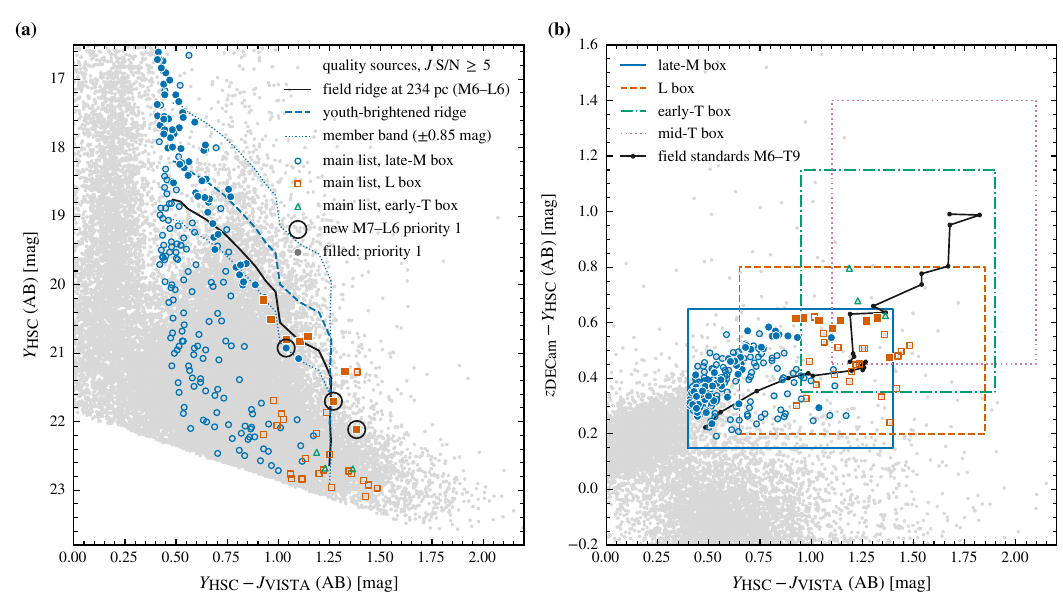}
\caption{Selection of the deep HSC search of the core region. (a) HSC $Y$ against $Y_\mathrm{HSC}-J$ for sources with $J$
S/N\,$\geq5$ (grey), with the field-dwarf ridge at 234\,pc (solid line), the ridge brightened for youth (dashed) and the %
member band (dotted). Open symbols are main-list sources in the late-M (circles), L (squares) and early-T (triangles)
boxes, and filled symbols the priority-1 candidates; large open circles mark the three new M7--L6 candidates. (b) The
colour boxes in $z_\mathrm{DECam}-Y_\mathrm{HSC}$ against $Y_\mathrm{HSC}-J$, with the synthetic colours of the field
spectral standards from M6 to T9. In the axis labels $z_\mathrm{DECam}$ is the Legacy Surveys DR10 $z$ and
$J_\mathrm{VISTA}$ the VIKING $J$, both AB.}
\label{fig:deep_cmd}
\end{figure*}

Proper motions combine VIKING, HSC, archival DECam \citep{Flaugher2015} and Rubin DP2 \citep{RubinDP2} positions from %
2011 to 2025, tied to Gaia DR3 as in Section~\ref{sec:offeuclid_pm}. A DECam frame is not used when the source is %
elongated in it (axis ratio above 1.5) or flagged as truncated by the source extractor \citep{BertinArnouts1996,Barbary2016}. A frame %
whose position lies more than 4$\sigma$ from those of the other frames of its night is also rejected. For the 194 main-list sources with both components measured %
to better than 5\,\masyr, the median baseline is 13\,yr and the median error 1.0\,\masyr. Gaia DR3 stars fitted %
without their Gaia epoch give an error scale factor of 1.02 (400 stars). We compute membership probabilities as in Section~\ref{sec:offeuclid_pm}, using the Gaia and velocity-shifted field models. The photometric term measures the $Y_\mathrm{HSC}$ offset from the field-dwarf ridge, the counterpart of the Euclid offset $D$. The member band spans 0.85\,mag either side of the ridge brightened for youth %
(Figure~\ref{fig:deep_cmd}). The prior of each colour family is its expected number of selected members, from the initial mass function (IMF) and the
completeness below, divided by its number of candidates. A priority-1 candidate has both motion components measured to %
better than 5\,\masyr\ and $P\geq0.5$ under both field models; all of them also have at least three epoch groups over %
at least 5\,yr.

Synthetic stationary sources reach priority~1 at a rate of 1 in 3000 (95\% upper limit %
$1.6\times10^{-3}$). To count the field objects that pass the test by chance, we applied it to 32 velocity-shifted %
control clusters, displaced from the cluster motion by 14, 16, 18 and 20\,\kms\ in eight directions. The alternative profile calibrations and frame-rejection criteria in Appendix~\ref{app:deep_versions} select comparable mean numbers of field objects by chance, while changing the recovery of bright stars.

\begin{figure}
\centering
\includegraphics{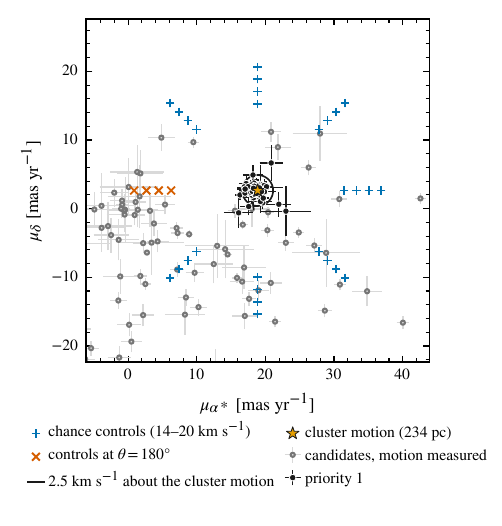}
\caption{Proper motions of the candidates of the main and low-S/N lists with motions measured to better than 5\,\masyr\ %
(grey) and of the 74 priority-1 %
candidates (black). The star marks the cluster motion at 234\,pc and the circle a radius of 2.5\,\kms\ about it. %
Crosses mark the centres of the 32 velocity-shifted control clusters at 14--20\,\kms; the four toward zero proper %
motion are drawn in vermilion.}
\label{fig:deep_vpd}
\end{figure}

Of the 74 priority-1 candidates, 35 have photometric types of M7 or later, at $Y_\mathrm{HSC}=18.0$--22.1 %
(Table~\ref{tab:deep_candidates}, Figures~\ref{fig:deep_cmd} and \ref{fig:deep_vpd}). They form the M7--L6 subset, named
after the range of their colour types.

Eleven of them have L-type colour types, L0.8--L6.6, which are later than their %
magnitudes allow. For single members at 234\,pc, $Y_\mathrm{HSC}$ implies M9.6--L3.6, so by magnitude the latest %
members of the subset are near L3--L4, and for the eight in the L box the median difference is four subtypes. Two %
effects produce it. The Legacy Surveys $z-Y$ of the five brightest in the L box is 0.17--0.20\,mag redder than that %
of early-L standards and close to that of T1 standards. Their colours therefore also fit a type near T0, and the mean %
type falls between the two solutions. Their VIKING $Z-Y$ colours, 0.86--1.18, are close to the L standards at 0.94--1.02 and far from T1 at 1.67. Using $Z$ instead of $z$ gives types M9.8--L3. The three faintest in the L box have %
$J-K_s=1.55$--1.64 (Vega), the colour of field L5--L6 dwarfs. Pleiades L dwarfs, of similar age, are marginally redder %
in $J-K$ than field dwarfs of the same type \citep{Bihain2010,2014Zapatero}, and the same method types the seven Pleiades dwarfs %
of types L0--L4.5 in the UltracoolSheet \citep{Best2025} at L5.2--L6.6. Table~\ref{tab:deep_candidates} therefore gives %
these eleven the range of types implied by their magnitudes. 

\begin{figure*}
\centering
\includegraphics{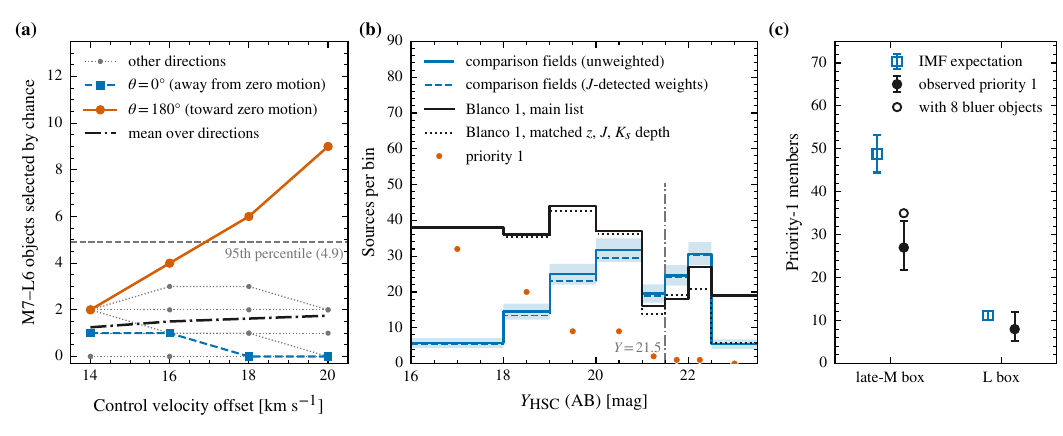}
\caption{Chance selections, field counts and the IMF expectation for the deep search. (a) Number of M7--L6 objects
selected by each velocity-shifted control cluster against its velocity offset. Vermilion marks the direction toward
zero proper motion, blue dashed the opposite direction and grey dotted the other six; the dash-dot line is the mean over
directions and the grey dashed line the 95th percentile of all 32 counts. (b) Main-list sources per magnitude bin in %
Blanco~1 (black solid) and in Blanco~1 re-evaluated at the $z$, $J$ and $K_s$ depths of the comparison fields (black
dotted). Blue lines give the expectation from the three HSC-SSP comparison fields, without weights (solid, with its error
as shading) and with weights from $J$-detected sources (dashed). Vermilion dots are the priority-1 candidates per bin. The dash-dot line
at $Y_\mathrm{HSC}=21.5$ marks the faint limit of the calibrated comparison. (c) Priority-1 members predicted by the %
Chabrier system IMF with the photometric completeness and the motion-test efficiency (open squares, ranges over age and
normalisation), against the observed late-M and L priority-1 counts with 68\% Poisson intervals (filled circles). The %
open circle adds the eight bluer priority-1 candidates at $Y_\mathrm{HSC}>18$ to the late-M count.} %
\label{fig:deep_counts}
\end{figure*}

For this subset, the velocity-shifted controls select between 0 and 9 objects, with a mean of 1.5, giving a kinematic membership purity of 0.96. The chance counts increase toward zero proper motion, where the shifted velocities approach those of distant field stars (Figure~\ref{fig:deep_counts}a). In that direction, offsets of 14 and 16\,\kms\ select 2 and 4 objects; at 18 and 20\,\kms, the counts increase to 6 and 9. Neighbouring controls select many of the same objects, so the 32 %
counts are worth about nine independent ones, and the mean chance count is uncertain by 0.6. At least 20 of the 35 are %
kinematic members of Blanco~1 at 95\% confidence. The bound is the one-sided 95\% Poisson lower limit on 35, which is 25.9, minus %
the 95th percentile of the control counts, 4.9. We chose the M7 boundary after inspecting the control counts. Halving the prior, a larger change than the %
late-M shortfall found below, or doubling it leaves the 35 unchanged and moves the purity to 0.97 or 0.92. %

The other 39 priority-1 candidates have colours at the blue edge of the type grid, which starts at M6, and lie %
0.3--2.2\,mag above the field ridge at $Y_\mathrm{HSC}=16.6$--18.5. They are consistent with M4--M6 members, and 38 %
of them were known, 25 as Gaia members \citep{HuntReffert2023}.

Of the 35 M7--L6 candidates, 32 were proposed %
earlier: 20 by \citet{Moraux2007}, 18 by \citet{Casewell2012}, and 30 by the VIKING-based search. Of these 30, 25 are %
priority-1 targets (Table~\ref{tab:offeuclid}), and 5 were priority-3 candidates of that search, whose motions it had %
not measured. Relative to the priority-1 targets of the VIKING-based search, the subset adds ten candidates: these %
five, two known only from earlier lists and the three new ones. Two of the 35 have lithium spectra
\citep{Cargile2010,Juarez2014}, and none is in the proper-motion catalogues of \citet{Platais2011}.

The same %
motions reject 19 earlier candidates of these types, which have $P<0.01$ under both field models and motion errors of %
0.3--2.5\,\masyr. Eleven of them are candidates of \citet{Moraux2007}, one also of \citet{Casewell2012}, and eight were %
listed only by the VIKING-based search, all at priority~4. %

The three new M7--L6 candidates need additional astrometric epochs because their motions depend on a few measurements. An L-box candidate at $Y_\mathrm{HSC}=21.7$ has the %
colours of the known L members and is point-like in the HSC, Legacy Surveys and VIKING images. Three of its six epoch groups determine the motion: VIKING 2012, HSC 2016 and DECam 2024. Omitting either end leaves the motion unmeasured. A second %
L-box candidate, at $Y_\mathrm{HSC}=22.1$, has three epoch groups and is selected by 6 of the 32 control clusters. A %
late-M-box candidate at $Y_\mathrm{HSC}=20.9$ lies 5\arcsec\ from a diffuse galaxy. The Legacy Surveys model it with %
a de~Vaucouleurs profile, so its $z$ is a model flux, and VIKING classes it as extended. Its motion agrees with the
cluster only after the frame rejection, which removes two of its four DECam frames of 2024, one hit by an image %
artefact and one elongated (Appendix~\ref{app:deep_versions}). Each of the three %
needs a further epoch, and the third also a test of its profile (Figure~\ref{fig:deep_finders}).

\begin{figure*}
\centering
\includegraphics{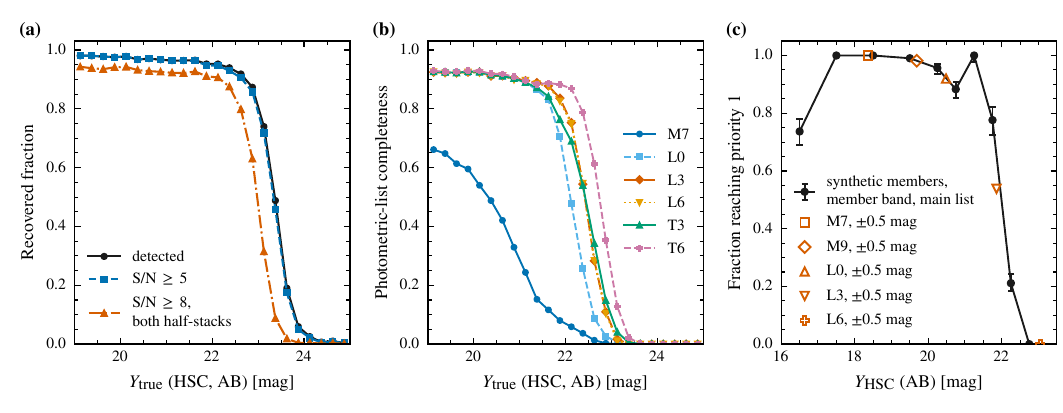}
\caption{Completeness of the deep search. (a) Fraction of injected point sources detected, measured at S/N\,$\geq5$, %
and measured at S/N\,$\geq8$ in the full coadd with at least 3$\sigma$ in both half-stacks, against their input %
magnitude. (b) Completeness of the photometric lists (main list and low-S/N list) for single members of six spectral
classes at 234\,pc, against input magnitude. (c) Fraction of synthetic members on the member band in the main list that %
reach priority~1, with Jeffreys 68\% intervals (filled circles). Open symbols give the values within 0.5\,mag of the magnitudes of M7, M9, L0, L3 and L6 members used in the IMF expectation.}
\label{fig:deep_completeness}
\end{figure*}

For synthetic members on the member band, the motion test recovers 97--100\% at M7--M9. Recovery decreases to 81--92\% at L0 and 52--54\% at L3, and reaches zero at L6 (Figure~\ref{fig:deep_completeness}). The photometric %
completeness of the main and low-S/N lists is 0.66 at M7, 0.92 at M9, 0.91 at L0, 0.76 at L3 and 0.24 at L6. 

We apply both completeness factors to the system IMF of \citet{Chabrier2003}, scaled to the 358--361 members of \citet{HuntReffert2023} in the HSC area. The expectation for late-M priority-1 members is 49, with a range of 44--53. For L members, it is 11 (10--12). The ranges cover ages of 100--150\,Myr %
\citep{Cargile2010,Juarez2014} in the models of \citet{Phillips2020} and the IMF normalisation. We find 27 late-M candidates, with a probability of $5\times10^{-4}$ for this count or fewer under the IMF prediction. The shortfall is 3.3$\sigma$, or 2.7--3.9$\sigma$ over the range (Figure~\ref{fig:deep_counts}c). Counting the eight bluer candidates at %
$Y_\mathrm{HSC}>18$ as well gives 35, still short of 49 ($P=0.024$, or 0.085 at the lower end of the range). The %
deficit is already present in the photometric
list. The member band contains 65 late-M-box sources with types of M7 or later. Of the 56 with measured motions, 27 are priority~1 and 29 have $P<0.1$. The L box, 8 against 11, agrees with the prediction ($P=0.21$). The shortfall is specific to the late-M population: its inferred abundance depends on the IMF normalisation and photometric completeness. A lower abundance per stellar member could follow from mass segregation \citep{Moraux2007} or evaporation; Section~\ref{sec:disc_field} discusses this interpretation alongside the completeness uncertainty.

\begingroup\setlength{\tabcolsep}{2.6pt}\renewcommand{\arraystretch}{0.93}
\begin{deluxetable*}{lrrrcrrrrll}
\tabletypesize{\scriptsize}
\tablecaption{Priority-1 candidates of types M7 or later, and sources without a measured motion whose photometry favours membership, in the deep search of the core region.\label{tab:deep_candidates}}
\tablehead{\colhead{Designation} & \colhead{$Y_\mathrm{HSC}$} & \colhead{$Y_\mathrm{HSC}-J$} & \colhead{$z-Y_\mathrm{HSC}$} & \colhead{Type} & \colhead{$\mu_{\alpha*}$} & \colhead{$\mu_\delta$} & \colhead{$P$} & \colhead{$N_\mathrm{grp}$} & \colhead{Earlier lists} & \colhead{Notes}\\
\colhead{(HSC, 2016.8)} & \colhead{(mag)} & \colhead{(mag)} & \colhead{(mag)} & \colhead{(phot.)} & \colhead{(\masyr)} & \colhead{(\masyr)} & \colhead{} & \colhead{} & \colhead{} & \colhead{}}
\startdata
\multicolumn{11}{c}{Priority-1 candidates with photometric types of M7 or later} \\
\hline
J000329.77$-$293610.0 & 17.96 & 0.66 & 0.39 & M7.5 $\pm$ 0.6 & 18.2 $\pm$ 0.7 & 2.6 $\pm$ 0.6 & 1.00 & 10 & HR23, V1 &  \\ %
J000148.79$-$303806.8 & 18.03 & 0.66 & 0.40 & M7.8 $\pm$ 0.6 & 18.0 $\pm$ 0.6 & 2.1 $\pm$ 0.5 & 0.99 & 11 & M07, C12, V10 &  \\ %
J000323.66$-$295517.8 & 18.04 & 0.65 & 0.43 & M7.6 $\pm$ 0.6 & 19.8 $\pm$ 0.9 & 3.4 $\pm$ 0.9 & 0.99 & 12 & M07, C12, V20 &  \\ %
J000135.72$-$300310.0 & 18.46 & 0.66 & 0.29 & M7.1 $\pm$ 0.6 & 16.3 $\pm$ 0.7 & 2.0 $\pm$ 0.7 & 0.98 & 13 & M07, C12, Li & e \\ %
J000428.92$-$302037.3 & 18.48 & 0.64 & 0.45 & M7.8 $\pm$ 0.6 & 18.5 $\pm$ 0.7 & 2.1 $\pm$ 0.7 & 0.99 & 10 & M07, Li, V2 &  \\ %
J000206.54$-$294459.9 & 18.61 & 0.78 & 0.49 & M8.6 $\pm$ 0.9 & 18.7 $\pm$ 0.9 & 3.4 $\pm$ 0.9 & 0.99 & 11 & V30 &  \\ %
J000301.17$-$293015.6 & 18.66 & 0.61 & 0.45 & M7.4 $\pm$ 0.6 & 19.6 $\pm$ 0.8 & 0.8 $\pm$ 0.8 & 0.98 & 10 & V13 &  \\ %
J000407.65$-$295918.9 & 18.71 & 0.62 & 0.44 & M7.5 $\pm$ 0.6 & 20.2 $\pm$ 0.8 & 3.2 $\pm$ 0.8 & 0.99 & 12 & M07, C12, V23 &  \\ %
J000216.14$-$301840.7 & 18.72 & 0.78 & 0.49 & M8.9 $\pm$ 1.2 & 17.6 $\pm$ 0.8 & 2.7 $\pm$ 0.7 & 0.99 & 11 & M07, C12, V8 &  \\ %
J000326.17$-$295107.2 & 18.73 & 0.66 & 0.45 & M7.8 $\pm$ 0.6 & 19.4 $\pm$ 0.9 & 2.8 $\pm$ 0.9 & 0.99 & 11 & V24 &  \\ %
J000144.73$-$293611.1 & 18.75 & 0.68 & 0.41 & M7.7 $\pm$ 0.6 & 18.6 $\pm$ 0.8 & 3.2 $\pm$ 0.8 & 0.99 & 10 & V17 &  \\ %
J000215.13$-$300952.4 & 18.80 & 0.67 & 0.44 & M7.7 $\pm$ 0.6 & 18.3 $\pm$ 0.8 & 2.6 $\pm$ 0.8 & 0.99 & 11 & M07, C12, V4 &  \\ %
J000341.61$-$293408.7 & 18.94 & 0.67 & 0.46 & M8.1 $\pm$ 0.6 & 18.8 $\pm$ 0.8 & 2.3 $\pm$ 0.8 & 0.99 & 11 & V22 &  \\ %
J000427.50$-$303519.9 & 19.11 & 0.69 & 0.50 & M8.3 $\pm$ 0.8 & 17.8 $\pm$ 0.9 & 3.0 $\pm$ 0.9 & 0.99 & 10 & M07, V6 &  \\ %
J000340.20$-$300341.1 & 19.26 & 0.72 & 0.49 & M8.6 $\pm$ 0.9 & 17.8 $\pm$ 0.9 & 3.3 $\pm$ 0.9 & 0.99 & 12 & M07, C12, V7 &  \\ %
J000355.65$-$303410.8 & 19.31 & 0.70 & 0.50 & M8.5 $\pm$ 0.9 & 18.9 $\pm$ 0.9 & 3.5 $\pm$ 0.9 & 0.99 & 11 & M07, V21 &  \\ %
J000408.11$-$294425.0 & 19.49 & 0.70 & 0.56 & M8.5 $\pm$ 0.9 & 18.3 $\pm$ 1.1 & 3.7 $\pm$ 1.1 & 0.99 & 11 & M07, V31 &  \\ %
J000456.71$-$300842.4 & 19.69 & 0.86 & 0.54 & M9.3 $\pm$ 1.6 & 19.1 $\pm$ 1.0 & 2.0 $\pm$ 1.0 & 0.99 & 11 & V25 &  \\ %
J000505.19$-$294955.6 & 19.71 & 0.83 & 0.54 & M9.3 $\pm$ 1.4 & 17.9 $\pm$ 1.0 & 2.4 $\pm$ 1.0 & 0.99 & 10 & M07, C12, V37 &  \\ %
J000202.70$-$293838.8 & 19.75 & 0.81 & 0.56 & M9.5 $\pm$ 1.7 & 19.4 $\pm$ 1.0 & 2.8 $\pm$ 1.0 & 0.99 & 11 & V36 &  \\ %
J000545.85$-$300346.0 & 19.85 & 0.83 & 0.54 & M9.4 $\pm$ 1.5 & 18.2 $\pm$ 1.4 & 3.2 $\pm$ 1.4 & 0.99 & 11 & M07, C12, V39 &  \\ %
J000608.95$-$294425.1 & 19.96 & 0.84 & 0.54 & M9.6 $\pm$ 1.7 & 17.0 $\pm$ 1.2 & 2.8 $\pm$ 1.2 & 0.99 & 10 & M07, C12, V34 &  \\ %
J000343.81$-$300401.9 & 20.00 & 0.90 & 0.54 & M9.9 $\pm$ 2.3 & 17.2 $\pm$ 1.1 & 3.6 $\pm$ 1.1 & 0.98 & 12 & M07, C12, V38 &  \\ %
J000404.54$-$293102.5 & 20.01 & 0.84 & 0.55 & M9.6 $\pm$ 1.7 & 17.6 $\pm$ 1.3 & 0.3 $\pm$ 1.3 & 0.96 & 10 & V44 &  \\ %
J000317.89$-$301141.1 & 20.18 & 0.95 & 0.53 & M9--L1 & 18.5 $\pm$ 1.2 & 3.1 $\pm$ 1.2 & 0.99 & 11 & C12, V77 (P3) & r \\ %
J000442.04$-$300433.3 & 20.22 & 0.94 & 0.60 & M9--L1 & 18.6 $\pm$ 1.0 & 3.4 $\pm$ 1.0 & 0.99 & 12 & M07, C12, V78 (P3) & r \\ %
J000306.65$-$302953.7 & 20.51 & 0.98 & 0.60 & M9--L1 & 18.0 $\pm$ 1.4 & 2.0 $\pm$ 1.4 & 0.99 & 10 & M07, C12, V92 (P3) & r \\ %
J000457.76$-$301401.8 & 20.75 & 1.16 & 0.59 & L0--L2 & 19.3 $\pm$ 1.6 & 1.0 $\pm$ 1.6 & 0.99 & 7 & M07, C12, V96 (P3) & r \\ %
J000406.55$-$292942.1 & 20.79 & 1.06 & 0.59 & L0--L2 & 19.9 $\pm$ 1.6 & 2.5 $\pm$ 1.6 & 0.99 & 8 & V86 (P3) & r \\ %
J000350.90$-$300158.2 & 20.82 & 1.12 & 0.56 & L0--L2 & 18.2 $\pm$ 1.4 & 4.9 $\pm$ 1.4 & 0.99 & 10 & C12, V63 & r \\ %
J000429.49$-$302951.7 & 20.93 & 1.05 & 0.28 & L0--L2 & 22.0 $\pm$ 1.8 & 0.6 $\pm$ 1.8 & 0.91 & 8 & none & m,r \\ %
J000205.69$-$300123.5 & 21.08 & 1.11 & 0.53 & L0--L3 & 18.3 $\pm$ 1.8 & 1.6 $\pm$ 1.8 & 0.98 & 9 & M07, C12, V65 & r \\ %
J000632.58$-$294605.8 & 21.26 & 1.34 & 0.60 & L0--L3 & 16.1 $\pm$ 2.2 & $-$0.6 $\pm$ 2.2 & 0.98 & 5 & M07, C12 & r \\ %
J000332.51$-$302048.6 & 21.70 & 1.28 & 0.59 & L1--L4 & 19.3 $\pm$ 3.4 & 3.0 $\pm$ 3.4 & 0.99 & 6 & none & e,r \\ %
J000341.17$-$294624.4 & 22.12 & 1.40 & 0.46 & L2--L5 & 23.0 $\pm$ 3.7 & $-$0.4 $\pm$ 3.7 & 0.96 & 3 & none & e,r \\ %
\hline
\multicolumn{11}{c}{Sources on the member band with a photometric likelihood ratio above 1 and no measured motion (6 in the L box, 3 in the early-T box)} \\
\hline
J000340.48$-$293556.6 & 21.87 & 1.25 & 0.43 & L & \nodata & \nodata & \nodata & \nodata & V275 (P4) & m \\ %
J000338.45$-$300752.2 & 22.06 & 1.01 & 0.44 & L & \nodata & \nodata & \nodata & \nodata & none &  \\ %
J000439.24$-$295549.2 & 22.19 & 0.94 & 0.28 & L & \nodata & \nodata & \nodata & \nodata & none & m \\ %
J000424.90$-$300326.6 & 22.45 & 1.20 & 0.78 & T & \nodata & \nodata & \nodata & \nodata & none &  \\ %
J000312.36$-$302940.3 & 22.69 & 1.38 & 0.61 & T & \nodata & \nodata & \nodata & \nodata & none &  \\ %
J000536.96$-$294606.5 & 22.83 & 1.08 & 0.52 & L & \nodata & \nodata & \nodata & \nodata & none &  \\ %
J000543.01$-$294348.1 & 22.93 & 1.45 & 0.48 & L & \nodata & \nodata & \nodata & \nodata & none &  \\ %
J000401.59$-$295746.5 & 22.97 & 1.50 & 0.50 & L & \nodata & \nodata & \nodata & \nodata & none &  \\ %
J000703.10$-$301920.4 & 22.98 & 1.51 & 0.62 & T & \nodata & \nodata & \nodata & \nodata & none &  \\ %
\enddata
\tablecomments{Magnitudes are AB; $z$ is from the Legacy Surveys DR10 and $J$ from VIKING. Designations omit the prefix HSCB1. Types are photometric, from the colours of field spectral standards, with errors in subtypes; sources of the lower part are listed by their colour box, L or T (early-T box), because a type from their magnitude would presuppose membership. $P$ is the smaller of the membership probabilities under the Gaia and control models, and $N_\mathrm{grp}$ the number of epoch groups in the motion fit. Earlier lists: M07, \citet{Moraux2007}; C12, \citet{Casewell2012}; Li, the lithium samples of \citet{Cargile2010} and \citet{Juarez2014}; HR23, the Gaia members of \citet{HuntReffert2023}; V$n$, rank $n$ in the VIKING-based search (Table~\ref{tab:offeuclid} for priority 1, otherwise the priority in parentheses); none, in no earlier list. The lower part lists the sources of the main and low-S/N lists on the member band whose photometric term favours membership (likelihood ratio above 1) and whose motions are not measured to 5\,\masyr; they are follow-up targets and are not counted as members. Notes: e, the motion is not comoving or not measured when one epoch group is left out; m, the $z$ magnitude is a model flux of an extended Legacy Surveys profile; r, L-dwarf colours, whose colour type runs late (Section~\ref{sec:deep_core}); the type given is the range implied by $Y_\mathrm{HSC}$ for a single member at 234\,pc (types M9, L0, L3 and L6 at $Y_\mathrm{HSC}=19.7$, 20.5, 21.9 and 23.0, scatter 0.55\,mag), and an equal-mass binary would be 1.4--1.9 subtypes later. All 74 priority-1 candidates, these 9 sources and the other 10 candidates of the T list are given in machine-readable form, with their colour types.}
\end{deluxetable*}
\endgroup

We estimate contamination by field dwarfs and galaxies using three comparison fields from the Wide layer of the HSC Subaru Strategic Program (HSC-SSP). They cover 5.2\,deg$^2$ at Galactic latitudes of $+57\arcdeg$ to %
$+64\arcdeg$. We degrade their coadds from the third public data release (PDR3; \citealt{Aihara2022}), made with the HSC pipeline \citep{Bosch2018}, to the depth and exposure pattern of Blanco~1 and apply the same measurements and selection. At $Y_\mathrm{HSC}\leq21.5$, where the $z$ and $J$ data of all four fields are complete, the comparison fields predict 90--96 %
field sources in the main list (Figure~\ref{fig:deep_counts}b). Blanco~1 has 171 (166 at the depths of the comparison %
fields), an excess of 70--81 over the %
weighting and depth-matching variants, with a Poisson error of 14. The 72 priority-1 candidates in that range, divided %
by the efficiency of the motion test for members on the member band, imply 75 members, in agreement.

Fainter than %
$Y_\mathrm{HSC}=21.5$ the comparison is not calibrated. There the comparison fields are 0.3\,mag shallower in $J$ at %
$J_\mathrm{AB}>21.6$, which lowers their counts. They are also 1.3--1.9 times as rich in point sources at
$Y_\mathrm{HSC}=19.5$--21.5, mostly distant thick-disc and halo stars, which raises them. At $Y_\mathrm{HSC}\leq21.5$ %
that richness has no net effect within the Poisson error, because the excess there agrees with the members implied %
by the motions. At
$Y_\mathrm{HSC}=21.5$--22.5 they predict 25 late-M-box sources with $J_\mathrm{AB}\leq21.2$, where $J$ is complete in all %
fields, against 14 in Blanco~1. Over all magnitudes, with Blanco~1 re-evaluated at the $z$, $J$ and $K_s$ depths of %
the comparison fields and with the $J$-detected weights, the excess is $61\pm16$. The choice of depth matching and %
of weights moves it between 55 %
and 72.

In the L box at $Y_\mathrm{HSC}\leq22$ the excess is $8.6\pm3.6$, which matches the seven L-box priority-1 %
candidates there. On the member band at $Y_\mathrm{HSC}=22$--23 the L box holds 17 main-list sources, of which two %
have measured motions: the new candidate at 22.1 and one rejected source. The comparison fields predict 14 field %
sources among the 17, and at least as many as Blanco~1 holds when it is evaluated at their depths, so most of the %
L-box sources there without measured motions are field objects. Five of them have a photometric term that favours %
membership and are listed in Table~\ref{tab:deep_candidates}. %

The search does not reach the T dwarfs. The completeness of the main and low-S/N lists is 0.24 at L6, that of the T %
list is at most 0.17 for T dwarfs, and %
no T candidate has a measurable motion. The 13 candidates of the T list compare with 10.5 expected from the comparison %
fields, and under any depth matching the excess stays below 1.9$\sigma$. The IMF prediction of about one T member (1.0--1.3 over the age and IMF ranges) is below the sensitivity of this comparison. Down to about L4, the result is mainly a test of the earlier candidates: the %
motions support 32, reject 19 and add three that need confirmation. In the 1.8\,deg$^2$ covered by the %
HSC images, at least 20 of the 35 M7--L6 priority-1 candidates are kinematic members of Blanco~1 at 95\% confidence. %

\subsection{Steps toward a substellar census}\label{sec:offeuclid_next}

Spectra are the first step from these lists to a census. Near-infrared types would remove the field
dwarfs at other distances and the bluer stars that make up the photometric contamination, and optical spectra could measure the levels of chromospheric activity and lithium abundance, which members below the lithium depletion boundary of Blanco~1 retain \citep{Cargile2010,Juarez2014},
as rank 2 already does.
The priority-1 targets inside $r_J$, at $J=16.4$--19.0, are bright enough for low-resolution near-infrared %
spectroscopy on large telescopes. Radial velocities would test the third component of the motion, against the
cluster velocity projected at each target, 5.4--6.5\,\kms\ inside $r_J$ and 0.3--7.7\,\kms\ over all priority-1 %
targets. Together with parallaxes they would also sharpen the traceback, which they could make decisive for
individual candidates (Section~\ref{sec:tb_future}).

Deeper near-infrared imaging from the
Euclid Wide Survey, whose reference footprint includes the cluster and its nominal tails \citep{Scaramella2022}, %
would extend the selection beyond $J\approx19.5$, where
the VIKING completeness falls to 12\%, to the mid- and late-L and T members. In the core region, the HSC search of %
Section~\ref{sec:deep_core} already measures motions of candidates to $Y_\mathrm{HSC}=22.1$, near L3 for   members, but %
does not reach the T dwarfs. The ten-year Rubin survey \citep{Ivezic2019} will extend motion measurements across the whole area, compared with 119 of the 246\,deg$^2$ covered by DP2. Those measurements will test the 162 priority-3 candidates, 131 of them in the tail zones. %

\clearpage
\begingroup\setlength{\tabcolsep}{2.7pt}
\startlongtable
\begin{deluxetable*}{rlrrcllrrrrrrll}
\tabletypesize{\scriptsize}
\tablecaption{Priority-1 targets of the photometric search beyond the Euclid field.\label{tab:offeuclid}}
\tablehead{\colhead{Rank} & \colhead{Designation} & \colhead{$\alpha$} & \colhead{$\delta$} & \colhead{Epoch} &
\colhead{Zone} & \colhead{$J$} & \colhead{$c_\mathrm{pred}$} & \colhead{$D_V$} & \colhead{$\mu_{\alpha*}$} &
\colhead{$\mu_\delta$} & \colhead{$P_\mathrm{min}$} & \colhead{Purity} & \colhead{Lit.} & \colhead{Notes}\\
\colhead{} & \colhead{(VIKING)} & \colhead{(deg)} & \colhead{(deg)} & \colhead{(yr)} & \colhead{} &
\colhead{(mag)} & \colhead{(mag)} & \colhead{(mag)} & \colhead{(\masyr)} & \colhead{(\masyr)} & \colhead{} &
\colhead{} & \colhead{} & \colhead{}}
\startdata
1 & J000329.7$-$293609.9 & 0.87398 & $-$29.60276 & 2013.13 & $<r_t$ & 16.40 & 2.71 & $-$1.03 & 18.0 $\pm$ 0.6 & 2.7 $\pm$ 0.5 & 1.00 & 0.91 & HR23 &  \\ %
2 & J000428.9$-$302037.2 & 1.12043 & $-$30.34368 & 2011.91 & $<r_t$ & 16.97 & 2.72 & $-$0.48 & 18.6 $\pm$ 0.6 & 2.1 $\pm$ 0.6 & 0.99 & 0.91 & M07 & g \\ %
3 & J235922.3$-$305925.6 & 359.84329 & $-$30.99046 & 2009.94 & $<r_t$ & 16.51 & 2.78 & $-$1.01 & 18.3 $\pm$ 0.8 & 2.7 $\pm$ 0.8 & 0.99 & 0.91 &  &  \\ %
4 & J000215.1$-$300952.4 & 0.56296 & $-$30.16456 & 2011.91 & $<r_t$ & 17.22 & 2.80 & $-$0.33 & 19.7 $\pm$ 0.6 & 2.7 $\pm$ 0.6 & 0.99 & 0.91 & M07 &  \\ %
5 & J000639.8$-$292947.9 & 1.66620 & $-$29.49667 & 2013.13 & $<r_t$ & 17.33 & 2.85 & $-$0.28 & 19.0 $\pm$ 1.4 & 2.3 $\pm$ 1.4 & 0.99 & 0.91 &  &  \\ %
6 & J000427.4$-$303519.8 & 1.11455 & $-$30.58884 & 2011.91 & $<r_t$ & 17.50 & 2.98 & $-$0.28 & 18.4 $\pm$ 0.9 & 3.0 $\pm$ 0.9 & 0.99 & 0.90 & M07 &  \\ %
7 & J000340.1$-$300341.0 & 0.91746 & $-$30.06140 & 2011.92 & $<r_t$ & 17.63 & 3.04 & $-$0.23 & 18.1 $\pm$ 0.7 & 2.6 $\pm$ 0.7 & 0.99 & 0.88 & M07 &  \\ %
8 & J000216.1$-$301840.6 & 0.56717 & $-$30.31129 & 2011.91 & $<r_t$ & 17.03 & 3.12 & $-$0.94 & 16.9 $\pm$ 0.8 & 3.3 $\pm$ 0.8 & 0.99 & 0.98 & M07 &  \\ %
9 & J001133.3$-$293714.9 & 2.88897 & $-$29.62082 & 2013.57 & $<r_t$ & 17.81 & 3.17 & $-$0.22 & 19.1 $\pm$ 1.4 & 3.8 $\pm$ 1.4 & 0.99 & 0.97 &  & a \\ %
10 & J000148.7$-$303806.7 & 0.45325 & $-$30.63521 & 2011.92 & $<r_t$ & 16.45 & 2.81 & $-$1.11 & 18.1 $\pm$ 0.6 & 2.9 $\pm$ 0.6 & 0.99 & 0.91 & M07 & g \\ %
11 & J000737.7$-$292434.6 & 1.90734 & $-$29.40962 & 2013.13 & $<r_t$ & 17.30 & 2.72 & $-$0.15 & 18.0 $\pm$ 1.1 & 2.8 $\pm$ 1.1 & 0.99 & 0.91 &  &  \\ %
12 & J000649.3$-$294041.3 & 1.70561 & $-$29.67815 & 2013.13 & $<r_t$ & 17.85 & 3.18 & $-$0.20 & 18.3 $\pm$ 0.7 & 3.1 $\pm$ 0.7 & 0.99 & 0.97 & M07 &  \\ %
13 & J000301.1$-$293015.6 & 0.75481 & $-$29.50434 & 2013.13 & $<r_t$ & 17.13 & 2.67 & $-$0.25 & 18.6 $\pm$ 0.7 & 2.2 $\pm$ 0.7 & 0.99 & 0.84 &  &  \\ %
14 & J000520.7$-$312443.2 & 1.33659 & $-$31.41201 & 2010.63 & $<r_t$ & 18.10 & 3.29 & $-$0.11 & 16.6 $\pm$ 1.1 & 2.7 $\pm$ 1.1 & 0.99 & 0.94 &  &  \\ %
15 & J000803.1$-$303514.8 & 2.01311 & $-$30.58746 & 2012.15 & $<r_t$ & 16.56 & 2.84 & $-$1.04 & 18.6 $\pm$ 1.0 & 3.0 $\pm$ 1.0 & 0.99 & 0.91 &  &  \\ %
16 & J000038.7$-$290539.8 & 0.16125 & $-$29.09440 & 2014.01 & $<r_t$ & 17.37 & 3.35 & $-$0.92 & 18.1 $\pm$ 1.2 & 2.9 $\pm$ 1.2 & 0.99 & 0.98 &  &  \\ %
17 & J000144.7$-$293610.8 & 0.43628 & $-$29.60301 & 2014.01 & $<r_t$ & 17.15 & 2.75 & $-$0.33 & 18.4 $\pm$ 0.8 & 2.7 $\pm$ 0.8 & 0.99 & 0.91 &  &  \\ %
18 & J235856.7$-$292520.4 & 359.73631 & $-$29.42235 & 2014.01 & $<r_t$ & 17.10 & 2.70 & $-$0.32 & 20.0 $\pm$ 1.0 & 1.8 $\pm$ 1.0 & 0.99 & 0.84 &  &  \\ %
19 & J000742.9$-$304243.7 & 1.92885 & $-$30.71216 & 2012.15 & $<r_t$ & 18.14 & 3.33 & $-$0.13 & 16.4 $\pm$ 2.1 & 3.8 $\pm$ 2.0 & 0.98 & 0.94 &  &  \\ %
20 & J000323.6$-$295517.6 & 0.84847 & $-$29.92158 & 2013.13 & $<r_t$ & 16.50 & 2.76 & $-$1.00 & 18.8 $\pm$ 0.7 & 2.9 $\pm$ 0.8 & 0.99 & 0.91 & M07 & b,g \\ %
21 & J000355.6$-$303410.6 & 0.98183 & $-$30.56964 & 2011.91 & $<r_t$ & 17.67 & 3.02 & $-$0.16 & 17.9 $\pm$ 1.2 & 3.1 $\pm$ 1.2 & 0.99 & 0.88 & M07 &  \\ %
22 & J000341.5$-$293408.5 & 0.92333 & $-$29.56905 & 2013.13 & $<r_t$ & 17.36 & 2.94 & $-$0.36 & 18.4 $\pm$ 1.2 & 1.1 $\pm$ 1.2 & 0.99 & 0.91 &  &  \\ %
23 & J000407.6$-$295918.8 & 1.03183 & $-$29.98858 & 2013.13 & $<r_t$ & 17.18 & 2.72 & $-$0.27 & 19.2 $\pm$ 0.7 & 3.4 $\pm$ 0.7 & 0.99 & 0.91 & M07 & g \\ %
24 & J000326.1$-$295107.0 & 0.85896 & $-$29.85197 & 2013.13 & $<r_t$ & 17.17 & 2.81 & $-$0.38 & 18.4 $\pm$ 0.5 & 2.1 $\pm$ 0.5 & 0.99 & 0.91 &  &  \\ %
25 & J000456.6$-$300842.4 & 1.23622 & $-$30.14513 & 2011.92 & $<r_t$ & 17.97 & 3.23 & $-$0.15 & 19.3 $\pm$ 1.6 & 2.5 $\pm$ 1.5 & 0.99 & 0.96 &  &  \\ %
26 & J000426.6$-$301001.5 & 1.11089 & $-$30.16709 & 2011.91 & $<r_t$ & 16.93 & 2.54 & $-$0.29 & 18.3 $\pm$ 0.8 & 1.6 $\pm$ 0.8 & 0.99 & 0.84 &  &  \\ %
27 & J235959.1$-$285652.2 & 359.99647 & $-$28.94785 & 2015.20 & $<r_t$ & 17.03 & 2.56 & $-$0.21 & 19.1 $\pm$ 1.1 & 3.0 $\pm$ 1.1 & 0.99 & 0.84 &  &  \\ %
28 & J000103.1$-$290906.0 & 0.26307 & $-$29.15167 & 2014.01 & $<r_t$ & 17.95 & 3.42 & $-$0.45 & 19.4 $\pm$ 1.6 & 1.6 $\pm$ 1.6 & 0.99 & 0.96 &  & a \\ %
29 & J235952.6$-$295426.9 & 359.96949 & $-$29.90748 & 2014.01 & $<r_t$ & 17.73 & 3.07 & $-$0.18 & 19.3 $\pm$ 0.9 & 2.5 $\pm$ 0.9 & 0.99 & 0.97 & M07 &  \\ %
30 & J000206.5$-$294459.8 & 0.52716 & $-$29.74995 & 2014.01 & $<r_t$ & 16.76 & 3.05 & $-$1.11 & 19.9 $\pm$ 0.5 & 2.3 $\pm$ 0.5 & 0.98 & 0.91 & M07 & a \\ %
31 & J000408.0$-$294424.8 & 1.03373 & $-$29.74024 & 2013.13 & $<r_t$ & 17.87 & 3.17 & $-$0.18 & 19.5 $\pm$ 1.3 & 5.8 $\pm$ 1.2 & 0.96 & 0.96 & M07 &  \\ %
32 & J000801.1$-$304950.0 & 2.00483 & $-$30.83057 & 2012.15 & $<r_t$ & 18.34 & 3.48 & $-$0.13 & 17.6 $\pm$ 1.6 & 2.5 $\pm$ 1.6 & 0.99 & 0.90 &  &  \\ %
33 & J000641.7$-$293423.4 & 1.67390 & $-$29.57317 & 2013.13 & $<r_t$ & 17.74 & 3.51 & $-$0.79 & 20.8 $\pm$ 1.9 & 2.9 $\pm$ 1.8 & 0.98 & 0.97 &  &  \\ %
34 & J000608.9$-$294424.9 & 1.53727 & $-$29.74027 & 2013.13 & $<r_t$ & 18.20 & 3.34 & $-$0.08 & 17.3 $\pm$ 1.3 & 3.2 $\pm$ 1.3 & 0.99 & 0.93 & M07 &  \\ %
35 & J000827.3$-$285312.4 & 2.11395 & $-$28.88678 & 2014.64 & $<r_t$ & 17.22 & 2.81 & $-$0.34 & 20.6 $\pm$ 1.2 & 1.3 $\pm$ 1.2 & 0.98 & 0.91 &  &  \\ %
36 & J000202.6$-$293838.6 & 0.51119 & $-$29.64406 & 2014.01 & $<r_t$ & 17.91 & 3.25 & $-$0.25 & 20.3 $\pm$ 1.4 & 1.6 $\pm$ 1.4 & 0.99 & 0.96 &  & a \\ %
37 & J000505.1$-$294955.6 & 1.27161 & $-$29.83211 & 2013.13 & $<r_t$ & 17.98 & 3.27 & $-$0.20 & 18.1 $\pm$ 1.0 & 2.3 $\pm$ 1.0 & 0.99 & 0.96 & M07 &  \\ %
38 & J000343.8$-$300401.8 & 0.93250 & $-$30.06719 & 2011.92 & $<r_t$ & 18.19 & 3.33 & $-$0.07 & 18.3 $\pm$ 1.0 & 3.5 $\pm$ 1.0 & 0.99 & 0.93 & M07 &  \\ %
39 & J000545.8$-$300346.0 & 1.44102 & $-$30.06279 & 2012.15 & $<r_t$ & 18.12 & 3.23 & $-$0.01 & 19.7 $\pm$ 1.2 & 4.2 $\pm$ 1.2 & 0.98 & 0.94 & M07 &  \\ %
40 & J000034.7$-$300251.2 & 0.14492 & $-$30.04757 & 2011.91 & $<r_t$ & 18.49 & 3.36 & 0.19 & 19.9 $\pm$ 1.6 & 3.0 $\pm$ 1.6 & 0.97 & 0.84 & M07 &  \\ %
41 & J000943.3$-$300736.7 & 2.43073 & $-$30.12688 & 2012.15 & $<r_t$ & 17.04 & 2.60 & $-$0.26 & 16.1 $\pm$ 0.9 & 1.5 $\pm$ 0.9 & 0.97 & 0.84 &  &  \\ %
42 & J000426.2$-$291620.0 & 1.10934 & $-$29.27222 & 2013.13 & $<r_t$ & 18.13 & 3.43 & $-$0.28 & 22.3 $\pm$ 1.6 & 5.7 $\pm$ 1.6 & 0.92 & 0.94 &  & f \\ %
43 & J000342.8$-$320926.6 & 0.92849 & $-$32.15740 & 2010.68 & $r_t$--$r_J$ & 17.39 & 2.95 & $-$0.35 & 18.5 $\pm$ 1.4 & 3.2 $\pm$ 1.4 & 0.90 & 0.91 &  &  \\ %
44 & J000404.5$-$293102.4 & 1.01882 & $-$29.51734 & 2013.13 & $<r_t$ & 18.26 & 3.34 & $-$0.01 & 19.5 $\pm$ 1.6 & $-$0.4 $\pm$ 1.6 & 0.95 & 0.92 &  &  \\ %
45 & J225624.0$-$271634.1 & 344.10016 & $-$27.27616 & 2015.47 & leading & 16.28 & 2.60 & $-$1.01 & 17.1 $\pm$ 0.8 & $-$1.3 $\pm$ 0.7 & 0.65 & 0.84 &  & c,d,e \\ %
46 & J000115.3$-$305149.6 & 0.31387 & $-$30.86379 & 2011.92 & $<r_t$ & 18.53 & 3.57 & $-$0.08 & 15.4 $\pm$ 1.5 & 1.7 $\pm$ 1.5 & 0.96 & 0.82 &  &  \\ %
47 & J000258.4$-$285115.0 & 0.74372 & $-$28.85418 & 2014.64 & $<r_t$ & 18.07 & 3.31 & $-$0.17 & 21.0 $\pm$ 2.2 & 6.9 $\pm$ 2.1 & 0.92 & 0.95 &  &  \\ %
48 & J001253.3$-$300102.6 & 3.22214 & $-$30.01739 & 2013.57 & $r_t$--$r_J$ & 16.98 & 2.67 & $-$0.40 & 17.2 $\pm$ 0.6 & 3.5 $\pm$ 0.6 & 0.93 & 0.84 &  &  \\ %
49 & J000958.1$-$312205.0 & 2.49243 & $-$31.36808 & 2010.66 & $r_t$--$r_J$ & 17.98 & 3.17 & $-$0.06 & 15.7 $\pm$ 1.9 & 2.3 $\pm$ 1.9 & 0.76 & 0.96 &  &  \\ %
50 & J235826.2$-$283018.8 & 359.60928 & $-$28.50525 & 2015.20 & $r_t$--$r_J$ & 17.50 & 3.32 & $-$0.75 & 21.3 $\pm$ 2.0 & 5.7 $\pm$ 2.0 & 0.66 & 0.98 &  & a,e \\ %
51 & J002315.4$-$320451.3 & 5.81420 & $-$32.08093 & 2010.79 & trailing & 17.86 & 3.19 & $-$0.20 & 18.4 $\pm$ 1.5 & 4.5 $\pm$ 1.5 & 0.72 & 0.97 &  &  \\ %
52 & J001428.2$-$322711.5 & 3.61756 & $-$32.45322 & 2011.69 & trailing & 17.23 & 2.77 & $-$0.28 & 18.8 $\pm$ 1.4 & 4.2 $\pm$ 1.4 & 0.67 & 0.91 &  &  \\ %
53 & J002338.7$-$330053.3 & 5.91157 & $-$33.01483 & 2010.79 & trailing & 17.60 & 2.97 & $-$0.16 & 17.6 $\pm$ 1.3 & 4.0 $\pm$ 1.3 & 0.74 & 0.89 &  &  \\ %
54 & J002458.2$-$312325.4 & 6.24262 & $-$31.39041 & 2010.72 & trailing & 17.41 & 3.08 & $-$0.51 & 16.6 $\pm$ 1.7 & 3.4 $\pm$ 1.7 & 0.73 & 0.98 &  &  \\ %
55 & J003227.5$-$344507.1 & 8.11461 & $-$34.75199 & 2013.58 & trailing & 18.06 & 3.46 & $-$0.40 & 17.7 $\pm$ 2.3 & 3.6 $\pm$ 2.3 & 0.67 & 0.95 &  &  \\ %
56 & J003216.6$-$335844.0 & 8.06929 & $-$33.97889 & 2013.58 & trailing & 17.57 & 3.03 & $-$0.28 & 17.8 $\pm$ 2.0 & 3.9 $\pm$ 1.9 & 0.68 & 0.89 &  &  \\ %
57 & J002144.0$-$310955.6 & 5.43353 & $-$31.16545 & 2010.72 & trailing & 17.45 & 2.99 & $-$0.35 & 16.6 $\pm$ 1.2 & 2.1 $\pm$ 1.2 & 0.61 & 0.90 &  & d,e \\ %
58 & J234231.5$-$274808.2 & 355.63127 & $-$27.80228 & 2015.60 & leading & 17.44 & 3.14 & $-$0.57 & 20.7 $\pm$ 2.4 & 1.8 $\pm$ 2.4 & 0.56 & 0.98 &  & e,h \\ %
63 & J000350.8$-$300158.2 & 0.96202 & $-$30.03285 & 2011.91 & $<r_t$ & 18.87 & 3.54 & 0.30 & 20.1 $\pm$ 2.9 & 5.6 $\pm$ 2.9 & 0.91 & 0.55 & C12 & c,h \\ %
65 & J000205.6$-$300123.4 & 0.52364 & $-$30.02318 & 2011.92 & $<r_t$ & 19.05 & 3.46 & 0.60 & 20.5 $\pm$ 2.3 & 3.3 $\pm$ 2.3 & 0.80 & 0.35 & M07 & c,h \\ %
\enddata
\tablecomments{Rank is the order of the search by $P_\mathrm{min}$, the smaller of the membership probabilities under
the Gaia and control field models; ranks 58, 63 and 65 were priority 2 in the search. Motions and $P_\mathrm{min}$
include the archival epochs of Section~\ref{sec:offeuclid_archival}; for ranks 1, 20 and 45 the fitted motion is
combined with Gaia DR3. Positions are VIKING catalogue positions on the Gaia DR3 frame at the epoch given. $J$ is
VIKING (Vega); $c_\mathrm{pred}$ is the $(\IE-\JE)_0$ colour predicted from VIKING and KiDS photometry, and $D_V$ the
offset from the field ridge at 234.4\,pc in VIKING $J$ (negative is brighter). Zones: inside $r_t$, between $r_t$
and $r_J$, and the leading and trailing tail zones. The Purity column gives photometric selection purity: the probability that Euclid would identify a source of this $J$ and $c_\mathrm{pred}$ as a compact ultracool dwarf on the cluster sequence, calibrated in the overlap field (Section~\ref{sec:offeuclid_purity}). Membership is described by $P_\mathrm{min}$. Lit.\ marks an earlier listing within
2\arcsec: M07, a candidate of \citet{Moraux2007} (rank 30 at 2.2\arcsec, an offset in the catalogued position;
Section~\ref{sec:offeuclid_targets}); C12, a member in Table 3 of \citet{Casewell2012}; and HR23, a Gaia member of
\citet{HuntReffert2023}. Designations truncate the coordinates. The machine-readable version lists all 76
priority-1 and priority-2 targets of the search with their search and adopted motions and the outcome of the image
inspection. Notes: (a) The search motion rested on one VIKING epoch group; with the VIKING epochs left out, the archival epochs keep priority 1. (b) Astrometric fit $\chi^2=92$ for 36 degrees of freedom with the archival epochs ($\chi^2=61.4$ for 12 in the search); a saturated star 11\arcsec\ east of the target spreads spikes and a halo across the field. (c) No DP2 epoch. (d) Priority 1 depends on the zone boundaries. Rank 45 lies 14.99\arcdeg\ from the centre and 29.6\arcdeg\ from the leading-tail axis, at the corner of a zone that ends at 15\arcdeg\ and has a half-width of 30\arcdeg; rank 57 lies 25.3\arcdeg\ from the trailing-tail axis. (e) Falls below $P=0.5$ when the prior of its zone is halved. (f) Bright-star ghosts cover the KiDS images; a point source in VIKING $J$, $K_s$ and Legacy Surveys $z$. (g) Published optical spectrum, typed M7--M8 by \citet{Moraux2007}; rank 2 also shows lithium \citep{Cargile2010,Juarez2014}. (h) Priority 2 in the search; priority 1 with the archival epochs.}
\end{deluxetable*}
\endgroup

\section{Conclusions}\label{sec:conclusions}

A 36\,deg$^2$ Euclid DR1 calibration field that lies 1.0--8.7\arcdeg\ west of the young open cluster Blanco~1 was used %
to test whether Euclid can find substellar members of a nearby cluster outside its core. The field covers none of the
core, 11.3\% of the area inside the tidal radius and the base of the leading tidal tail. %

\begin{enumerate}
\item Euclid photometry selects 13,590 red sources in the field, NISP spectra type 3,002 of them, and VIKING, KiDS, %
Euclid and Rubin DP2 positions give proper motions over a median baseline of 13\,yr, with errors of 1.1--1.4\,\masyr\ %
for $\JE\leq20.5$. %
\item The comoving late-M and L population exceeds all 64 velocity-shifted controls, with a likelihood statistic more %
than five times the largest control value. We identify ten late-M and L candidates (R4, C1--C8 and B2), including C8 conditional on its photometric type, with membership probabilities of at least 0.5 under both field models. The adopted model predicts about two field dwarfs among them. If they %
are single members, their masses are 0.030--0.061\,$M_\odot$
at 100--150\,Myr. R4 and C1--C4 stay above 0.5 in every variant of the model; the other five depend on the fitted %
location of the member sequence, and C8 also on its photometric type.
\item The fitted member sequence lies within about 0.2\,mag of the field ridge, 0.53\,mag fainter than the %
youth-corrected ridge. We leave its physical interpretation open, for the reasons discussed in Section~\ref{sec:disc_field}.
\item The fitted number of late-M and L members is 2.2 (1.5--2.9) times the number that a lognormal mass function %
predicts when scaled to the field's Gaia stars, a ratio that depends on that normalisation. Of the fitted members of
all classes, $53\pm17$\% lie inside the projected Jacobi radius, against 45--59\% of the field's Gaia members, the same %
within the errors. The fitted number of L-T and T members is zero, against 1.9 predicted among the quality-passing L-T and T sources. %
\item SPHEREx detects 21 of the 22 named objects. None of the late-M and L objects shows 3--5\,\micron\ emission above its photosphere, which excludes an excess above about 35\% of the photosphere for a typical object. %
\item Of the nine ultracool dwarfs in our spectroscopic selection using the method of \citet{DominguezTagle2025}, only R4 moves with the cluster. The other eight
are field dwarfs, three of them foreground T dwarfs at 30--67\,pc. In tracebacks over 150\,Myr with free radial %
velocities, no draw takes any of the eight within the Jacobi radius slower than 10\,\kms, which makes them inconsistent %
with slow evaporation from the cluster. The distance-independent transverse velocities also exclude slow evaporation: seven have floors of 9--23\,\kms\ relative to the cluster over the tested distance range. R1 attains its lower floor only at a distance where its track misses the cluster (Section~\ref{sec:tb_rejected}). %
\item For R4 and the comoving candidates the traceback leaves members and recent escapees indistinguishable. Distance
errors of 0.15\,mag would let 18\% of true escapees reach a slow-passage fraction of 0.5, and with radial velocities %
good to 2\,\kms\ as well, a slow-passage fraction of 0.1 would keep about 30 times more escapees than comoving field %
stars.
\item Rubin DP2 extends the baseline by about two years and adds an optical colour that flagged two inconsistent %
spectral types, after correction of a 13\,mas frame offset in right ascension; it changes no membership category. %
\item Applying the selection to VIKING and KiDS photometry over 246\,deg$^2$ beyond the Euclid field gives 60 priority-1 spectroscopic targets with the archival astrometry included. Images from 1999--2026 reduce the median motion error from 2.1 to 1.2\,\masyr; the search controls predict about seven chance selections. Of these targets, 51 lie inside the Jacobi %
radius, 19 are earlier candidates of \citet{Moraux2007}, one more is a member listed by \citet{Casewell2012}, one is %
a Gaia member and four already have optical spectra. %
\item Archival Subaru HSC $Y$-band images, 1.7\,mag deeper than VIKING, cover 1.8\,deg$^2$ that contain the core. With %
motions from 2011 to 2025, 35 candidates of photometric types M7 or later reach priority~1 there, against 1.5 %
expected by chance, and at least 20 of them are kinematic members at 95\% confidence. The motions support 32 earlier candidates, reject 19 and add three that need further epochs. The bound applies to these photometric types; T members lie beyond the sensitivity of the search. The late-M priority-1 count there, 27, falls %
3.3$\sigma$ short of the 49 that the system mass function predicts. Interpreting that shortfall requires the completeness assessment discussed in Section~\ref{sec:disc_field}.
\end{enumerate}

In this partial field, we identify L-dwarf candidates at 234\,pc using Euclid photometry, NISP types for nine of the ten candidates, and a decade-long astrometric baseline anchored by Euclid positions. The reference footprint of the Euclid Wide Survey includes the core and nominal tail directions of Blanco~1, so later data could extend this test across the cluster. Parallaxes and radial velocities for the candidates are the next step
towards confirming individual members and testing whether any of them have escaped.

\begin{acknowledgments}
We are grateful to Ivan Baldry, René Laureijs, Yannick Mellier, Ismael Tereno and Roland Vavrek for their help in collecting the Euclid data. E.L.M., %
M.{\v Z}., N.S., S.T., and N.V. are supported by the European Research Council Advanced grant SUBSTELLAR, project number 101054354. 
This paper is the first of a series in the study of Euclid ultracool dwarfs making intense use of our custom AI platform, developed in the framework of a Proof of Concept grant linked to the Substellar ERC project (https://doi.org/10.3030/101331394).
NL acknowledge support from the Agencia  Estatal de Investigaci\'on del Ministerio de Ciencia e Innovaci\'on (AEI-MCINN) under grants PID2022-137241NB-C41 and PID2025-176415NB-I00\@.
The authors acknowledge financial support from the Agencia Estatal de Investigación (AEI) through the Severo Ochoa Centre of Excellence accreditation awarded to the Instituto de Astrofísica de Canarias, grant CEX2025-001609-S, funded by MICIU/AEI/10.13039/501100011033 . We acknowledge the European Space Agency and a number of agencies and institutes that have
supported the development of Euclid, in particular the Agenzia Spaziale Italiana, the Austrian
Forschungsf\"orderungsgesellschaft funded through BMIMI, the Belgian Science Policy, the Canadian Euclid Consortium,
the Deutsches Zentrum f\"ur Luft- und Raumfahrt, the DTU Space and the Niels Bohr Institute in Denmark, the French
Centre National d'Etudes Spatiales, the Funda\c{c}\~{a}o para a Ci\^{e}ncia e a Tecnologia, the Hungarian Academy of
Sciences, the Ministerio de Ciencia, Innovaci\'{o}n y Universidades, the National Aeronautics and Space
Administration, the National Astronomical Observatory of Japan, the Netherlandse Onderzoekschool Voor Astronomie, the
Norwegian Space Agency, the Research Council of Finland, the Romanian Space Agency, the Swiss Space Office (SSO) at
the State Secretariat for Education, Research, and Innovation (SERI), and the United Kingdom Space Agency. A complete
and detailed list is available on the Euclid web site (\url{www.euclid-ec.org/consortium/community/}). This work has
made use of the Euclid Data Release 1 (DR1) from the European Space Agency (ESA) mission Euclid, available at
\url{https://www.cosmos.esa.int/web/euclid/data-release-dr1}. This work has made use of the Euclid Quick Release Q1
data from the Euclid mission of the European Space Agency (ESA), 2025, \url{https://doi.org/10.57780/esa-2853f3b}
\citep{EuclidQ1data}. This research makes use of ESA Datalabs (\url{datalabs.esa.int}), an initiative by ESA's Data
Science and Archives Division in the Science and Operations Department, Directorate of Science \citep{Navarro2024}.

This material is based upon work supported in part by the National Science Foundation through Cooperative Agreements
AST-1258333 and AST-2241526 and Cooperative Support Agreements AST-1202910 and 2211468 managed by the Association of
Universities for Research in Astronomy (AURA), and the Department of Energy under Contract No.\ DE-AC02-76SF00515 with
the SLAC National Accelerator Laboratory managed by Stanford University. Additional Rubin Observatory funding comes
from private donations, grants to universities, and in-kind support from LSST-DA Institutional Members. This
publication is based in part on proprietary Rubin Observatory Legacy Survey of Space and Time (LSST) data, and was
prepared in accordance with the Rubin Observatory data rights and access policies. All authors of this publication
meet the requirements for co-authorship of proprietary LSST data. This research uses services or data provided by the
Rubin Science Platform at NSF-DOE Vera C. Rubin Observatory, which is jointly funded by the U.S. National Science
Foundation and the U.S. Department of Energy, Office of Science.

Based on data obtained from the ESO Science Archive Facility with DOIs \url{https://doi.org/10.18727/archive/37}
(KiDS), \url{https://doi.org/10.18727/archive/59} (VIKING, ESO programme 179.A-2004) and
\url{https://doi.org/10.18727/archive/57} (VHS, ESO programme 179.A-2010), and on data products produced by the KiDS
consortium \citep{ESOKiDS,ESOVIKING,ESOVHS}. The KiDS production team acknowledges support from:
Deutsche Forschungsgemeinschaft, ERC, NOVA and NWO-M grants; Target; the University of Padova, and the University
Federico II (Naples).

This work has made use of data from the European Space Agency (ESA) mission {\it Gaia}
(\url{https://www.cosmos.esa.int/gaia}), processed by the {\it Gaia} Data Processing and Analysis Consortium (DPAC,
\url{https://www.cosmos.esa.int/web/gaia/dpac/consortium}). Funding for the DPAC has been provided by national
institutions, in particular the institutions participating in the {\it Gaia} Multilateral Agreement.

This work uses the Gaia-ESO Survey membership catalogue of \citet{Jackson2022}, based on data products from
observations made with ESO Telescopes at the La Silla Paranal Observatory under programme ID 188.B-3002. These data
products have been processed by the Cambridge Astronomy Survey Unit (CASU) at the Institute of Astronomy, University
of Cambridge, and by the FLAMES/UVES reduction team at INAF/Osservatorio Astrofisico di Arcetri.

This work made use of the Fourth Data Release of the GALAH Survey \citep{Buder2025}. The GALAH Survey is based on
data acquired through the Australian Astronomical Observatory, under programs: A/2013B/13 (The GALAH pilot survey);
A/2014A/25, A/2015A/19, A2017A/18 (The GALAH survey phase 1); A2018A/18 (Open clusters with HERMES); A2019A/1
(Hierarchical star formation in Ori OB1); A2019A/15, A/2020B/23, R/2022B/5, R/2023A/4, R2023B/5 (The GALAH survey
phase 2); A/2015B/19, A/2016A/22, A/2016B/10, A/2017B/16, A/2018B/15 (The HERMES-TESS program); A/2015A/3, A/2015B/1,
A/2015B/19, A/2016A/22, A/2016B/12, A/2017A/14, A/2020B/14 (The HERMES K2-follow-up program); R/2022B/02 and
A/2023A/09 (Combining asteroseismology and spectroscopy in K2); A/2023A/8 (Resolving the chemical fingerprints of
Milky Way mergers); and A/2023B/4 (s-process variations in southern globular clusters). We acknowledge the
traditional owners of the land on which the AAT stands, the Gamilaraay people, and pay our respects to elders past
and present. This paper includes data that has been provided by AAO Data Central (datacentral.org.au).

This publication makes use of data products from the Wide-field Infrared Survey Explorer, which is a joint project of
the University of California, Los Angeles, and the Jet Propulsion Laboratory/California Institute of Technology,
funded by the National Aeronautics and Space Administration \citep{WrightEL2010}. This publication makes use of data
products from the Near-Earth Object Wide-field Infrared Survey Explorer (NEOWISE), which is a joint project of the Jet
Propulsion Laboratory/California Institute of Technology and the University of California, Los Angeles. NEOWISE is
funded by the National Aeronautics and Space Administration. This research has made use of the NASA/IPAC Infrared
Science Archive, which is funded by the National Aeronautics and Space Administration and operated by the California
Institute of Technology.
This research has made use of SPHEREx Quick Release spectral images \citep{SPHERExQR2} obtained from the
NASA/IPAC Infrared Science Archive; SPHEREx is a NASA Medium Explorer mission.

This work has benefited from The UltracoolSheet at \url{http://bit.ly/UltracoolSheet}, maintained by Will Best, Trent
Dupuy, Michael Liu, Aniket Sanghi, Rob Siverd, and Zhoujian Zhang, and developed from compilations by
\citet{DupuyLiu2012}, \citet{DupuyKraus2013}, \citet{Deacon2014}, \citet{Liu2016}, \citet{Best2018},
\citet{Best2021}, \citet{Sanghi2023}, and \citet{Schneider2023}.

The Legacy Surveys consist of three individual and complementary projects: the Dark Energy Camera Legacy Survey
(DECaLS; Proposal ID \#2014B-0404; PIs: David Schlegel and Arjun Dey), the Beijing-Arizona Sky Survey (BASS; NOAO
Prop. ID \#2015A-0801; PIs: Zhou Xu and Xiaohui Fan), and the Mayall z-band Legacy Survey (MzLS; Prop. ID
\#2016A-0453; PI: Arjun Dey). DECaLS, BASS and MzLS together include data obtained, respectively, at the Blanco
telescope, Cerro Tololo Inter-American Observatory, NSF's NOIRLab; the Bok telescope, Steward Observatory, University
of Arizona; and the Mayall telescope, Kitt Peak National Observatory, NOIRLab. Pipeline processing and analyses of the
data were supported by NOIRLab and the Lawrence Berkeley National Laboratory (LBNL). The Legacy Surveys project is
honored to be permitted to conduct astronomical research on Iolkam Du'ag (Kitt Peak), a mountain with particular
significance to the Tohono O'odham Nation. NOIRLab is operated by the Association of Universities for Research in
Astronomy (AURA) under a cooperative agreement with the National Science Foundation. LBNL is managed by the Regents of
the University of California under contract to the U.S. Department of Energy. This project used data obtained with the
Dark Energy Camera (DECam), which was constructed by the Dark Energy Survey (DES) collaboration. Funding for the DES
Projects has been provided by the U.S. Department of Energy, the U.S. National Science Foundation, the Ministry of
Science and Education of Spain, the Science and Technology Facilities Council of the United Kingdom, the Higher
Education Funding Council for England, the National Center for Supercomputing Applications at the University of
Illinois at Urbana-Champaign, the Kavli Institute of Cosmological Physics at the University of Chicago, Center for
Cosmology and Astro-Particle Physics at the Ohio State University, the Mitchell Institute for Fundamental Physics and
Astronomy at Texas A\&M University, Financiadora de Estudos e Projetos, Fundacao Carlos Chagas Filho de Amparo,
Financiadora de Estudos e Projetos, Fundacao Carlos Chagas Filho de Amparo a Pesquisa do Estado do Rio de Janeiro,
Conselho Nacional de Desenvolvimento Cientifico e Tecnologico and the Ministerio da Ciencia, Tecnologia e Inovacao,
the Deutsche Forschungsgemeinschaft and the Collaborating Institutions in the Dark Energy Survey. The Collaborating
Institutions are Argonne National Laboratory, the University of California at Santa Cruz, the University of
Cambridge, Centro de Investigaciones Energeticas, Medioambientales y Tecnologicas-Madrid, the University of Chicago,
University College London, the DES-Brazil Consortium, the University of Edinburgh, the Eidgenossische Technische
Hochschule (ETH) Zurich, Fermi National Accelerator Laboratory, the University of Illinois at Urbana-Champaign, the
Institut de Ciencies de l'Espai (IEEC/CSIC), the Institut de Fisica d'Altes Energies, Lawrence Berkeley National
Laboratory, the Ludwig Maximilians Universitat Munchen and the associated Excellence Cluster Universe, the University
of Michigan, NSF's NOIRLab, the University of Nottingham, the Ohio State University, the University of Pennsylvania,
the University of Portsmouth, SLAC National Accelerator Laboratory, Stanford University, the University of Sussex, and
Texas A\&M University. BASS is a key project of the Telescope Access Program (TAP), which has been funded by the
National Astronomical Observatories of China, the Chinese Academy of Sciences (the Strategic Priority Research Program
``The Emergence of Cosmological Structures'' Grant \# XDB09000000), and the Special Fund for Astronomy from the
Ministry of Finance. The BASS is also supported by the External Cooperation Program of Chinese Academy of Sciences
(Grant \# 114A11KYSB20160057), and Chinese National Natural Science Foundation (Grant \# 12120101003, \# 11433005).
The Legacy Survey team makes use of data products from the Near-Earth Object Wide-field Infrared Survey Explorer
(NEOWISE), which is a project of the Jet Propulsion Laboratory/California Institute of Technology. NEOWISE is funded
by the National Aeronautics and Space Administration. The Legacy Surveys imaging of the DESI footprint is supported by
the Director, Office of Science, Office of High Energy Physics of the U.S. Department of Energy under Contract No.
DE-AC02-05CH1123, by the National Energy Research Scientific Computing Center, a DOE Office of Science User Facility
under the same contract; and by the U.S. National Science Foundation, Division of Astronomical Sciences under Contract
No. AST-0950945 to NOAO.

This research is based in part on data collected at the Subaru Telescope, which is operated by the National
Astronomical Observatory of Japan. We are honored and grateful for the opportunity of observing the Universe from
Maunakea, which has the cultural, historical, and natural significance in Hawaii. Based in part on data collected at
Subaru Telescope and obtained from the SMOKA, which is operated by the Astronomy Data Center, National Astronomical
Observatory of Japan. Based in part on observations obtained with CFH12K at the Canada-France-Hawai`i Telescope
(CFHT) which is operated by the National Research Council of Canada, the Institut National des Sciences de l'Univers
of the Centre National de la Recherche Scientifique of France, and the University of Hawai`i. CFHT is located on
Maunakea on Hawai`i Island, a mountain of considerable cultural, natural, and ecological significance. Maunakea is a
sacred site to Native Hawaiians, also known as K\={a}naka `\={O}iwi. This research used the facilities of the
Canadian Astronomy Data Centre operated by the National Research Council of Canada with the support of the Canadian
Space Agency. This research is based in part on data obtained from the Astro Data Archive at NSF NOIRLab. These data
are associated with observing programs 2006B-0569 (PI: Barkhouse), 2012B-0001 (PI: Frieman), 2012B-0003 (PI: Frieman), 2012B-3013 (PI: James), 2013A-0617 (PI: Mackey), 2014A-0634 (PI: James), 2014B-0404 (PI: Schlegel), 2015A-0617 (PI: Nataf), 2016B-0124 (PI: Frieman), 2017A-0260 (PI: Soares-Santos), 2017B-0078 (PI: Bouy), 2018B-0340 (PI: Bouy), 2018B-0905 (PI: Oh), 2019A-0305 (PI: Drlica-Wagner), 2019A-0308 (PI: Dell'Antonio), 2019B-0042 (PI: Bouy), 2019B-1004 (PI: Chanam\'e), 2019B-1014 (PI: Olivares), 2020B-0040 (PI: Prato), 2020B-0909 (PI: Chanam\'e), 2021B-0905 (PI: Assef), 2022B-297190 (PIs: Palmese, Wang), 2022B-715089 (PIs: Andreoni, Palmese), 2023B-851374 (PIs: Palmese, Andreoni), 2024B-305587 (PIs: Bouy, Points), 2024B-591619 (PI: Demarco), 2024B-826402 (PI: Assef), 2025A-729671 (PI: Palmese). NOIRLab is managed by the Association of Universities for
Research in Astronomy (AURA) under a cooperative agreement with the U.S. National Science Foundation. This research
draws upon DECam data as distributed by the Astro Data Archive at NSF NOIRLab. Based on observations at NSF Cerro
Tololo Inter-American Observatory, a program of NOIRLab (NOIRLab Prop. IDs and PIs listed above), which is managed by
the Association of Universities for Research in Astronomy (AURA) under a cooperative agreement with the U.S. National
Science Foundation. %

The Hyper Suprime-Cam (HSC) collaboration includes the astronomical communities of Japan and Taiwan, and Princeton
University. The HSC instrumentation and software were developed by the National Astronomical Observatory of Japan
(NAOJ), the Kavli Institute for the Physics and Mathematics of the Universe (Kavli IPMU), the University of Tokyo, the
High Energy Accelerator Research Organization (KEK), the Academia Sinica Institute for Astronomy and Astrophysics in
Taiwan (ASIAA), and Princeton University. Funding was contributed by the FIRST program from the Japanese Cabinet
Office, the Ministry of Education, Culture, Sports, Science and Technology (MEXT), the Japan Society for the Promotion
of Science (JSPS), Japan Science and Technology Agency (JST), the Toray Science Foundation, NAOJ, Kavli IPMU, KEK,
ASIAA, and Princeton University. This paper makes use of software developed for Vera C. Rubin Observatory. We thank
the Rubin Observatory for making their code available as free software at \url{http://pipelines.lsst.io/}. This paper
is based on data collected at the Subaru Telescope and retrieved from the HSC data archive system, which is operated by
the Subaru Telescope and Astronomy Data Center (ADC) at NAOJ. Data analysis was in part carried out with the
cooperation of Center for Computational Astrophysics (CfCA), NAOJ.

This research has made use of the SIMBAD database, CDS, Strasbourg Astronomical Observatory, France
\citep{Wenger2000}. This research has made use of the VizieR catalogue access tool, CDS, Strasbourg Astronomical
Observatory, France (DOI: 10.26093/cds/vizier) \citep{Ochsenbein2000}. This research made use of
hips2fits,\footnote{\url{https://alasky.cds.unistra.fr/hips-image-services/hips2fits}} a service provided by CDS.
This research has made use of the Astrophysics Data System, funded by NASA under Cooperative Agreement
80NSSC21M0056.

\end{acknowledgments}

\facilities{Euclid, ESO:VISTA (VIRCAM), VST (OmegaCAM), Gaia, WISE, NEOWISE, SPHEREx, Rubin:Simonyi (LSSTCam), Rubin:USDAC, VLT:Kueyen (FLAMES), AAT (HERMES),
CFHT (CFH12K), Blanco (DECam, Mosaic II), Subaru (Hyper Suprime-Cam), IRSA, CDS, CADC, Astro Data Archive}

\software{Astropy \citep{Astropy2013,Astropy2018,Astropy2022}, NumPy \citep{Harris2020}, SciPy \citep{Virtanen2020},
Matplotlib \citep{Hunter2007}, pandas \citep{McKinney2010}, galpy \citep{Bovy2015}, PyVO \citep{Graham2014},
astroquery \citep{Ginsburg2019}, SEP \citep{Barbary2016,BertinArnouts1996}, reproject \citep{Robitaille2020},
SPLAT \citep{Burgasser2017}, healpy \citep{Zonca2019,Gorski2005}, pyarrow}

\section*{Data availability}
The Euclid DR1 images, catalogues and spectra used here will become public through the ESA Euclid Science Archive
with the ESA release of DR1 \citep{EuclidDR1}; the Euclid Q1 data used for the colour calibration are public
\citep{EuclidQ1data}. The VIKING and KiDS catalogues and images are available from the ESO Science Archive Facility
\citep{ESOVIKING,ESOKiDS}, Gaia DR3 from the ESA Gaia Archive, and CatWISE2020 from IRSA. Rubin Observatory DP2 data
\citep{RubinDP2data} are available to Rubin data-rights holders through the Rubin Science Platform, and Legacy
Surveys DR10 from the Legacy Surveys web site \citep{Dey2019}. The archival images are available from the NOIRLab Astro Data Archive (DECam and Mosaic-II), the
CADC (CFH12K), SMOKA (HSC) and the ESO Science Archive Facility (VHS; \citealt{ESOVHS}), and the HSC-SSP PDR3 images of
the comparison fields of Section~\ref{sec:deep_core} from the HSC-SSP data release site \citep{Aihara2022}. Tables~\ref{tab:named}, \ref{tab:traceback},
\ref{tab:fulla}, \ref{tab:fullb} and \ref{tab:offeuclid} will be provided in machine-readable form, the last for all
76 targets of Section~\ref{sec:prospects}, with the finder charts of Figure~\ref{fig:offeuclid_finders} as a figure set.
Table~\ref{tab:deep_candidates} will be provided in machine-readable form for all 74 priority-1 candidates of
Section~\ref{sec:deep_core}, the 9 sources on the member band whose photometry favours membership but whose motions are
not measured, and the other 10 candidates of the T list. %

\appendix
\section{Astrometric validation}\label{app:astrometry}

\begin{figure*}
\centering
\includegraphics{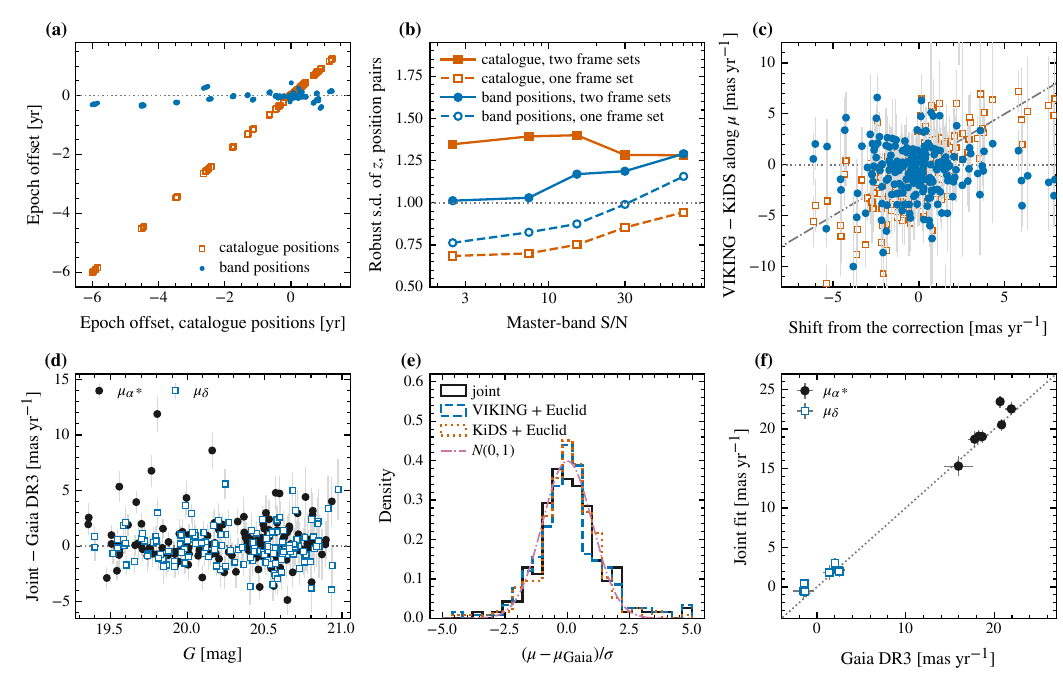}
\caption{Validation of the joint VIKING, KiDS and Euclid proper motions. (a) Apparent epoch offset of each
VIKING frame from its tie to Gaia DR3, for the catalogue positions (vermilion squares) and the band positions (circles).
(b) Robust standard deviation of the normalised position differences $z$ between two detections of a source in one
VIKING frame set, the band-merged images of one tile observation (open symbols, dashed), and in two frame sets (filled,
solid), against master-band S/N. (c) VIKING minus KiDS proper motion along the direction of motion against the shift
introduced by the correction; the slope falls from $0.96\pm0.07$ for the catalogue positions to $-0.04\pm0.04$ for the %
band positions (dash-dotted line: slope 1; symbols as in a). (d) Joint minus Gaia DR3 proper motion for the 145 stars with unsaturated %
VIS photometry ($G=19.4$ to 21.0). (e) Distributions of the normalised differences from Gaia DR3 for the joint, %
VIKING plus Euclid and KiDS plus Euclid fits, with a unit Gaussian; the joint fit has robust standard deviations of
1.07 and 1.15 in the two coordinates. (f) Joint against Gaia DR3 proper motions for the seven Blanco~1 Gaia members %
and candidates with DR1 counterparts. %
}
\label{fig:astrometry}
\end{figure*}

Three external samples test the joint proper motions (Figure~\ref{fig:astrometry}). The first is 145 Gaia DR3 stars %
that are unsaturated in VIS and as faint as the candidates ($G=19.4$--21.0). Fitted with their Gaia parallaxes, their %
joint motions differ from Gaia's by a median of $(+0.19\pm0.14,\,-0.14\pm0.13)$\,\masyr\ in $(\mu_{\alpha*},\mu_\delta)$, %
which bounds any common zero-point offset at about 0.3\,\masyr. The normalised differences have robust standard %
deviations of 1.07 and 1.15: the formal errors of these stars are 7--15\% too small; the errors of the faint %
candidates were not rescaled. The second sample is 130 background galaxies, whose median apparent motion in the %
VIKING+Euclid solution is $(-0.03\pm0.77,\,+0.14\pm0.74)$\,\masyr, consistent with zero. The third is the seven %
Gaia-selected Blanco~1 members and candidates with DR1 counterparts (Section~\ref{sec:field}). All seven satisfy the comoving test at 95\% with their joint motions. Their joint-minus-Gaia motions average $(+0.67,\,+0.59)$\,\masyr, with a scatter of (1.2, 1.0)\,\masyr. One star differs by $+2.8$\,\masyr\ in right ascension, a normalised residual of 3.7. %

The VIKING band positions of Section~\ref{sec:astrometry} remove the epoch errors of the catalogue positions. For the catalogue positions, 33\% of the 403 frames have effective epochs more than 0.3\,yr from nominal, with a maximum error of 6.0\,yr. With the band positions, the fraction is 3.5\% and the maximum error is 0.43\,yr. The tie residual of bright stars fell from 13.0 to 10.9\,mas. The %
error calibration on 4,016 position pairs gives normalised scatters of 1.28--1.40 for the catalogue positions at all %
signal-to-noise ratios. The adopted centroid scales increase with master-band signal-to-noise ratio (Figure~\ref{fig:astrometry}b): 1.01 at S/N 0--5 and 1.03 at S/N 5--10. At S/N 10--20 the scale is 1.19; it rises to 1.28 at 20--40 and 1.87 above 40. %

\section{Gravity indices}\label{app:gravity}

\begin{figure*}
\centering
\includegraphics{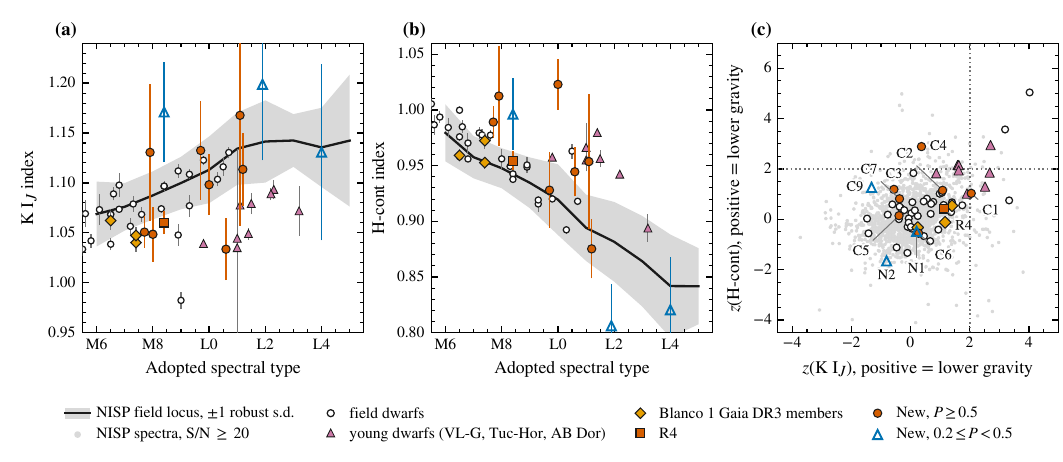}
\caption{NISP gravity indices. (a) KI$_J$ and (b) $H$-cont index against adopted spectral type, with the median
NISP field locus and its $\pm1$ robust standard deviation (line and grey band). Field dwarfs (open circles), young
dwarfs in Tucana-Horologium and AB~Doradus or classed VL-G (triangles), Blanco~1 Gaia DR3 members (diamonds), R4 and
the new candidates are shown. (c) Departures from the field locus in units of its scatter, $z(\mathrm{KI}_J)$ against
$z(H\mbox{-cont})$, for the candidates and for 1,908 NISP spectra with $\mathrm{S/N}\geq20$ (grey dots); positive %
values indicate lower gravity, and the dotted lines mark $z=2$. C8, whose type comes from its colours, is not shown; B2 has no
$H$-cont measurement and appears only in panel (a).}
\label{fig:gravity}
\end{figure*}

We measured KI$_J$ and $H$-cont on the co-added spectra using the definitions of \citet{Allers2013} implemented in SPLAT \citep{Burgasser2017}. The NISP field locus is built from 1,908 spectra with signal-to-noise %
ratios of at least 20 and types M6--L5, most of them field dwarfs, and each index is expressed as a z-score against it, %
positive for lower gravity (Figure~\ref{fig:gravity}).

Of the 3,002 typed objects, 1,389 have types outside M6--L5, where the locus is undefined, and 1,300 have spectra too %
noisy for a classification. The remaining 313 are classed as field-like (157), inconclusive (124), weak low-gravity hint (17) and %
low gravity indicated (15). Only two of the 15 sources classed as low gravity are compact with grade-A motions, %
and these move at 200 and 157\,\masyr, about 130 and 90\,\kms\ at their spectrophotometric distances: they are field %
dwarfs, possibly metal-poor.

Among the candidates, C1 shows a weak low-gravity hint ($z=2.0$ in KI$_J$ and 1.0 in $H$-cont, mean 1.54). C4's weak hint rests %
on $H$-cont alone ($z=2.9$) and coincides with a spike in the continuum window of that index. C2 and B2 are %
inconclusive, R4 and C6 read as field-like, and the spectra of C3, C5, C7, C9, N1 and N2 are too noisy for a %
classification. The three Gaia-selected Blanco~1 members with spectra, at M6.5--M7.4, read as field-like, with %
$z(\mathrm{KI}_J)=1.4$, 1.2 and 0.2. %

\section{Finder charts}\label{app:finders}

\begin{figure*}
\centering
\includegraphics{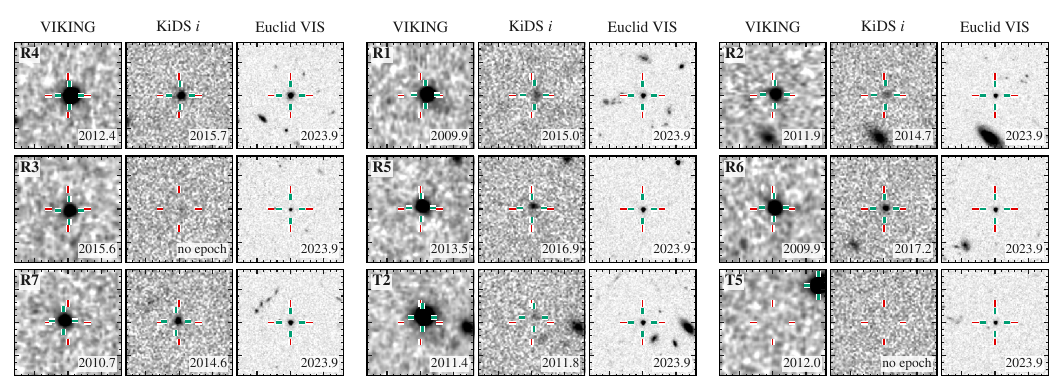}
\caption{Finder charts of R4 and the eight spectroscopically selected objects that move unlike the cluster. Each panel is
16\arcsec\ on a side, north up and east left, with major ticks every 5\arcsec; the three panels of an object share
one grid. VIKING: the CDS HiPS image of the survey in $J$ ($Y$ for R7), which can combine several epochs, retrieved
with hips2fits \citep{Fernique2015}. KiDS: one DR5 $i$-band image. Euclid: the DR1 VIS image. The red crosshair at
the centre marks the Euclid reference position of the proper-motion fit. The green crosshair marks the position at
the epoch printed in the panel, the mean epoch of that survey's positions in the fit, propagated with the proper
motion of Table~\ref{tab:named}; ``no epoch'' marks a survey without positions in the fit. Both crosshairs have open centres, so the
source beneath remains visible.}
\label{fig:findersa}
\end{figure*}

\begin{figure*}
\centering
\includegraphics{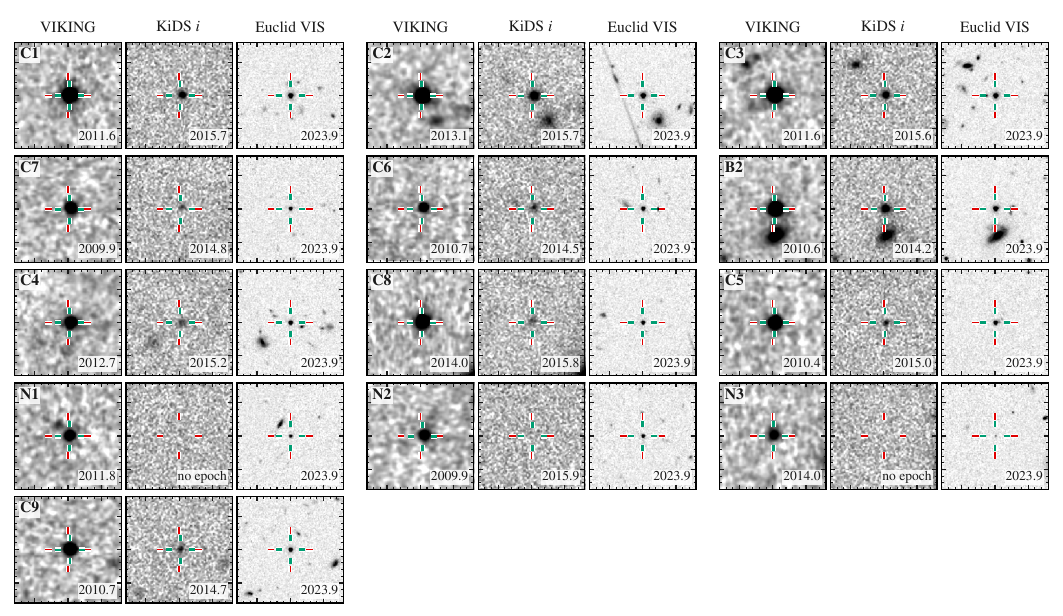}
\caption{Finder charts of the thirteen new candidates, as in Figure~\ref{fig:findersa}. The VIKING image of C6 is in
$H$.}
\label{fig:findersb}
\end{figure*}

We compare VIKING, KiDS and Euclid VIS images for each object on a common grid, with the same field of view, scale and orientation (Figures~\ref{fig:findersa} and \ref{fig:findersb}). The green crosshair follows the measured proper motion back from the Euclid epoch. For a member it
stays on the source at every epoch, while for T5, which moves at 0.74\arcsec\ per year, it lies near the corner of %
the VIKING panel, 12\,yr before the Euclid epoch. %

\section{The named objects}\label{app:table}

Tables~\ref{tab:fulla} and \ref{tab:fullb} give the positions, photometry, spectral measurements and membership terms
of the 22 named objects. The likelihood ratios include the Rubin epochs; the column
labelled ``no Rubin'' gives \Pg\ from the VIKING+KiDS+Euclid motions alone.

\begin{deluxetable*}{lrrrrrrrllrrr}
\tabletypesize{\scriptsize}
\tablecaption{Positions, photometry and spectral measurements of the 22 objects in Table~\ref{tab:named}.\label{tab:fulla}}
\tablehead{\colhead{Label} & \colhead{$\alpha$} & \colhead{$\delta$} & \colhead{Sep.} & \colhead{$I_\mathrm{E}$} & \colhead{$J_\mathrm{E}$} &
\colhead{$H_\mathrm{E}$} & \colhead{S/N} & \colhead{Template} & \colhead{Index} & \colhead{Gravity reading} & \colhead{$z_{\mathrm{KI}_J}$} & \colhead{$z_{H\mathrm{cont}}$}\\
\colhead{} & \colhead{(deg)} & \colhead{(deg)} & \colhead{(deg)} & \colhead{(AB)} & \colhead{(AB)} & \colhead{(AB)} & \colhead{} &
\colhead{type} & \colhead{type} & \colhead{} & \colhead{} & \colhead{}}
\startdata
R4 & 358.82850 & $-$29.96270 & 1.81 & 21.23 & 18.28 & 18.17 & 43.5 & M7 & M8.4 & field-like & 1.1 & 0.4 \\ %
R1 & 354.51550 & $-$31.09230 & 5.62 & 22.23 & 18.98 & 18.87 & 19.1 & L1 & \nodata & field-like & $-$0.9 & 1.1 \\ %
R2 & 352.87070 & $-$31.81900 & 7.13 & 23.06 & 19.50 & 19.17 & 11.1 & L3 & L2.4 & S/N too low & $-$1.2 & 2.1 \\ %
R3 & 356.54400 & $-$28.20590 & 4.22 & 23.46 & 19.73 & 19.47 & 9.7 & T0 & \nodata & type outside M6--L5 & \nodata & \nodata \\ %
R5 & 355.86180 & $-$29.34630 & 4.44 & 22.17 & 19.21 & 19.11 & 8.6 & L1 & M9.1 & S/N too low & $-$1.2 & 1.1 \\ %
R6 & 355.10490 & $-$31.21400 & 5.14 & 22.35 & 19.25 & 19.11 & 14.6 & L0 & M8.4 & S/N too low & $-$0.7 & $-$0.3 \\ %
R7 & 354.36830 & $-$33.55420 & 6.59 & 22.06 & 18.81 & 18.60 & 18.0 & L1 & L0.8 & field-like & $-$1.0 & 0.7 \\ %
T2 & 351.96240 & $-$30.91240 & 7.77 & 22.38 & 17.94 & 17.73 & 42.9 & T2 & \nodata & type outside M6--L5 & \nodata & \nodata \\ %
T5 & 353.11860 & $-$30.30990 & 6.75 & 23.40 & 18.54 & 18.89 & 18.7 & T5 & \nodata & type outside M6--L5 & \nodata & \nodata \\ %
\hline
C1 & 356.21060 & $-$30.65820 & 4.11 & 21.77 & 18.73 & 18.58 & 17.1 & L1 & L0.6 & weak low-gravity hint & 2.0 & 1.0 \\ %
C2 & 358.19090 & $-$29.76420 & 2.37 & 21.17 & 18.31 & 18.19 & 31.0 & M8 & M7.7 & inconclusive & 1.1 & 1.2 \\ %
C3 & 356.09630 & $-$30.34430 & 4.18 & 21.81 & 18.62 & 18.43 & 10.0 & L0 & M7.9 & S/N too low & $-$0.6 & 1.2 \\ %
C7 & 359.42840 & $-$31.75570 & 2.16 & 23.10 & 19.71 & 19.56 & 7.1 & M7 & L1.1 & S/N too low & $-$0.4 & 0.8 \\ %
C6 & 353.24800 & $-$33.16400 & 7.25 & 22.84 & 19.58 & 19.36 & 14.3 & M8 & L1.2 & field-like & 0.3 & $-$0.6 \\ %
B2 & 352.85250 & $-$32.59850 & 7.36 & 21.49 & 18.78 & 18.69 & 15.5 & M8 & \nodata & inconclusive & 1.0 & \nodata \\ %
C4 & 358.93890 & $-$30.07120 & 1.71 & 22.50 & 19.23 & 19.06 & 18.4 & M7 & L0 & weak low-gravity hint & 0.4 & 2.9 \\ %
C8 & 352.93240 & $-$29.66650 & 6.93 & 22.64 & 19.41 & 19.20 & 8.5 & L7 & \nodata & type outside M6--L5 & \nodata & \nodata \\ %
C5 & 354.54210 & $-$31.99480 & 5.81 & 22.64 & 19.47 & 19.26 & 11.4 & M8 & M9.7 & S/N too low & $-$0.4 & 0.1 \\ %
N1 & 358.04100 & $-$30.79570 & 2.60 & 24.01 & 20.32 & 20.02 & 6.8 & L4 & \nodata & S/N too low & 0.2 & $-$0.5 \\ %
N2 & 358.86600 & $-$31.49800 & 2.30 & 23.73 & 20.08 & 19.74 & 8.9 & L4 & L1.9 & S/N too low & $-$0.8 & $-$1.7 \\ %
N3 & 352.90680 & $-$29.67480 & 6.95 & 26.03 & 21.24 & 21.54 & \nodata & \nodata & \nodata & type outside M6--L5 & \nodata & \nodata \\ %
C9 & 356.42420 & $-$32.08890 & 4.37 & 22.73 & 19.57 & 19.41 & 12.7 & M8 & M8.4 & S/N too low & $-$1.3 & 1.3 \\ %
\enddata
\tablecomments{Positions (ICRS) and magnitudes from the Euclid DR1 MER catalogue: $I_\mathrm{E}$ from the VIS point-spread-function flux,
$J_\mathrm{E}$ and $H_\mathrm{E}$ from NISP template fits. Sep.: angular separation from the cluster centre. S/N: signal-to-noise
ratio of the co-added NISP spectrum. Template and index types as in Table~\ref{tab:named}; the index type is listed only where it is adopted.
$z$: gravity-index z-scores against the NISP field locus (positive for lower gravity; Appendix~\ref{app:gravity}).}
\end{deluxetable*}

\begin{deluxetable*}{lrrrrrrrrrrr}
\tabletypesize{\scriptsize}
\tablecaption{Membership terms for the 22 objects in Table~\ref{tab:named}.\label{tab:fullb}}
\tablehead{\colhead{Label} & \colhead{$\mathrm{LR}_\mu$} & \colhead{$\mathrm{LR}_\mathrm{type}$} & \colhead{$\mathrm{LR}_d$} &
\colhead{$\mathrm{LR}_\mathrm{grav}$} & \colhead{$P_\mathrm{gaia}$} & \colhead{$P_\mathrm{ctrl}$} & \colhead{$P_\mathrm{gaia}$} &
\colhead{$P_\mathrm{gaia}$} & \colhead{$P_\mathrm{gaia}$} & \colhead{Rubin} & \colhead{$i-J_\mathrm{E}$}\\
\colhead{} & \colhead{} & \colhead{} & \colhead{} & \colhead{} & \colhead{} & \colhead{} & \colhead{half prior} &
\colhead{no binaries} & \colhead{no Rubin} & \colhead{visits} & \colhead{(AB)}}
\startdata
R4 & 115 & 1.2 & 0.92 & 1.6 & 0.86 & 0.93 & 0.76 & 0.83 & 0.85 & 7 & 2.68 \\ %
R1 & $3.3\times10^{-15}$ & 1 & 1 & 1 & $<10^{-4}$ & $<10^{-4}$ & $<10^{-4}$ & $<10^{-4}$ & $<10^{-4}$ & 7 & 3.26 \\ %
R2 & $7.1\times10^{-4}$ & 1.4 & 1.3 & 1 & $<10^{-4}$ & $<10^{-4}$ & $<10^{-4}$ & $<10^{-4}$ & $<10^{-4}$ & 1 & 3.62 \\ %
R3 & 0.013 & 0.020 & 3.8 & 1 & $<10^{-4}$ & $<10^{-4}$ & $<10^{-4}$ & $<10^{-4}$ & $<10^{-4}$ & 0 & 3.91 \\ %
R5 & $2.2\times10^{-3}$ & 1.1 & 1.1 & 1 & $<10^{-4}$ & $<10^{-4}$ & $<10^{-4}$ & $<10^{-4}$ & $<10^{-4}$ & 8 & 2.90 \\ %
R6 & $1.4\times10^{-6}$ & 1.2 & 0.76 & 1 & $<10^{-4}$ & $<10^{-4}$ & $<10^{-4}$ & $<10^{-4}$ & $<10^{-4}$ & 3 & 3.10 \\ %
R7 & $1.5\times10^{-59}$ & 1.1 & 1.4 & 0.69 & $<10^{-4}$ & $<10^{-4}$ & $<10^{-4}$ & $<10^{-4}$ & $<10^{-4}$ & 3 & 3.22 \\ %
T2 & $3.3\times10^{-3}$ & 1 & 1 & 1 & $<10^{-4}$ & $<10^{-4}$ & $<10^{-4}$ & $<10^{-4}$ & $<10^{-4}$ & 0 & \nodata \\ %
T5 & 0.66 & 1 & 1 & 1 & $<10^{-4}$ & $<10^{-4}$ & $<10^{-4}$ & $<10^{-4}$ & $<10^{-4}$ & 0 & \nodata \\ %
\hline
C1 & 134 & 0.90 & 1.9 & 2.0 & 0.94 & 0.97 & 0.89 & 0.94 & 0.94 & 4 & 2.98 \\ %
C2 & 91 & 1.2 & 1.2 & 2.1 & 0.90 & 0.93 & 0.82 & 0.91 & 0.90 & 4 & 2.73 \\ %
C3 & 119 & 1.1 & 1.3 & 1 & 0.89 & 0.94 & 0.82 & 0.89 & 0.89 & 4 & 3.09 \\ %
C7 & 97 & 1.4 & 1.5 & 1 & 0.82 & 0.89 & 0.67 & 0.83 & 0.82 & 0 & 3.38 \\ %
C6 & 102 & 1.0 & 2.5 & 0.69 & 0.79 & 0.87 & 0.62 & 0.79 & 0.79 & 0 & \nodata \\ %
B2 & 111 & 1.2 & 1.0 & 1.5 & 0.68 & 0.77 & 0.48 & 0.68 & 0.68 & 0 & \nodata \\ %
C4 & 31 & 1.3 & 1.1 & 1.0 & 0.68 & 0.78 & 0.51 & 0.72 & 0.68 & 7 & 3.19 \\ %
C8 & 68 & 1 & 1 & 1 & 0.66 & 0.73 & 0.47 & 0.67 & 0.66 & 3 & 3.21 \\ %
C5 & 71 & 1.1 & 0.81 & 1 & 0.56 & 0.70 & 0.36 & 0.57 & 0.60 & 2 & 3.11 \\ %
N1 & 14 & 1.1 & 2.6 & 1 & 0.45 & 0.62 & 0.27 & 0.46 & 0.45 & 0 & \nodata \\ %
N2 & 30 & 1.4 & 0.58 & 1 & 0.41 & 0.44 & 0.24 & 0.44 & 0.41 & 0 & 3.69 \\ %
N3 & 3.6 & 1 & 1 & 1 & 0.29 & 0.35 & 0.16 & 0.21 & 0.29 & 0 & \nodata \\ %
C9 & 38 & 1.2 & 0.47 & 1 & 0.22 & 0.32 & 0.11 & 0.22 & 0.25 & 3 & 3.04 \\ %
\enddata
\tablecomments{Likelihood ratios (member over field) for proper motion (Gaia field model), spectral type, spectrophotometric distance and gravity indices; omitted terms are set to 1. Half prior: member normalisation halved.
No binaries: binary component removed from the member model. No Rubin: VIKING+KiDS+Euclid motions alone (approximate: the prior normalisation is applied as an odds factor). %
Rubin visits: retained DP2 $i$-band positions; $i-J_\mathrm{E}$ from the DP2 catalogue means.}
\end{deluxetable*}

\onecolumngrid
\section{Finder charts of the targets beyond the Euclid field}\label{app:offeuclid_finders}

\begin{figure*}
\centering
\includegraphics{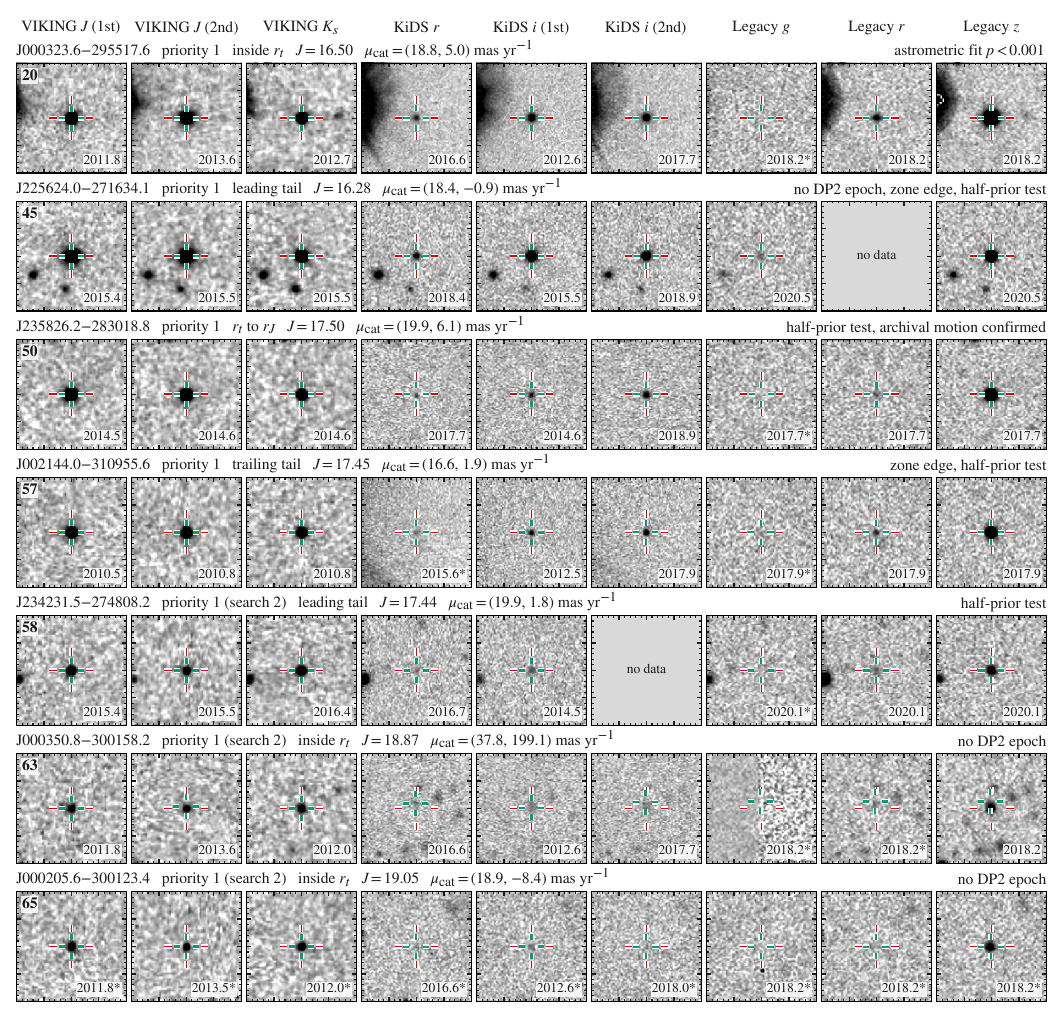}
\caption{Finder charts of the seven priority-1 targets with notes b--e in Table~\ref{tab:offeuclid}. Each row shows VIKING
$J$ at two epochs and $K_s$, KiDS $r$ and both $i$ passes, and Legacy Surveys DR10 $g$, $r$ and $z$, resampled onto
one 20\arcsec\ $\times$ 20\arcsec\ grid, north up and east left, with major ticks every 5\arcsec, centred on the
VIKING catalogue position (red crosshair). The green crosshair marks that position moved with the catalogue proper
motion of the search, $\mu_\mathrm{cat}$, to the epoch printed in the panel, in years; for the Legacy Surveys coadds the epoch is the midpoint of the
contributing exposures. Panels are tied to the VIKING catalogue frame through Legacy Surveys sources, and an asterisk
marks a panel with too few of them to be tied. Each row is labelled with the VIKING designation, priority including archival epochs (Section~\ref{sec:offeuclid_archival}), zone, VIKING $J$ and $\mu_\mathrm{cat}$. The survey-only priority is also given where it differs. Labels on the right identify the limitations listed in Table~\ref{tab:offeuclid}: an astrometric fit with $p<0.001$ (b), missing DP2 epoch (c), proximity to a zone edge (d), and a probability below the priority-1 threshold when the zone prior is halved (e). Both crosshairs have open centres, so the source beneath remains visible. Legacy
Surveys images: Legacy Surveys / D. Lang (Perimeter Institute). The complete figure set (10 images) covers all 76
priority-1 and priority-2 targets of the search and gives the outcome of the image inspection for each and, for the targets of note
a, that the archival epochs confirm the motion.}
\label{fig:offeuclid_finders}
\end{figure*}

Figure~\ref{fig:offeuclid_finders} shows the seven priority-1 targets with notes b--e in Table~\ref{tab:offeuclid}, %
whose priority depends on a poor astrometric fit, a missing DP2 epoch, a zone edge or the zone prior. The figure set in
the online journal shows all 76 priority-1 and %
priority-2 targets of the search in rank order, eight per image. In each row the nine panels share one grid, so the epochs can be
compared directly. For a source that moves with its catalogue motion the green crosshair stays on it in every panel,
as for ranks 3, 7, 22 and 69. For ranks 63 and 68 it drifts away from a source that stays at the VIKING position, %
which is how the images refute their catalogue motions; the archival epochs give rank 63 a cluster-like motion %
(Section~\ref{sec:offeuclid_archival}).

\figsetstart
\figsetnum{\ref{fig:offeuclid_finders}}
\figsettitle{Finder charts of the priority-1 and priority-2 targets beyond the Euclid field}
\figsetend

\clearpage
\onecolumngrid
\section{Images of the deep-search candidates}\label{app:deep_finders}

\begin{figure*}
\centering
\includegraphics{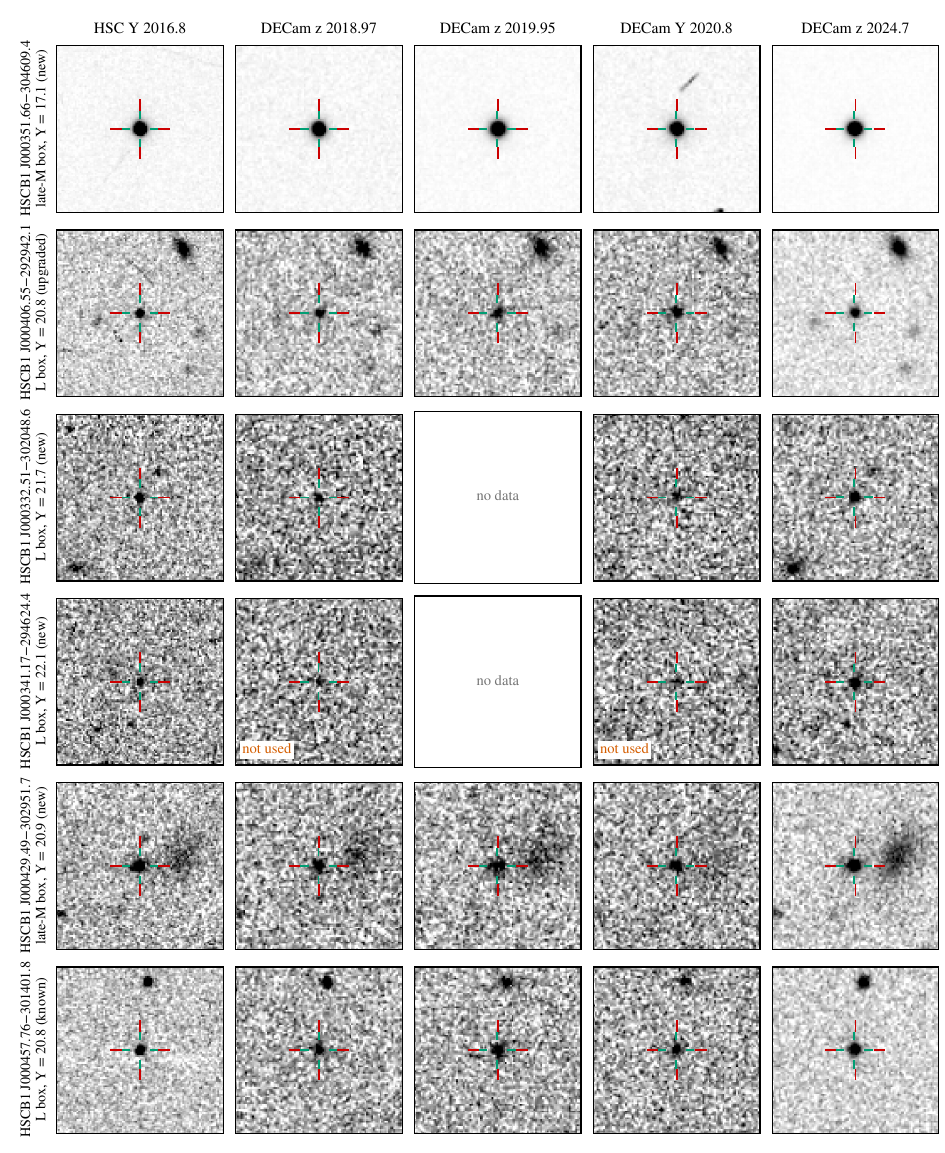}
\caption{HSC and DECam images of the four priority-1 candidates without earlier identification, the one candidate that
only the VIKING-based search had listed, at priority~3, and one known L-box candidate for comparison. Each row shows a %
20\arcsec\ field on a common grid (north up, east left) at the HSC epoch 2016.8 and four DECam epochs. The red %
crosshair marks the HSC position at the centre, and the green crosshair the position that the measured motion predicts
at the epoch of the image; both leave the source uncovered. A stamp marked ``not used'' shows a DECam frame that the
motion fit does not use, and one marked ``no data'' a night without a DECam measurement of the source. The label of each row gives the designation (with the prefix HSCB1 of %
Table~\ref{tab:deep_candidates}), the colour box, $Y_\mathrm{HSC}$ and whether the candidate is new, known or upgraded from
the VIKING-based search. The late-M-box candidate at $Y_\mathrm{HSC}=20.9$ lies 5\arcsec\ east of the centre of a %
diffuse galaxy.}
\label{fig:deep_finders}
\end{figure*}

Figure~\ref{fig:deep_finders} shows the four priority-1 candidates of Section~\ref{sec:deep_core} without earlier %
identification, the candidate that only the VIKING-based search had listed, and one known L-box candidate. One of the
four, at $Y_\mathrm{HSC}=17.1$, is among the 39 bluer candidates. For a source that moves with its measured motion, the %
green crosshair stays on it in every panel.

\section{Versions of the deep selection}\label{app:deep_versions}

\begin{deluxetable}{lcccc}[ht]
\tablecaption{Sensitivity of the deep selection to profile calibration and frame rejection.\label{tab:deep_versions}}
\tablehead{\colhead{Quantity} & \colhead{1} & \colhead{2} & \colhead{3} & \colhead{4}}
\startdata
Priority-1 candidates & 36 & 51 & 74 & 74 \\ %
\quad of types M7--L6 & 27 & 30 & 35 & 35 \\ %
Chance count, all & 2.31 & 2.66 & 2.56 & 2.44 \\ %
Purity, all & 0.936 & 0.948 & 0.965 & 0.967 \\ %
Chance count, M7--L6 & \nodata & 1.81 & 1.75 & 1.53 \\ %
95th percentile, M7--L6 & \nodata & 5.9 & 5.9 & 4.9 \\ %
M7--L6 lower bound & \nodata & 15.69 & 19.97 & 20.97 \\ %
\enddata
\tablecomments{Chance counts are means over the 32 velocity-shifted control clusters. The bound is the one-sided 95\% Poisson lower limit on the M7--L6 count minus the 95th percentile of the control counts, before rounding down. Selection 4 gives the adopted bound of at least 20, while selection 3 gives at least 19; M7--L6 control counts are unavailable for selection 1.}
\end{deluxetable}

The alternatives in Table~\ref{tab:deep_versions} measure how stellar-profile calibration and DECam frame rejection affect the core sample. In selection 1, cosmic-ray rejection clips the cores of bright stars in the coadds. Selection 2 avoids this clipping but calibrates the stellar-profile interval only on injected point sources, rejecting 8.5--72\% of Gaia stars at S/N 100--3000. Both omit the bright-star calibration used in selections 3 and 4. These profile losses reduce the priority-1 counts to 36 and 51, compared with 74 when the calibration includes Gaia stars; the M7--L6 counts are 27, 30 and 35, respectively. Most of the additional objects have brightnesses consistent with M4--M6 members.

We calibrate the bright-star profile interval by adding a term proportional to S/N, measured on 1842 Gaia stars. This interval keeps 97--99\% of Gaia stars from S/N 60 to above 3000, and 96--98\% from S/N\,100 for stars excluded from the PSF models. Of the calibration stars, 1731 also contribute to the PSF models and 31 are priority-1 candidates. Restricting the calibration to the 108 stars in neither group gives the same priority-1 list. Five M7--L6 targets from the VIKING-based search are recovered with the Gaia-calibrated interval but lost with the injection-only interval. Of the 27 VIKING-based priority-1 targets inside 0.75\arcdeg\ (Table~\ref{tab:offeuclid}), 26 are recovered at priority~1; because these targets helped diagnose the profile losses, their recovery is coupled to the calibration.

The profile calibration also changes the expected contamination. Selections 2 and 3 predict 138 and 157 field sources from the comparison fields, while their velocity-shifted controls select 2.7 and 2.6 objects on average. The increase in purity from 0.948 to 0.965 partly reflects the lower late-M prior, 0.24 instead of 0.28. For the candidate at $Y_\mathrm{HSC}=19.4$, this prior change lowers $P$ from 0.530 to 0.479, below priority~1. Across the four selections, the mean chance counts are 2.3, 2.7, 2.6 and 2.4, despite the larger differences in bright-star recovery.

The adopted selection 4 rejects elongated or truncated DECam measurements and positional outliers. Among nights with at least three frames, 2.3\% of measurements lie more than 4$\sigma$ from the other frames of the night. An axis-ratio limit of 1.5, chosen from the frame outlier rates, removes 1052 of the 20\,096 frames for elongation or truncation, and the positional test rejects 84 more. Using positional rejection alone gives 75 priority-1 candidates, 36 in the M7--L6 subset, a mean chance count of 2.75 and a lower bound of 20.8. Adding the elongation cut gives 74, 35, 2.44 and 21.0, respectively. The price is reduced sensitivity at the faint end: the elongation cut removes 40\% of frames at S/N 5--7 and lowers the L3 motion-test efficiency from 0.67 to 0.54.

Frame rejection matters particularly for the faint candidates. With it, the L-box source at $Y_\mathrm{HSC}=22.1$ qualifies, while the source at 21.9, of priority~4 in the VIKING-based search, loses its measured motion. The late-M-box candidate at 20.9 becomes comoving after two of its four DECam frames from 2024 are removed: one contains an image artefact and the other has an axis ratio of 2.6. The remaining pair is combined without an outlier test. Removing only the artefact frame gives a marginally comoving motion ($\chi^2=5.7$ for a limit of 6.2) and $P=0.82$, with the motion changing when an epoch group is omitted. This dependence on the frame selection motivates the additional epoch requested in Section~\ref{sec:deep_core}.

The IMF comparison uses the M7--L6 subset, the lower-bound construction of Section~\ref{sec:deep_core}, and sensitivity ranges in age, normalisation and motion-test efficiency. The latter includes magnitude windows of 0.1 and 0.25\,mag; the comparison also includes the bluer candidates as a separate case. For the spatial comparison, matching $z$ depth in selection 3 gives a main-list excess of 67 without weights and 80 with weights from all sources. We use the unweighted comparison and weights from $J$-detected sources in Section~\ref{sec:deep_core}; the all-source weighting is a sensitivity case.

\clearpage
\bibliography{references}{}
\bibliographystyle{aasjournalv7}

\end{document}